\documentclass[%
reprint,
twocolumn,
superscriptaddress,
showpacs,preprintnumbers,
nofootinbib,
aps,
prd,
]{revtex4-1}

\usepackage{axodraw2,pix}
\usepackage{epsfig}
\usepackage{amsfonts}
\usepackage{bbm,bm}

\usepackage[breaklinks,urlbordercolor={1 1 1}]{hyperref}

\AtBeginDvi{
  
}

\usepackage[utf8]{inputenc}

\usepackage{graphicx}
\usepackage[dvips]{xcolor}
\graphicspath{ {./figures/} }

\usepackage{comment}
\usepackage{amsmath}

\usepackage{multirow}

\newcommand\RE{\operatorname{Re}}

\newcommand\MSbar{$\overline{\text{MS}}$ } 
\newcommand\Veff{V_\text{eff}} 
\newcommand\he[1]{#1^\dagger}
\newcommand\gr[1]{\mathrm{#1}}

\let\vec\mathbf 

\newcommand\sumint[1]{\int\kern-1.5em\sum\nolimits_{#1}}

\newcommand{\nn}{\nonumber \\}
\newcommand{\rmi}[1]{{\mbox{\scriptsize #1}}}
\newcommand{\rmii}[1]{{\mbox{\tiny\rm{#1}}}}

\newcommand{\nf}{n_{\rm f}}
\newcommand{\Nc}{N_{\rm c}}

\newcommand{\Tc}{T_{\rm c}}
\newcommand{\mD}{m_\rmii{D}}
\newcommand{\gY}{y_t}

\newcommand{\gs}{g_\rmi{s}}

\newcommand{\Tint}[1]{{\hbox{$\sum$}\!\!\!\!\!\!\!\int\,}_{\!\!\!\!\raise-0.9ex\hbox{$\scriptstyle{#1}$}}}
\newcommand{\Tinti}[1]{{{\Sigma}\!\!\!\!\raise0.3ex\hbox{$\int$}_\rmii{${#1}$}}}
\newcommand{\Tintip}[1]{{{\Sigma'}\!\!\!\!\!\raise0.3ex\hbox{$\int$}_\rmii{${#1}$}}}

\newcommand\ct{c_{\bar{\theta}}}
\newcommand\st{s_{\bar{\theta}}}

\def\scfc{0.7}  
\def\phgt{21}   
\def\pwc{21}    
\def\pwcc{42} 
\newcommand{\PIC}[4]{\;\parbox[c]{#2 pt}{\begin{picture}(#2,#3)(0,0)
\SetWidth{1.0}\SetScale{#4} #1 \end{picture}}\;}
\renewcommand{\pic}[1]{\PIC{#1}{\pwc}{\phgt}{\scfc}}
\renewcommand{\picb}[1]{\PIC{#1}{\pwcb}{\phgt}{\scfc}}
\renewcommand{\picc}[1]{\PIC{#1}{\pwcc}{\phgt}{\scfc}}

\begin{document}

\newcommand{\HEL}{\affiliation{%
Department of Physics and Helsinki Institute of Physics,
PL 64, FI-00014 University of Helsinki,
Finland }}

\newcommand{\NOR}{\affiliation{Nordita,
KTH Royal Institute of Technology and Stockholm University, \\
Roslagstullsbacken 23,
SE-106 91 Stockholm,
Sweden }}

\newcommand{\TSU}{\affiliation{Tsung-Dao Lee Institute \& School of Physics and Astronomy, Shanghai Jiao Tong University, Shanghai 200240, China }}

\newcommand{\SHA}{\affiliation{Shanghai Key Laboratory for Particle Physics and Cosmology, Key Laboratory for Particle Astrophysics and Cosmology (MOE), Shanghai Jiao Tong University, Shanghai 200240, China}}

\title{Singlet-assisted electroweak phase transition at two loops}

\preprint{HIP-2021-8/TH}
\preprint{NORDITA 2021-011}

\author{Lauri~Niemi} \email{lauri.b.niemi@helsinki.fi} \HEL

\author{Philipp Schicho} \email{philipp.schicho@helsinki.fi} \HEL

\author{Tuomas V. I.~Tenkanen} \email{tuomas.tenkanen@su.se} \NOR \TSU \SHA

\begin{abstract}

We investigate the electroweak phase transition in the real-singlet extension of the Standard Model at two-loop level, building upon existing one-loop studies.
We calculate the effective potential in the high-temperature approximation and detail the required resummations at two-loop order.
In typical strong-transition scenarios, we find deviations of order $20\% - 50\%$ from one-loop results in transition strength and critical temperature for both one- and two-step phase transitions.
For extremely strong transitions, the discrepancy with one-loop predictions is even larger, presumably due to sizable scalar couplings in the tree-level potential.
Along the way, we obtain a dimensionally-reduced effective theory applicable for non-perturbative lattice studies of the model.

\end{abstract}

\maketitle

\newpage

\section{Introduction}

Phase transitions occurring in various extensions of the Standard Model (SM) have
attracted a lot of attention in recent years.
Many candidate theories for physics beyond the SM (BSM) involve scalar fields in
addition to the SM Higgs, and it is interesting to ask if there can be
phase transitions associated with the scalar potential.
A cosmological first-order phase transition around temperatures comparable to
the electroweak (EW) scale could produce gravitational waves with observational
prospects at future detectors~\cite{Weir:2017wfa,Caprini:2019egz}, and may
provide the necessary conditions for EW baryogenesis~\cite{Kuzmin:1985mm,Shaposhnikov:1987tw} (see~\cite{Morrissey:2012db} for a review).

In the SM, the phase structure of the EW sector is well established:
there is no phase transition associated with the Higgs mechanism~\cite{Kajantie:1996mn,Csikor:1998eu}.
Instead, the early universe smoothly interpolates between
the high-temperature ``symmetric'' and ``broken'' Higgs regimes, and a first-order discontinuity between
the two would require the Higgs particle to be unphysically light
(mass $\lesssim 70$ GeV).
For any BSM scenario to change this conclusion, drastic modifications to
the Higgs potential are necessary. First-order electroweak phase transition (EWPT) can be achieved by coupling the Higgs to new scalars at sufficient strength, or the phase structure itself can be more complicated with the theory undergoing multiple phase transitions at temperatures comparable to the EW scale. In both cases the new particles cannot be much heavier than the Higgs, making the EWPT a potential target at present and planned collider experiments \cite{Ramsey-Musolf:2019lsf}.
For cosmology, one is mainly interested in strong transitions with large latent heat available for the production of gravitational waves, or with significant discontinuity in the Higgs VEV ($v / T \gtrsim 1$) as required for efficient baryogenesis.

Unfortunately, studying the EWPT with perturbation theory is complicated because (i) first-order transitions in simple scalar extensions typically require $\mathcal{O}(1)$ couplings among the scalars, and (ii) finite-temperature perturbation theory is sensitive to infrared (IR) physics and has a different structure than at zero temperature, often leading to slow convergence. Because of the second point, perturbation theory breaks down near second-order transitions where the Higgs correlation length diverges, making it impossible to determine the order of the EWPT with perturbative methods alone. Hence, when applying perturbation theory to first-order transitions one simply assumes that the transition remains first order also at the non-perturbative level, and that predictions are not significantly altered by higher order effects. Because of points (i) and (ii), it is far from obvious whether such assumptions are justified in typical BSM settings. 

The situation can be contrasted to that of the $\gr{SU(2)}$-Higgs theory relevant for the SM, where the dominant contributions to the free energy come from gauge loops. These are under perturbative control for $v/T \gtrsim 1$, although there are still quantitatively important corrections at two-loop order \cite{Arnold:1992rz}. In many BSM scenarios, corrections from the scalar sector dominate over the gauge loops, and predictions may well be sensitive to higher-order scalar diagrams even when the transition is strong.

The standard tool for studying the EWPT in BSM settings is the one-loop effective potential with resummation of thermal ``daisy'' diagrams. The purpose of this paper is to extend this calculation to two-loop level, and scrutinize the effects of the two-loop corrections on basic characteristics of the EWPT (critical temperature, latent heat, $v/T$). We work in the real-singlet extension of the Standard Model which we shall refer to as xSM, following~\cite{Profumo:2014opa}. Several one-loop analyses of the EWPT exist in this minimalistic yet realistic BSM model, see~\cite{Profumo:2007wc,Ashoorioon:2009nf,Espinosa:2011ax,Profumo:2014opa,Curtin:2014jma,Fuyuto:2014yia,Kozaczuk:2015owa,Chala:2016ykx,Beniwal:2017eik,Kurup:2017dzf,Chen:2017qcz,Chiang:2018gsn,Carena:2019une,Papaefstathiou:2020iag}
and references therein, but we are not aware of existing two-loop calculations in the finite-$T$ context. Two-loop studies in other scalar extensions and in the MSSM have reported on considerable deviations from the one-loop behavior~\cite{Espinosa:1996qw,Bodeker:1996pc,Laine:2000rm, Laine:2017hdk, Kainulainen:2019kyp,Niemi:2020hto}, motivating a detailed two-loop investigation of the xSM as well.

Our calculation proceeds in two stages.
First, we apply effective field theory (EFT) methodology to integrate out
hard thermal loops (loops involving thermal excitations at the scale $\pi T$)
in the static limit, leaving a three-dimensional (3d) EFT describing
the long-distance physics of the phase transition~\cite{Appelquist:1981vg,Ginsparg:1980ef}. This dimensional reduction automatically incorporates the required resummations while also summing a subset of higher-order corrections using the renormalization group (RG).
We then calculate the effective potential in the resulting EFT.
The 3d EFT constructed in this paper can also serve as a starting point for future
non-perturbative lattice studies \cite{Farakos:1994xh} of the xSM phase structure.

Applying the two-loop effective potential, we compute the strength and critical temperature of the EWPT in different scenarios, including two-step and purely radiatively generated transitions. The analysis is concentrated on a handful of benchmark points to better illustrate the impact of two-loop corrections on physical predictions. The main takeaway is that the corrections are significant even if the transition occurs through a tree-level barrier, and should be taken into account if reliable predictions of \textit{e.g.}\ the associated gravitational-wave background are sought.

The outline of this paper is as follows. We begin by introducing the model in Section~\ref{sec:model}. Section~\ref{sec:review} reviews some formal aspects of thermal resummation and high-$T$ dimensional reduction that form the backbone for understanding the remainder of the paper. In Section~\ref{sec:power-counting} we introduce a formal power counting scheme for various parameters in the theory and discuss subtleties specific to the high-$T$ behavior of the xSM. Sections~\ref{sec:DR} and \ref{sec:A0} contain details of the EFT construction and collect the relevant 4d $\rightarrow$ 3d matching relations. In Section~\ref{sec:Veff-main} we calculate the resummed effective potential using the EFT approach. Finally, our results concerning the importance of two-loop corrections are presented in section~\ref{sec:results}. Section~\ref{sec:conclusions} concludes. There are two appendices:
Appendix~\ref{sec:renormalization} details our renormalization prescription and loop-corrected relations to physical observables.
Appendix~\ref{sec:Veff-2loop} calculates the two-loop correction to the effective potential.

\section{Model}
\label{sec:model}

We work on a model consisting of the usual SM field content and an additional real scalar $S$, that is a singlet under the SM gauge group. The Lagrangian in Euclidean signature is
\begin{align}
\label{eq:Lag-4d}
\mathcal{L} =& \mathcal{L}_\text{YM} + \mathcal{L}_\text{fer} + y_t (\bar{q_t} i \sigma_2 \phi^* t + h.c.) \nonumber \\
& + |D_\mu \phi|^2 + \frac12 (\partial_\mu S)^2 + V(\phi, S)
\end{align}
where $\mathcal{L}_\text{YM}$ is the Yang-Mills part for the SM gauge fields (including hypercharge), $\mathcal{L}_\text{fer}$ contains the fermion kinetic terms and $y_t$ is the top Yukawa coupling. Yukawa couplings of the light fermions are orders of magnitude smaller than other parameters in the theory and are neglected. Covariant derivative for the Higgs doublet $\phi$ is
$D_\mu \phi = (\partial_\mu - \frac12 i g \sigma_a A_\mu^a - \frac12 i g' B_\mu)\phi$,
where $g$ and $g'$ are the $\gr{SU(2)}_L$ and $U(1)_Y$ gauge couplings respectively.
The scalar potential is
\begin{align}
\label{eq:pot}
V(\phi, S) =& m^2_\phi \he\phi\phi + \lambda (\he\phi\phi)^2 + b_1 S + \frac12 m^2_S S^2 \nonumber \\
& + \frac13 b_3 S^3 + \frac14 b_4 S^4  + \frac12 a_1 S \he\phi\phi + \frac12 a_2 S^2 \he\phi\phi.
\end{align}

As usual, the zero-temperature theory can be studied by assuming $m_\phi^2 < 0$ and
specifying components of the doublet as
\begin{align}
\label{eq:Higgs_param}
\phi = \begin{pmatrix} 
G^+ \\
\frac{1}{\sqrt{2}} \left( v + h + i G\right) 
\end{pmatrix}
\end{align}
where $v^2  = -m^2_\phi / \lambda$ is the gauge-fixed Higgs vacuum expectation value (VEV) at tree level, and $G^\pm, G$ are the would-be Goldstone modes. The singlet VEV and hence the linear coupling $b_1$ can be absorbed into a redefinition of the tree-level parameters, \textit{i.e.}\ the theory contains five free parameters instead of six. At tree level, the condition $\langle S\rangle = 0$ fixes $b_1 = -a_1 v^2 / 4$.
In the limit $b_1, a_1, b_3 \rightarrow 0$ the theory becomes symmetric under $S \rightarrow -S$; we shall call this the $Z_2$ symmetric case. Outside the $Z_2$ symmetric limit the physical scalar excitations $h_1, h_2$ are mixtures of $h$ and $S$:
\begin{align}
\begin{pmatrix} 
h_1 \\ h_2
\end{pmatrix}
= 
\begin{pmatrix} 
\cos\theta & -\sin\theta \\
\sin\theta & \cos\theta
\end{pmatrix}
\begin{pmatrix} 
h \\ S
\end{pmatrix}.
\end{align}
The angle $\theta$ is chosen to diagonalize the mass matrix at tree level and we will identify the lighter $h_1$ as the SM-Higgs-like excitation with mass $125$ GeV.

The input parameters we use for our analysis are $b_3, b_4, a_2, \sin\theta$ and the masses of $h_1, h_2$ (the VEV $v$ can be fixed using the Fermi constant). Inverting the defining tree-level equations for these leads to the following expressions for the parameters in (\ref{eq:pot}):
\begin{align}
\label{eq:params-tree1}
m^2_\phi =& -\frac14 \left( m^2_{h_1} + m^2_{h_2} - (m^2_{h_2} - m^2_{h_1})\cos2\theta \right)  \\
m^2_S =& \frac12 \left( m^2_{h_1} + m^2_{h_2} + (m^2_{h_2} - m^2_{h_1})\cos2\theta - a_2 v^2 \right) \\
b_1 =& -\frac14 |v| (m^2_{h_2} - m^2_{h_1}) \sin2\theta \\
a_1 =& |v|^{-1} (m^2_{h_2} - m^2_{h_1}) \sin2\theta \\
\label{eq:params-tree2}
\lambda =& \frac14 v^{-2} \left( m^2_{h_1} + m^2_{h_2} - (m^2_{h_2} - m^2_{h_1})\cos2\theta \right).
\end{align}
Without loss of generality, $\sin\theta$ can be restricted to values $|\sin\theta| < 1/\sqrt{2}$~\cite{Profumo:2007wc}.

These relations obtain corrections at higher orders of perturbation theory. At one-loop, the corrections are parametrically of the same order as two-loop corrections to the thermal effective potential. Hence, a two-loop investigation of the finite-temperature behavior should be accompanied by a one-loop renormalization of the zero-temperature theory.
This is a separate calculation from the main focus of the paper, and we perform it in Appendix \ref{sec:renormalization} using the \MSbar scheme. 
This fixes our renormalized parameters at the $Z$ boson pole. The main effect of the one-loop improvement is to shift the singlet mass parameter $m_S^2$, by as much as $50\%$ relative to the tree-level result in some of our benchmark points below. The parameters $b_1$ and $\lambda$ also get considerable corrections.

The input parameters are constrained experimentally by electroweak precision measurements and direct searches for new physics~\cite{Robens:2015gla,Adhikari:2020vqo}. We will study the theory near its $Z_2$ symmetric limit, see the discussion in Section~\ref{sec:power-counting} below, taking $|\sin\theta| < 0.1$ and assuming the new excitation to be heavier than the SM Higgs.
The remaining parameter space is then relatively unconstrained by current experiments~\cite{Curtin:2014jma,Chen:2017qcz,Gould:2019qek,Papaefstathiou:2020iag}.
We will not consider experimental constraints further in this paper as our focus is on the finite-$T$ behavior of the theory.

\section{Review of thermal resummations and high-$T$ effective theories}
\label{sec:review}

Moving on to the finite-temperature theory, let us briefly review the premise of thermal corrections that are important for phase transitions. Much of the discussion here is collected from refs.~\cite{Arnold:1992rz,Farakos:1994kx,Kajantie:1995dw,Bodeker:1996pc}. Readers familiar with thermal resummations and dimensionally-reduced theories may want to skip this section.

Equilibrium quantum field theory is equivalent to a Euclidean field theory with a compact time direction, where the periodicity of the imaginary time coordinate $\tau$ is identified with $1/T$. The fields have momentum-space representations of the form
\begin{align}
\phi(x) = T \sum_{n=-\infty}^{\infty} \int \frac{d^3 p}{(2\pi)^3} \phi(\omega_n, \mathbf{p}) e^{i \omega_n \tau} e^{-i \mathbf{p} \cdot \mathbf{x}},
\end{align}
where the Matsubara frequency is $\omega_n = 2\pi n T$ for bosons and $\omega_n = (2n + 1)\pi T$ for fermions. The bosonic zero modes with $\omega_n = 0$ carry no momentum in the time direction, so their dynamics is effectively three-dimensional. At length scales $\gg 1/T$ the zero-mode contributions dominate, leading to the well-known IR problems in perturbation theory because the leading interaction terms in 3d have dimensionful couplings~\cite{Jackiw:1980kv}. In contrast, modes with $\omega_n \neq 0$ (non-zero Matsubara modes) are safe in the IR due to additional factors of $(\pi T)^2$ in the propagators.

Qualitatively, the EWPT can occur because of a $T$-dependent correction to the Higgs mass parameter from thermal loops, schematically $m^2_\phi \rightarrow m^2_\phi(T) = m^2_\phi + g^2 T^2$. At large enough $T$ the VEV is relaxed to zero, and at the critical temperature the ``symmetric'' and ``broken'' phases have equal free energies.
In the absence of a singlet VEV, this happens, at the mean field level, when $m^2_\phi(T) = 0$ and requires a cancellation between the vacuum and thermal masses.\footnote{In the general xSM, radiative corrections will generate a non-zero VEV even if we set $\langle S \rangle = 0$ at tree level.
  Consequently, the condition for the critical temperature changes.
  The arguments here are not qualitatively affected by such loop effects.
}
Such large corrections to the tree-level behavior call for resummation of specific thermal corrections. For the one-loop effective potential this is achieved by the ring (``daisy'') resummation where the propagators are replaced with thermally-corrected ones~\cite{Carrington:1991hz}. 

At two-loop level, a consistent prescription has been given in ref.~\cite{Arnold:1992rz}: thermal masses for the zero modes are generated by integrating out the heavier modes with $\omega_n^2 \geq (\pi T)^2$, and resummation of the non-zero modes is not needed at all. Such prescription is possible in the formal high-$T$ limit where the IR degrees of freedom are lighter than $\pi T$. Outside this limit, there is no simple way of choosing thermal masses for resummation and the procedure becomes ambiguous beyond one-loop level~\cite{Laine:2017hdk}.
In many cases the high-$T$ assumption is justified because the (thermally corrected) Higgs mass is parametrically small compared to $T$ near the critical temperature, and possible BSM excitations cannot be arbitrarily heavy if they are to produce a first-order EWPT \cite{Ramsey-Musolf:2019lsf}.

To understand the necessary resummations beyond one-loop order, it is convenient to reformulate the problem in the language of effective field theory. The finite-$T$ theory contains several mass scales in addition to those present at zero temperature. The most obvious one is the scale $\pi T$, characterizing the non-zero Matsubara modes. There are also the scales associated with Debye screening of electric and magnetic fields, parametrically $gT$ and $g^2 T$, respectively. The latter appears in non-Abelian gauge theory and is inherently non-perturbative~\cite{Linde:1980ts}. Conventional terminology for the scales $\pi T$, $gT$,  and $g^2 T$ is ``hard'', ``soft'' and ``ultrasoft'' (or sometimes ``superheavy'', ``heavy'' and ``light''; we will stick to the former). In a weakly coupled theory, these are all distinct scales. For scalars, a similar hierarchy of scales between the zero- and non-zero Matsubara modes is present in a window around $\Tc$ where the thermal mass is comparable to the vacuum one.

The strategy is now as follows. Making use of the thermal scale hierarchy, we construct an effective theory for the Matsubara zero-modes by integrating out all field modes with $\omega_n \neq 0$, which includes all fermionic fields. This leaves a 3d theory for the zero modes, describing the static properties of the finite-$T$ theory at length scales $\gtrsim (\pi T)^{-1}$.
The effects of the hard scale are incorporated in tree-level parameters of the EFT, and the process of thermal resummation at \textit{any} order of perturbation theory is then equivalent to calculating the EFT parameters to corresponding accuracy. Consequently, the effective potential calculated \textit{within} the dimensionally-reduced EFT reproduces the resummed high-$T$ effective potential~\cite{Farakos:1994kx}.

The form of the EFT is governed by 3d gauge invariance. In particular, under static gauge transformations the gauge-field time components transform like adjoint scalar fields. They generate a thermal mass due to Debye screening of electric fields in the thermal medium. Therefore, the dimensionally-reduced theory corresponding to (\ref{eq:Lag-4d}) has an action of the form\footnote{The integration over hard modes also produces a $T$-dependent constant term, a ``cosmological constant", to the effective action, corresponding to a renormalization of the free energy~\cite{Braaten:1995cm}. It plays no role in our discussion of phase transitions where the relevant quantities depend only free-energy differences.}
\begin{align}
\label{eq:3d-schematic}
S_\text{3d} =& \frac{1}{T} \int d^3x \Big\{ \mathcal{L}_\text{YM}^\text{(3d)} + |D_i \phi|^2 + \frac12 (\partial_i S)^2 + V(\phi, S) \nonumber \\
& + \frac12 (D_i A_0^a)(D_i A_0^a) + \frac12 \mD^2 A^a_0 A^a_0  \nonumber \\ 
& + \text{(scalar interactions with }A_0) \nonumber \\ 
& + (\text{operators of dimension 5 and higher})
\Big\}
\end{align}
and is valid in the high-$T$ limit.
Here $\mathcal{L}_\text{YM}^\text{(3d)}$ is the Yang-Mills part in three spatial dimensions and $A_0$ is the adjoint scalar corresponding to the time component of the $\gr{SU(2)}_L$ gauge field; the mass $\mD$ is the Debye screening mass.
The corresponding scalars for the hypercharge field and QCD gluons are not shown for simplicity.
The potential $V(\phi, S)$ is of the same form as in (\ref{eq:pot}) but with modified parameters. The overall factor $1/T$ comes from a trivial integration over the imaginary time and can be absorbed in a rescaling of the fields and couplings. At leading order one matches masses at one loop and couplings at tree level, which is precisely the standard procedure for daisy resummation.

The higher-dimensional operators in (\ref{eq:3d-schematic}) are suppressed by
powers of the ultraviolet (UV) scale $\pi T$, but Ref.~\cite{Landsman:1989be} pointed 
out that their effects do not fully decouple even in the $T\rightarrow \infty$ limit.
In many cases it is still a good approximation to neglect them, provided that all mass scales appearing in
the EFT (including cubic couplings) are small compared to $\pi T$ \cite{Farakos:1994xh}. It should be emphasized that the 3d approach is not suited for describing heavy BSM fields if their zero-mode masses are comparable to $\pi T$. However, thermal effects from such fields are Boltzmann suppressed and can be incorporated by integrating out the field together with the hard Matsubara modes. It is also difficult to obtain a first-order EWPT in such a scenario, unless there are
non-negligible effects from the higher-dimensional operators arising from
integration over the heavy field~\cite{deVries:2017ncy,Gould:2019qek}.

If the higher-dimensional operators are dropped by truncation, the remaining EFT is super-renormalizable. In this case a non-perturbative approach to the phase transition is possible using relatively simple lattice simulations~\cite{Farakos:1994xh,Kajantie:1995kf}. Hence, our calculations in the following sections are a pre-requisite for future non-perturbative studies of the singlet-extended SM, but in the present paper we restrict ourselves to perturbation theory.

A further simplification can be achieved by explicitly integrating out the adjoint scalar $A_0$, whose mass is of the order $gT$. The resulting theory is formally valid for momenta at the ultrasoft scale, $k \lesssim g^2 T$, and has the same form as (\ref{eq:3d-schematic}) but without the $A_0$ field. This is often a good approximation even when parametric mass hierarchies are not strictly satisfied, because of small numerical factors involved in the integration over $A_0$~\cite{Kajantie:1995dw}.

The reader may wonder if purely perturbative studies of the effective potential actually benefit from the EFT approach, which does seem like an extra step compared to the more direct calculation described in~\cite{Arnold:1992rz}. 
The two approaches are, of course, equivalent, since the resummation of ref.~\cite{Arnold:1992rz} also utilizes approximate decoupling of hard Matsubara modes.
In the perturbative context, the main advantage of the EFT lies in factorization: physics at the hard thermal scale is accounted for in the EFT matching and can, in principle, be calculated to any order in perturbation theory, while the IR sensitive contributions are obtained by separate calculations within the EFT. There are also additional resummations of coupling constants and those related to the $A_0$ field that are typically not included in the 4d calculations but are straightforward to implement using EFT methods. In the context of hot QCD, where dimensionally-reduced theories are widely used, including these higher-order effects is known to improve the convergence of perturbation theory~\cite{Blaizot:2003iq,Laine:2006cp}.

\section{Power counting and parametric estimates}
\label{sec:power-counting}

Let us now discuss the accuracy goal of our calculation. Near a phase transition, there is competition between the vacuum and thermal masses, so we assume that our tree-level masses are not considerably heavier than $T$. We express this parametrically as $m_\phi, m_S \sim g T$, where $g$ is a formal expansion parameter that will be identified as the $\gr{SU(2)}_L$ gauge coupling for concreteness.
For the dimensionful cubic couplings, we consequently require $a_1, b_3 \sim g^2 T$; otherwise ratios such as $a_1/m_\phi$ are unsuppressed and perturbativity is questionable.
For other couplings, the renormalization structure of the theory suggests
the parametric countings
$\lambda, a_2, b_4 \sim g^2$ and
$y_t, g',\gs \sim g$ (here $\gs$ is the QCD coupling).
Adopting this schematic counting facilitates the discussion of loop corrections
where combinations of the various couplings appear.

The structure of perturbative expansion at finite $T$ is modified by resummations and differs from that of its zero-temperature counterpart. Loops involving hard Matsubara modes contribute to the effective potential at orders $g^2, g^4, \dots$ while the resummed zero modes provide corrections also at $g^3, g^5, \dots$. A one-loop calculation with daisy resummation gives the field-dependent terms in the effective potential at $\mathcal{O}(g^3)$ with partial $\mathcal{O}(g^4)$ contributions, but reaching $g^4$ accuracy in the quadratic terms requires a two-loop calculation. In particular the relative difference between a one-loop daisy resummed potential and a two-loop calculation is $\mathcal{O}(g)$, while at zero temperature the difference between one and two loops is $\mathcal{O}(g^2)$. Thus the two-loop improvement is relatively more important in the finite-$T$ theory. We will perform a full $\mathcal{O}(g^4)$ computation applying the 3d EFT approach.

The cubic couplings appearing in the xSM have implications for the parametric accuracy of dimensional reduction. In the absence of chemical potentials, the leading higher-dimensional operators appearing in the dimensionally-reduced EFT of the SM are of dimension six (in 4d units)~\cite{Kajantie:1997ky}. This is also true for the $Z_2$-symmetric case of xSM.
However, in the more general case there can be $Z_2$-breaking operators like $c_{4,1} (\he\phi\phi)^2 S$ at dimension five. The coefficient is parametrically $c_{4,1} \sim g^4 a_1 / T^2$. With this operator present in the 3d EFT, it contributes, among others, to the $\he\phi\phi S^2$ correlator via the diagram 
\begin{align}
\renewcommand{\picb}[1]{\PIC{#1}{45}{30}{1}}
\ToptVSS(fex(\Lsc1,\Lcs1,\Lsr1,\Lsr1),\Asc1,\Asc1,\Asc1,\Asc1,\Asc1,\Lsr1)
  \sim c_{4,1} a_1^3 \frac{T^2}{m^4}
  \sim \frac{g^4 a_1^4}{m^4}
.
\end{align}
Since the typical mass scale of the 3d EFT is $m \sim gT$, this diagram contributes at order $a_1^4 / T^4$. But there is a one-loop contribution of the same parametric order from integrating out the hard thermal modes. We conclude that that as far as arbitrary Green's functions are concerned, there is no reason to keep terms proportional to $a_1^4$ in the EFT matching relations unless dimension five operators are included as well.%
\footnote{This conclusion can also be reached via the general power-counting arguments
  given in~\cite{Kajantie:1997ky}.
}

To keep the discussion simple, we will only include operators up to dimension four in the EFT. In terms of our power counting, the maximal accuracy that can be reached for field-dependent terms in the effective potential is then $\mathcal{O}(g^4)$, sufficient for our purposes. The EFT reproduces hard-mode contributions at this accuracy provided that we match 1- and 2-point Green's functions at two-loop order, and 3- and 4-point functions at one loop. Terms such as $a_1^4$ and $g^2 a_1^2$ will be dropped, following the power counting described above. An additional advantage of dropping the higher-dimensional operators is that the resulting EFT admits a simple continuum limit, facilitating non-perturbative lattice studies in the future.

\section{Matching to the 3d EFT}
\label{sec:DR}

\begin{figure*}[t]
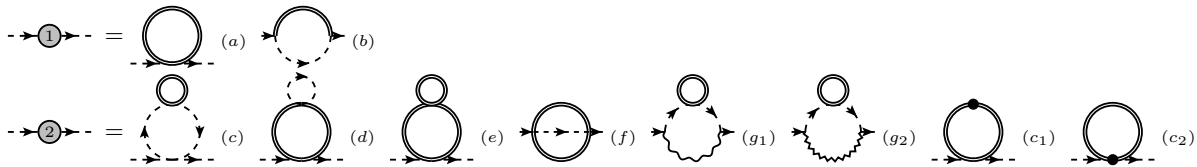

  \centering
  \begin{align*}
    \TopSi(\Lsc1,1) =&
     \TopoST(\Lsc1,\Asr1)_{(a)}\;
     \TopoSB(\Lsc1,\Asr1,\Asc1)_{(b)}\;
    \;
    \nn[4mm]
    \TopSi(\Lsc1,2) =&
     \ToptSTT(\Lsc1,\Acs1,\Acs1,\Asr1)_{(c)}\;
     \ToptSTT(\Lsc1,\Asr1,\Asr1,\Acs1)_{(d)}\;
     \ToptSTT(\Lsc1,\Asr1,\Asr1,\Asr1)_{(e)}\;
     \ToptSS(\Lsc1,\Asr1,\Asr1,\Lcs1)_{(f)}\;
     \ToptSBT(\Lsc1,\Acs1,\Acs1,\Agl1,\Asr1)_{(g_1)}\;
     \ToptSBT(\Lsc1,\Acs1,\Acs1,\Agl2,\Asr1)_{(g_2)}\;
     \TopoSTx(\Lsc1,\Asr1)_{(c_1)}\;
     \TopoSTc(\Lsc1,\Asr1)_{(c_2)}\;
  \end{align*}
  \caption{
    Singlet contributions to the $\he\phi\phi$ self energy.
    Oriented dashed lines refer to Higgs doublets, double lines to the singlet $S$,
    and the two types of wiggly lines describe $\gr{SU(2)}$ and $\gr{U(1)}_Y$ gauge propagators.
    Black blobs represent counterterm insertions.
    In our power counting, two-loop diagrams involving cubic couplings $a_1$ or $b_3$ contribute at higher orders and are not shown.
    The diagrams were drawn with {\tt Axodraw}~\cite{Collins:2016aya}.}
  \label{fig:phi:2pt:2loop}
\end{figure*}

Building on the discussion of section~\ref{sec:review}, the dimensionally reduced EFT for the full xSM has the Euclidean action
\begin{widetext}
\begin{align}
\label{3d-EFT1}
S_\text{3d} =& \int d^3x \Big\{
  \frac14 F^a_{ij} F^a_{ij}
+ \frac14 B_{ij} B_{ij}
+ |D_i \phi|^2
+ \frac12 (\partial_i S)^2
+ V^{\text{3d}}(\phi, S)
+ \frac12 (D_i A_0^a)(D_i A_0^a)
+ \frac12 \mD^2 A^a_0 A^a_0
+ \frac12 (\partial_i B_0)^2  \nonumber \\[1mm]
&
+ \frac12 (\mD')^2 B_0^2
+ \frac12 (D_i C_0^\alpha)(D_i C_0^\alpha)
+ \frac12 (\mD'')^2 C_0^\alpha C_0^\alpha
+ h_3 \he\phi\phi A_0^a A_0^a + h_3' \he\phi\phi B_0^2
+ h_3'' \he\phi A_0^a \sigma_a \phi B_0 \nonumber \\[2mm]
&
+ \omega_3 \he\phi\phi C_0^\alpha C_0^\alpha
+ x_3 S A_0^a A_0^a
+ x_3' S B_0^2
+ y_3 S^2 A_0^a A_0^a
+ y'_3 S^2 B_0^2
+  \text{interactions among } {A_0, B_0} \text{ and } C_0 \Big\}.
\end{align}
\end{widetext}

We use ``natural'' 3d units where all couplings are dimensionful and fields have mass dimension $1/2$. Here $\sigma_a$ are the Pauli matrices and $F_{ij}, B_{ij}$ are 3d field strengths for the $\gr{SU(2)}$ and $\gr{U(1)}_Y$ gauge fields whose couplings are denoted by $g_3$ and $g_3'$. The fields $A_0, B_0, C_0$ are scalars in adjoint representations of $\gr{SU(2)}, \gr{U(1)}_Y, \gr{SU(3)}$, respectively (\textit{i.e.}\ $B_0$ is a singlet).%
\footnote{3d gauge invariance also allows for operators with an odd number of the adjoint fields, such as $\he\phi A_0^a \sigma_a \phi$ and $\he\phi\phi B_0$. However, these can only appear in the presence of non-vanishing chemical potentials~\cite{Kajantie:1997ky}.
}
Their masses are parametrically $gT, g' T, \gs T$ due to Debye screening; we will shortly integrate these fields out to considerably simplify the EFT.

Couplings between the singlet and the adjoint fields are generated by $\phi$ loops in the 4d theory and the $\he\phi\phi C_0 C_0$ term by quark loops. In principle the $\gr{SU(3)}$ gluons are also present in the EFT, but these couple to the Higgs sector only through $C_0$ and any such contribution to the $\Veff$ is of higher order than our accuracy goal. Interaction vertices among the adjoint scalars are $\mathcal{O}(g^4)$ and contribute only at higher orders~\cite{Gorda:2018hvi}. The scalar potential for $\phi$ and $S$ in the EFT is
\begin{align}
\label{eq:EFT-pot}
V^\text{3d} &= m^2_{\phi,3} \he\phi\phi + \lambda_3 (\he\phi\phi)^2 + b_{1,3} S + \frac12 m^2_{S,3} S^2 \nonumber \\
& + \frac13 b_{3,3} S^3 + \frac14 b_{4,3} S^4  + \frac12 a_{1,3} S \he\phi\phi + \frac12 a_{2,3} S^2 \he\phi\phi.
\end{align}
To avoid clutter, we will use the same notation for fields in the original theory and those in (\ref{3d-EFT1}).

To relate the EFT parameters to those appearing in the full xSM in 4d, we identify the fields in (\ref{3d-EFT1}) as the Matsubara zero-modes of the original theory and integrate out all loops involving a non-zero (hard) Matsubara mode. Effects of the hard modes approximately decouple in the high-$T$ limit, apart from a $T$-dependent renormalization of parameters and fields in the zero-mode sector.
It is convenient to make these corrections explicit in the EFT by writing, for instance $m^2_{\phi,3} = m^2_\phi + \delta_T m^2_\phi$, where $\delta_T m^2_\phi$ is the $T$-dependent correction from hard modes (``thermal counterterm''). Here $m^2_{\phi,3}$ is the renormalized parameter; a separate counterterm $\delta m^2_{\phi,3}$ is introduced to absorb UV divergences in dimensional regularization. In the matching we treat both counterterms as perturbations. We then calculate 1PI Green's functions both in the EFT and in the full theory at soft external momenta, solving for the matching corrections order-by-order in perturbation theory by requiring that the Green's functions match (note that hard Matsubara modes cannot appear in reducible internal legs). 

The required matching relations have been worked out previously in ref.~\cite{Schicho:2021gca},
where terms such as $g^2 a_1^2$ and $a_1^4$ were included. As discussed above, one then has corrections of the same order from higher-dimensional operators, and the resulting EFT becomes complicated. Here we shall stick to the power counting of the previous section and omit such terms, whose effects are negligible unless the cubic couplings are unnaturally large compared to masses appearing in the quadratic terms. We have checked this numerically in the benchmark points discussed below. 

For the matching, we can expand the fields around any constant background value that is parametrically smaller than the UV scale $\pi T$. Any such field shifts affect only the IR sector and have no effect on the matching. Therefore we choose to expand around the origin so that no background fields are explicitly present in the calculation. The resulting EFT correctly describes the phase structure of the full xSM, provided that VEVs after the EWPT are not considerably larger than $\pi T$; otherwise there will be unsuppressed contributions from higher-dimensional operators.
We take the linear term $b_1 S$ to be an additional interaction rather than part of the free particle action. This is allowed because it only creates one-particle-reducible insertions that can be incorporated in the EFT by matching the singlet 1-point correlator at $\mathcal{O}(g^4)$.

Detailed discussions of matching calculations in the high-$T$ context can be found for instance in \cite{Braaten:1995cm,Kajantie:1995dw,Gould:2021dzl,Schicho:2021gca}.
Here we cut down on details, showing explicitly only the matching of singlet contributions to the Higgs mass parameter $m^2_{\phi,3}$.
The relevant diagrams are shown in figure~\ref{fig:phi:2pt:2loop}.
At two-loop level there are diagrams where both soft and hard momenta flow in loops, however these are absorbed in the EFT by one-loop diagrams containing a thermal counterterm insertion~\cite{Kajantie:1995dw}.%
\footnote{For example, diagram $(c)$ with hard momentum in the lower and soft momentum in the upper loop corresponds to a one-loop diagram of the type $(c_2)$ in the EFT, with the blob replaced by $\delta_T a_2$, the thermal correction to $a_2$.}
Hence, only diagrams with hard momenta in all loops are needed for the matching. The sum-integrals are calculated using high-$T$ expansion, \textit{i.e.}\ an expansion in $m^2 / (\pi T)^2$, and we assume $m \sim gT$. For $\mathcal{O}(g^4)$ accuracy we need to expand the one-loop diagrams to next-to-leading order, while at two-loop level the leading order suffices.

We denote sum-integrals over non-zero modes as
\begin{align}
\Tint{P}' \equiv T \sum_{\omega_n \neq 0} \Big( \frac{e^\gamma \bar{\mu}^2}{4\pi} \Big)^{\epsilon} \int \frac{d^dp}{(2\pi)^d} ,
\end{align}
where $P = (\omega_n, \vec{p})$, $d = 3-2\epsilon$, $\bar{\mu}$ is the \MSbar scale and $\gamma$ is the Euler-Mascheroni constant.
At one loop, the relevant bosonic sum-integrals are~\cite{Arnold:1994ps}
\begin{align}
\label{I4b1}
I^{4b}_1 =& \; \Tint{P}' \frac{1}{P^2} \nn
=& \frac{T^2}{12} \Big( 1 + \epsilon \left[ 2\gamma - \ln 4 + 2\ln\frac{\bar{\mu}}{T} - 2 \frac{\zeta'(2)}{\zeta(2)} \right] \Big) + \mathcal{O}(\epsilon^2) \\
\label{I4b2}
I^{4b}_2 =& \; \Tint{P}' \frac{1}{P^4} = \frac{1}{16\pi^2} \Big( \frac{1}{\epsilon} + 2\gamma + 2\ln \frac{\bar{\mu}}{4\pi T} \Big) + \mathcal{O}(\epsilon).
\end{align}
At vanishing external momentum, the one-loop diagrams in fig.~\ref{fig:phi:2pt:2loop} give
\begin{align}
(a) + (b) = -\frac12 a_2 \Big( I^{4b}_1 - m_S^2 I^{4b}_2 \Big) + \frac14 a_1^2 I^{4b}_2 + \mathcal{O}(a_1^2 m_S^2).
\end{align}
Here the $a_1^2 m_S^2$ term exceeds our desired accuracy. The first term is the $\mathcal{O}(g^2)$  singlet contribution to thermal Higgs mass,
$\delta_T^{ } m^2_\phi =  (\delta_T^{ } m^2_\phi)_\text{SM} + \frac{T^2}{24} a_2 + \mathcal{O}(g^4)$.

On the EFT side, we renormalize the field as
$\phi \rightarrow (1 + \delta_T Z_\phi)^{1/2} \phi$, where the counterterm arises from integration over the hard modes at soft external momenta in the full theory. Then, in the EFT Lagrangian we have $m^2_{\phi, 3}\he\phi\phi / T \rightarrow (m^2_\phi + \delta_T m^2_\phi) (1 + \delta_T Z_\phi)\he\phi\phi / T$, plus a mass counterterm needed to cancel $1/\epsilon$ poles in 3d. For $\mathcal{O}(g^4)$ matching, the field renormalization counterterm needs to be known at $\mathcal{O}(g^2)$ accuracy.
Momentum-dependent contributions from the singlet are $\sim k^2 a_1^2 / T^2 \sim g^6 T^2$ for soft $k \sim gT$, therefore the SM result is sufficient (equation (B.37) in~\cite{Croon:2020cgk}):
\begin{align}
\label{eq:field-renorm}
\delta_T Z_\phi = -\frac{1}{(4\pi)^2} \Big[ \frac{L_b}{4} \big(3(3 - \xi) g^2 + (3 - \xi) {g'}^2 \big) - 3 y_t^2 L_f \Big],
\end{align} 
which holds in a general covariant gauge ($\xi$ is the gauge-fixing parameter).
Here $L_b$ and $L_f$ are logarithms that arise frequently from one-loop bosonic and fermionic sum-integrals:
\begin{align}
L_b &= 2\ln \frac{\bar{\mu}}{T} - 2 (\ln 4\pi - \gamma) \\ 
L_f &= L_b + 4\ln 2 .
\end{align}
Gauge dependence of (\ref{eq:field-renorm}) cancels against two-loop contributions in the final result for $m^2_{\phi,3}$.

We then turn to the two-loop diagrams in fig.~\ref{fig:phi:2pt:2loop}, which we evaluate at vanishing external momentum and leading order in the high-$T$ expansion (\textit{i.e.}\ propagators can be taken massless). Diagrams $(c)$, $(d)$, $(e)$ are products of one-loop sum-integrals:
\begin{align}
(c) + (d) + (e) = \Big( 3\lambda a_2 + a_2^2 + \frac32 b_4 a_2 \Big) I^{4b}_1 I^{4b}_2.
\end{align}
The ``sunset'' diagram $(f)$ does not contribute to matching at $\mathcal{O}(g^4)$ because of the identity~\cite{Nishimura:2012ee}
\begin{align}
\Tint{P} \Tint{Q} \frac{1}{P^2 Q^2 (P+Q)^2} = 0.
\end{align}
Diagrams $(g_1), (g_2)$ contain gauge fields, but the resulting integrals are trivial because the gauge propagator $D_{\mu\nu}(P)$ with loop momentum $P$ gets contracted with $P_\mu P_\nu$, leaving only the longitudinal part.
In a general covariant gauge, the result is
\begin{align}
(g_1) + (g_2) = -\frac18 \left( 3g^2 + {g'}^2 \right) a_2 \xi I^{4b}_1 I^{4b}_2.
\end{align}

Finally, there are one-loop diagrams with counterterm insertions. Our counterterms for the \MSbar scheme are listed in Appendix~\ref{sec:renormalization}. Here we note that mass counterterms contribute finite parts at $\mathcal{O}(g^6)$ in our counting and that the singlet does not need field renormalization at one loop.
Thus diagram $(c_1)$ does not contribute while $(c_2)$ is proportional to $\delta a_2$. We also need an additional diagram proportional to $\delta \lambda$ (not shown in fig.~\ref{fig:phi:2pt:2loop}), because the Higgs self-interaction gets renormalized by singlet loops. The required diagrams are 
\begin{align}
\TopoSTc(\Lsc1,\Asr1)_{} + \TopoSTc(\Lsc1,\Acs1)_{} = \left( -\frac12 \delta a_2 - 6 \delta\lambda \right) I^{4b}_1
\end{align}
where, for the present discussion, we only include singlet contributions to $\delta \lambda$.

Summing all the diagrams, expanding in $\epsilon$ and discarding terms that go beyond $\mathcal{O}(g^4)$ gives 
\begin{align}
\label{eq:higgs_self_hard}
\Gamma^{(S)}_{\he\phi\phi} =& -\frac{T^2}{24} a_2 + \frac{L_b}{2(4\pi)^2} \left( a_2 m_S^2 + \frac12 a_1^2 \right) \nn
& + a_2  L_b \frac{T^2}{(4\pi)^2} \Big( 5 a_2 + 3 b_4 + 6\lambda - \frac{1}{4} \xi( 3g^2 + {g'}^2 )  \Big) \nn 
& + \frac{T^2}{(4\pi)^2}\frac12 a_2^2 \left( c + \ln \frac{3T}{\bar{\mu}}\right)
,
\end{align}
which is the $\mathcal{O}(g^4)$ correction to the Higgs self energy due to hard singlet loops. There are also leftover $1/\epsilon$ divergences that are canceled by contributions from the soft modes, but this is irrelevant for our present discussion. Here
\begin{align}
c = \frac12 \left( \ln\frac{8\pi}{9} + \frac{\zeta'(2)}{\zeta(2)} - 2\gamma \right) \approx -0.348723.
\end{align}

The self energy (\ref{eq:higgs_self_hard}) is to be reproduced by tree-level terms in the EFT, so we match as 
\begin{align}
-\delta_T m^2_\phi - (m^2_\phi + \delta_T m^2_\phi) \delta_T Z_\phi = \Gamma^{(S)}_{\he\phi\phi} 
\end{align}
from which the matching correction $\delta_T m^2_\phi$ can be solved order-by-order, \textit{i.e.}\ the one-loop result is used in the $\delta_T m^2_\phi \delta_T Z_\phi$ term (the $m^2_\phi$ term actually contains no singlet contributions). Accounting for the full SM part, given in eq.~(A28) of~\cite{Niemi:2018asa}, our result for the effective Higgs mass parameter is
\begin{align}
\label{eq:DR-matching-start}
m^2_{\phi,3} &= \Big(m^2_\phi \Big)_\rmi{SM}
  + \frac{T^2}{24} a_2
  - \frac{L_b}{2(4\pi)^2} \Big( a_2 m^2_S + \frac12 a_1^2 \Big) \nn
&  + \frac{a_2 T^2}{(4\pi)^2} \Big[ \frac{1}{24} L_b \Big( \frac34 (3g^2 + {g'}^2) - 6 \lambda - 5 a_2 - 3 b_4 \Big) \nn 
& - \frac18 y_t^2 L_f \Big] 
 - \frac{T^2}{(4\pi)^2} \frac{1}{2} a_{2}^2 \Big( c + \ln\Big(\frac{3T}{\bar{\mu}} \Big) \Big), 
\end{align}
where gauge dependence duly cancelled.
This is a two-loop generalization of the thermal Higgs mass.

Eq.~(\ref{eq:DR-matching-start}) can be improved by considering RG behavior of the 3d EFT. Applying RG evolution to the parameters in (\ref{eq:DR-matching-start}), one can check that apart from the $\ln (3T/\bar{\mu})$-term, the effective mass is RG invariant up to higher-order effects. This means that RG evolution down from the matching scale is governed by the $\ln (3T/\bar{\mu})$-term. But we can also solve the running directly in the super-renormalizable EFT, using the renormalization scale $\bar{\mu}_3$, to obtain an exact%
\footnote{In the running we actually neglect terms containing $\omega_3$ or couplings between
  the singlet and adjoint scalars. Because logarithmic divergences in 3d appear first at two-loop level,
  our RG running is exact up to terms of order $\omega_3^2 \sim g^8$.
  There are no further corrections to the running at higher orders.
}
RG equation at two-loop order~\cite{Farakos:1994kx,Kajantie:1995dw}. Replacing the $\ln (3T/\bar{\mu})$-term with the solution of this RG equation incorporates higher-order effects to the running of $m^2_{\phi,3}$, in this case we simply replace
\begin{align}
\frac{1}{2} T^2 a_{2}^2 \Big( c + \ln\Big(\frac{3T}{\bar{\mu}} \Big)\Big) \rightarrow \frac{1}{2} a_{2,3}^2 \Big( c + \ln\Big(\frac{3T}{\bar{\mu}_3} \Big)\Big)
,
\end{align}
and similarly in the SM part (\textit{c.f.}~\cite{Niemi:2018asa}). Written in this form, the effective mass is independent of the matching scale $\bar{\mu}$ at order $\mathcal{O}(g^4)$. We apply this same improvement to parameters $b_{1,3}^{ }$ and $m^2_{S,3}$ below, while all couplings are RG invariant in 3d. 
This is an example of a higher-order improvement that is difficult to incorporate directly in the 4d effective potential. Numerically the effect is small, for instance in benchmark points BM2 and BM6 below the difference between improved and unimproved results is less than $1\%$.

The matching calculation outlined above is straightforward to generalize to other parameters in the effective theory (\ref{3d-EFT1}). At $\mathcal{O}(g^4)$, the results are:
\begin{widetext}
\begin{align}
%
m^2_{S,3} &=
  m^2_S
  + T^2 \Big(\frac{1}{6}a_2+ \frac{1}{4}b_4 \Big)
  - \frac{L_b}{(4\pi)^2}\Big( 2b_3^2+ \frac{1}{2} a_1^2 + 2 a_2 m^2_\phi + 3 b_4 m^2_S \Big)
  \nn &
  + \frac{T^2}{(4\pi)^2}\bigg[
      \frac{2+3L_b}{24}(3g^2 + {g'}^2) a_2
  - L_b \bigg( \Big(\lambda + \frac{7}{12}a_2 + \frac{1}{2} b_4 \Big) a_2 + \frac94 b_4^2 \bigg)
  - \frac{1}{4} a_2\gY^{2}(3L_b - L_f) \bigg]
\nn
& + \frac{1}{(4\pi)^2} \Big( (3g^2_3 + g'^2_3) a_{2,3} - 2 a_{2,3}^2 - 6 b_{4,3}^2 \Big) \Big( c + \ln\Big(\frac{3T}{\bar{\mu}_3} \Big) \Big) \\
%
%
b_{1,3} &= \frac{1}{\sqrt{T}} \bigg[
    b_1
  + \frac{T^2}{12} \Big( b_3 + a_1 \Big)
  - \frac{L_b}{(4\pi)^2} \Big( a_1 m^2_\phi + b_3 m^2_S \Big)
  \nn &
  + \frac{T^2}{(4\pi)^2}\bigg[
      \frac{2+3L_b}{48}(3g^2 + {g'}^2) a_1
    - \frac{L_b}{2} \bigg(
      \Big( \lambda + \frac{7}{12} a_2 \Big) a_1
    + \Big( \frac{1}{3} a_2 + \frac{3}{2}b_4 \Big)b_3
  \bigg)
  - \frac{1}{8} a_1\gY^{2} \Big(3L_b - L_f \Big) 
  \bigg]
  \bigg] \nn
& - \frac{1}{(4\pi)^2} \bigg( 2 b_{3,3} b_{4,3}- \frac12 a_{1,3} \Big(3g^2_3 + g'^2_3 - 2 a_{2,3} \Big) \bigg) \Big( c + \ln\Big(\frac{3T}{\bar{\mu}_3} \Big) \Big) \\
\lambda_3 &= T \lambda + \frac{T}{(4\pi)^2} \bigg(
      \frac{2-3L_b}{16}\Big(3g^4 + 2g^2{g'}^2 + {g'}^4\Big)
      + 3 y_t^2 L_f \Big( y_t^2
      - 2 \lambda \Big)
  + L_b \Big(
      \frac{3}{2}(3g^2 + {g'}^2) \lambda
    - 12 \lambda^2
    - \frac{1}{4} a_2^2
  \Big)
  \bigg)
\\
b_{3,3} &= \sqrt{T} b_3
  - \frac{\sqrt{T}}{(4\pi)^2} 3 L_b \Big( \frac{1}{2} a_1 a_2 + 3 b_4 b_3 \Big)
\\
b_{4,3} &= T b_4
  - \frac{T}{(4\pi)^2} L_b \Big( a_2^2 + 9 b_4^2 \Big)
\\
a_{1,3} &= \sqrt{T} a_1
  + \frac{\sqrt{T}}{(4\pi)^2} \bigg( L_b \Big(\frac{3}{4} (3g^2 + {g'}^2) a_1 -2 b_3 a_2 - ( 6\lambda + 2a_2) a_1 \Big) -3 L_f y_t^2 a_1 \bigg) \\
\label{eq:a2match}
a_{2,3} &= T a_2
  + \frac{T}{(4\pi)^2} \bigg( L_b \Big(\frac{3}{4}(3g^2 + {g'}^2) - 6\lambda - 2 a_2 - 3 b_4 \Big)a_2  -3 L_f y_t^2 a_2 \bigg)
.
\end{align}
\end{widetext}
Matching the interactions between adjoint fields and the singlet at one-loop level gives:
\allowdisplaybreaks[1]
\begin{align}
\label{eq:match_x3}
x_3 &= \frac{\sqrt{T}}{(4\pi)^2} g^2 a_1 \\
\label{eq:match_x3p}
x_3' &= \frac{\sqrt{T}}{(4\pi)^2} {g'}^2 a_1 \\
y_3 &= 
    \frac{T}{(4\pi)^2} \frac{1}{2} g^2 a_2 \\
\label{eq:DR-matching-end}
y_3' &=  
    \frac{T}{(4\pi)^2} \frac{1}{2} {g'}^2 a_2 \; .
\end{align}
Couplings between $S$ and the temporal gluon $C_0$ do not appear at one loop.

The remaining parameters
$\mD^{ },\mD',\mD'',h_3^{ }, h_3',h_3'',\omega_3^{ }$
receive no contributions from the singlet at this order and the results can be
read from eqs.~(A10)--(A12) and (A18)--(A20) in ref.~\cite{Niemi:2018asa} (by setting $N_t \rightarrow 0, a_2 \rightarrow 0$ in those expressions).%
\footnote{Eq.~(A12) in~\cite{Niemi:2018asa} should read
  ${\mD''}^2 = \gs^2 T^2 \left( 1 + \frac{\nf}{3}\right)$ with $\nf = 3$.
}
For the Debye masses a one-loop matching is sufficient, as they contribute to the Higgs effective potential only through loops.

In practice, we have automated the matching calculation 
using in-house
{\tt FORM}~\cite{Ruijl:2017dtg}
software based on a standard Laporta algorithm~\cite{Laporta:2001dd}
for the reduction of sum-integrals.
All required integrals can be read from~\cite{Gorda:2018hvi}, and \cite{Schicho:2021gca} collects them in a reduced form.
While installing a general covariant gauge, we have verified that
all matching relations are manifestly gauge invariant and independent of the matching scale up to higher-order corrections.
For the ensuing numerical analysis we fix the matching scale
$\bar{\mu} = 4\pi e^{-\gamma} T \approx 7T$, for which the $L_b$ terms vanish identically and
which should be optimal for minimizing higher-order logarithms~\cite{Farakos:1994kx}.

For later convenience, we also match a set of dimension five and six operators
(in 4d language).
These are used merely to estimate the accuracy of our EFT and are excluded from
the numerical analysis below.
The effective potential is most sensitive to non-derivative operators constructed
from $\phi$ and $S$.
Up to dimension six, these are
\begin{align}
\label{eq:dim6}
V^\rmi{(3d)}_{\leq 6} &=
    c_{0,5} S^5 + c_{2,3} \he\phi\phi S^3
  + c_{4,1} (\he\phi\phi)^2 S
  + c_{6,0} (\he\phi\phi)^3
  \nonumber \\ &
  + c_{0,6} S^6
  + c_{4,2} (\he\phi\phi)^2 S^2
  + c_{2,4} \he\phi\phi S^4
  \;,
\end{align}
where the coefficients are to be matched at one-loop.
In the SM, the operator $c_{6,0} (\he\phi\phi)^3$ is dominated by contributions
from the top quark~\cite{Kajantie:1995dw}.
Here we include the top quark loop and pure scalar contributions at $\mathcal{O}(g^6)$
in our power counting.
Corrections from gauge fields are neglected, because we expect dominant effects
to come from the scalar sector.%
\footnote{In fact, gauge contributions to the dimension five and six operators are generally
  gauge dependent at one loop, whereas those from scalars and the top quark are not.
}
At leading order in high-$T$ expansion, the matching reads
\begin{align}
\label{eq:dim5-start}
c_{0,5} =& \frac{\zeta(3)}{(4\pi)^4 \sqrt{T}} \Big( 18 b_4^2 b_3 + \frac12 a_2^2 a_1 \Big) \\
c_{2,3} =& \frac{\zeta(3)}{(4\pi)^4 \sqrt{T}} \frac{a_2}{4} \Big( 8(a_2 + 3b_4)(2b_3 + a_1) + 24\lambda a_1 \Big) \\
c_{4,1} =& \frac{\zeta(3)}{(4\pi)^4 \sqrt{T}}  \Big( 12 a_1 a_2 \lambda + 2 a_2^2 (b_3 + a_1) + 24\lambda^2 a_1 \Big) \\
c_{6,0} =& \frac{\zeta(3)}{(4\pi)^4} \Big( 80 \lambda^3 - 28 y_t^6 + \frac13 a_2^3 \Big) \\
c_{0,6} =& \frac{\zeta(3)}{(4\pi)^4} \Big( \frac16 a_2^3 + 9 b_4^3 \Big) \\
c_{4,2} =& \frac{\zeta(3)}{(4\pi)^4} a_2 \Big( 24\lambda^2 + 2a_2^2 + a_2( 12 \lambda + 3 b_4) \Big) \\
\label{eq:dim5-end}
c_{2,4} =& \frac{\zeta(3)}{(4\pi)^4} \Big( a_2^3 + 9a_2 b_4^2 + 3a_2^2 \lambda + 6 a_2^2 b_4 \Big)
.
\end{align}
Since the Yukawa and BSM contributions to $c_{6,0}$ come with opposite signs,
the error from neglecting the $(\he\phi\phi)^3$ operator can be
{\it smaller} than in the minimal SM.

\section{Integrating out the Debye screened gauge fields}
\label{sec:A0}

As a final step of our EFT construction we integrate out the adjoint fields $A_0, B_0, C_0$, incorporating additional ring resummations at the soft scale. These are not typically included if calculating the potential directly in 4d. This procedure is possible because the scale $gT$ of electric Debye screening is parametrically larger than that of magnetic screening, $g^2T$. The resulting EFT is simply
\begin{align}
\label{3d-EFT2}
\bar{S}_\text{3d} =& \int d^3x \Big\{ \frac14 F^a_{ij} F^a_{ij} + \frac14 B_{ij} B_{ij} \nonumber \\
& \quad\quad\quad + |D_i \phi|^2 + \frac12 (\partial_i S)^2 + V^{\text{3d}}(\phi, S) \Big\}
\end{align}
where the scalar potential is as in (\ref{eq:EFT-pot}) but with modified parameters, which we denote by an additional overline (for instance $m^2_{\phi,3} \rightarrow \bar{m}^2_{\phi,3}$). Likewise, the gauge couplings are $\bar{g}_3, \bar{g}'_3$. Again, the notation is left unchanged for fields.

Matching proceeds in complete analogy to that of the dimensional reduction step. Because the singlet couples to the adjoint fields only at $\mathcal{O}(g^4)$, corrections to couplings $\bar{a}_{1,3}, \bar{a}_{2,3}, \bar{b}_{3,3}, \bar{b}_{4,3}$ start at $\mathcal{O}(g^5)$ and can be dropped.
For $S$ and $SS$ correlators at two-loop, we likewise neglect diagrams containing singlet-adjoint vertices. This leaves diagrams where the interaction with adjoint scalars occurs through the Higgs, but these are proportional to $\mD / m_{\phi,3}$ and are absorbed by one-loop diagrams involving a counterterm in the IR theory (this is analogous to the cancellation of mixed hard and soft loops due to a thermal counterterm in dimensional reduction). Thus one-loop matching is sufficient for $\bar{b}_{1,3}, \bar{m}^2_{S,3}$:
\begin{align}
\label{eq:A0-matching-start}
\bar{b}_{1,3} &= b_{1,3} - \frac{1}{4\pi} \Big( 3\mD x_3 + \mD' x_3' \Big) \nn
\bar{m}^2_{S,3} &= m^2_{S,3} - \frac{1}{2\pi} \Big( 3\mD y_3 + \mD' y_3' + \frac{3x_3^2}{2\mD} + \frac{{x'_3}^2}{2\mD'} \Big).
\end{align}
Even these are formally higher-order corrections, but we nevertheless retain them here.

For the Higgs mass parameter, we again neglect all two-loop contributions involving singlet-adjoint couplings. Then the result for $\bar{m}^2_{\phi,3}$ is the same as in the SM:
\begin{align}
\bar{m}^2_{\phi,3} &= m^2_{\phi,3} - \frac{1}{4\pi} \Big( 3h_3 \mD + h_3' \mD' + 8\omega_3 \mD'' \Big) \nn
& + \text{(two-loop)}.
\end{align}
The two-loop part can be read from equation (B.97) of~\cite{Croon:2020cgk}. In our power counting the $\gr{SU(3)}$ term $\omega_3 \mD''$ is $\mathcal{O}(g^5)$; the reason we include it here is because $\omega_3 \sim \gs^2 y_t^2$ is numerically large and affects the mass more than the hypercharge contribution.
The remaining parameters obtain no BSM corrections at one-loop:
\begin{align}
\bar{g}^2_3 &= g^2_3 \Big( 1 - \frac{g^2_3}{6(4\pi) \mD} \Big) \\
\bar{g'_3}^2 &= {g'_3}^2 \\
\label{eq:A0-matching-end}
\bar{\lambda}_3 &= \lambda_3 - \frac{1}{2(4\pi)} \Big( \frac{3h_3^2}{\mD} + \frac{{h'_3}^2}{\mD'} + \frac{{h''_3}^2}{\mD + \mD'} \Big)
.
\end{align}

Finally, corrections to the higher-dimensional operators in eqs.~(\ref{eq:dim5-start})--(\ref{eq:dim5-end}) are numerically small compared to the hard scale contributions~\cite{Kajantie:1995dw} and will be neglected in our error estimates.
This completes the construction of our EFT.

\section{The effective potential}
\label{sec:Veff-main}

We are now ready to construct the effective potential $\Veff$ using the EFT (\ref{3d-EFT2}), with resummation of hard thermal loops implemented by the 4d $\rightarrow$ 3d matching relations (\ref{eq:DR-matching-start})--(\ref{eq:DR-matching-end}) and the Debye screened adjoint fields integrated out. The effective potential can be obtained by integrating over all non-zero momentum modes in the path integral around constant background fields. In 3d this is essentially a zero-temperature calculation: temperature dependence appears only in the definitions of the renormalized 3d parameters.

We shift the Higgs field in analogy to eq.~(\ref{eq:Higgs_param}), denoting the associated background field in 3d units by $\bar{v}$, and the singlet is shifted as $S \rightarrow S + \bar{x}$. The background fields $\bar{v}$ and $\bar{x}$ are treated as free real parameters. As in the zero-temperature case, the neutral components of $\phi$ and $S$ mix to form eigenstates $h_1, h_2$, but now the mixing angle depends on the background fields. In the diagonal basis, obtaining the one-loop correction to $\Veff$ is straightforward~\cite{Farakos:1994kx}. Including the gauge fields, the 3d effective potential up to one-loop order is
\begin{widetext}
\begin{align}
\label{eq:Veff-01}
V_{0} + V_{1} &= \bar{b}_{1,3} \bar{x} + \frac12 \bar{m}^2_{\phi,3} \bar{v}^2 + \frac12 \bar{m}^2_{S,3} \bar{x}^2 + \frac13 \bar{b}_{3,3} \bar{x}^3 + \frac14 \bar{a}_{1,3} \bar{v}^2 \bar{x} + \frac14 \bar{\lambda}_3 \bar{v}^4 + \frac14 \bar{b}_{4,3} \bar{x}^4 + \frac14 \bar{a}_{2,3} \bar{v}^2 \bar{x}^2 \nn
& + 2(d - 1) J_3(\bar{m}_W) + (d - 1)J_3(\bar{m}_Z) + J_3(\bar{m}_{h,1}) + J_3(\bar{m}_{h,2}) + 3 J_3(\bar{m}_G),
\end{align}
\end{widetext}
where $d = 3 - 2\epsilon$, and we have used Landau gauge $\xi = 0$. The integral appearing at one-loop is UV finite and given by
\begin{align}
J_3(m) = \frac12 \int \frac{d^3 p}{(2\pi)^3} \ln (p^2 + m^2) = -\frac{m^3}{12\pi}.
\end{align}
Masses of the gauge and would-be Goldstone fields are
\begin{align}
\bar{m}_W^2 &= \frac14 \bar{g}^2_3 \bar{v}^2 , \quad\quad \bar{m}_Z^2 = \frac14 \left(\bar{g}^2_3 + (\bar{g}'_3)^2 \right) \bar{v}^2 \\
\bar{m}_G^2 &= \bar{m}^2_{\phi,3} + \bar{\lambda}_3 \bar{v}^2 + \frac12 \bar{a}_{1,3} \bar{x} + \frac12 \bar{a}_{2,3} \bar{x}^2,
\end{align}
while $\bar{m}_{h,1}$ and $\bar{m}_{h,2}$ are given in Appendix~\ref{sec:Veff-2loop}.

The two-loop correction to $\Veff$ requires more work but is straightforward, as there is no need for further resummations and the required integrals are known~\cite{Niemi:2020hto}.
The correction is obtained from two-loop vacuum diagrams in the shifted theory and is UV divergent, unlike the one-loop correction. Details of the calculation are relegated to Appendix~\ref{sec:Veff-2loop}.

A qualitative difference to 4d effective potentials is that in 3d, RG evolution starts only at two-loop order and can be solved exactly, with no corrections at higher orders (ignoring the running of field-independent additive terms). The only parameters requiring renormalization are $\bar{b}_{1,3}, \bar{m}^2_{\phi,3}, \bar{m}^2_{S,3}$, whose values depend on the 3d renormalization scale $\bar{\mu}_3$, see section~\ref{sec:DR}. Accounting for their running in the tree-level part yields a manifestly RG-invariant result for the two-loop $\Veff$~\cite{Farakos:1994kx}.
In practice, we use the RG-evolved parameters even inside the loop corrections.
This results in residual $\bar{\mu}_3$-dependence that would be cancelled by logarithms at 3- and 4-loop orders.

The vacuum structure can be studied by minimizing the effective potential perturbatively around the tree-level minima, \textit{i.e.}\ $\bar{v} = \bar{v}_0 + \hbar \bar{v}_1 + \hbar^2 \bar{v}_2 + \dots$ where $\hbar$ is a formal loop-counting parameter (and similarly for $\bar{x}$). This ``$\hbar$ expansion'' gives gauge-invariant results for $\Veff$ in its minima, order-by-order in $\hbar$~\cite{Fukuda:1975di,Patel:2011th}. Unfortunately, this construction can fail near phase transitions~\cite{Kajantie:1995kf}. For instance in the $Z_2$ symmetric xSM, the critical temperature for ``symmetric $\rightarrow$ Higgs'' type transitions is determined, at tree level, by the condition $\bar{m}^2_{\phi,3}(\Tc) = 0$. In the $\hbar$ expansion one needs the two-loop correction evaluated at the tree-level minimum, \textit{i.e.}\ at vanishing $\bar{m}^2_{\phi,3}$. This leads to an explicit IR divergence in the two-loop potential,
invalidating the method near $\Tc$~\cite{Laine:1994zq,Niemi:2020hto}.

Such divergences can be avoided by solving the equations $\partial \Veff / \partial \bar{v} = \partial \Veff / \partial \bar{x} = 0$ ``exactly'', in practice by minimizing the potential numerically. This sums a subset of one-particle-reducible corrections that render the potential IR-finite at its minima, and the difference to a strict $\hbar$ expansion is formally of higher order~\cite{Kajantie:1995kf}. However, the exact minimization generally introduces residual gauge dependence to the results that is only cancelled at higher loop orders.

In this paper, we will take the latter approach and minimize $\Veff$ numerically in Landau gauge. Our reasons for not implementing a gauge-invariant $\hbar$ expansion are as follows:
\begin{enumerate}

	\item We will study the xSM near the $Z_2$ symmetric limit, where the tree-level condition for EWPT is $\bar{m}^2_{\phi,3}(\Tc) \approx 0$ and performance of the $\hbar$ expansion is questionable.
	
	\item For small gauge parameters $|\xi| \lesssim 1$, one expects the remaining gauge dependence to be numerically comparable to the gauge-independent corrections from higher loop orders. Landau gauge $\xi = 0$ should minimize this effect, and earlier studies in other models support this assumption~\cite{Laine:1994bf,Martin:2018emo}.

	\item Many existing studies in the xSM apply Landau gauge, and since our goal is to investigate the convergence of perturbation theory in a typical BSM setting it makes sense to present our results in the same gauge. 

	\item For strong transitions in the xSM, we expect that dominant uncertainties come from residual RG scale dependence in scalar loops and not from gauge loops where the gauge dependence appears. This expectation is supported by the one-loop study of ref.~\cite{Chiang:2018gsn}.

\end{enumerate}
It is worth noting that the residual gauge dependence arises only in the effective potential, while our 4d $\rightarrow$ 3d matching relations are gauge invariant.

Of course, while the gauge-dependent approach avoids explicit IR divergences at two-loop, it does not remove the fundamental IR problem of high-$T$ perturbation theory, which is an intrinsic property of the Matsubara zero modes. There can still be divergences at higher orders due to massless gauge bosons in the symmetric phase, and slow convergence is to be expected if the scalars are light compared to the dimensionful couplings in the 3d scalar potential.

\section{Numerical results}
\label{sec:results}

\begin{figure*}[t]
    \begin{tabular}{c c c}
    \includegraphics[width=.3333\textwidth]{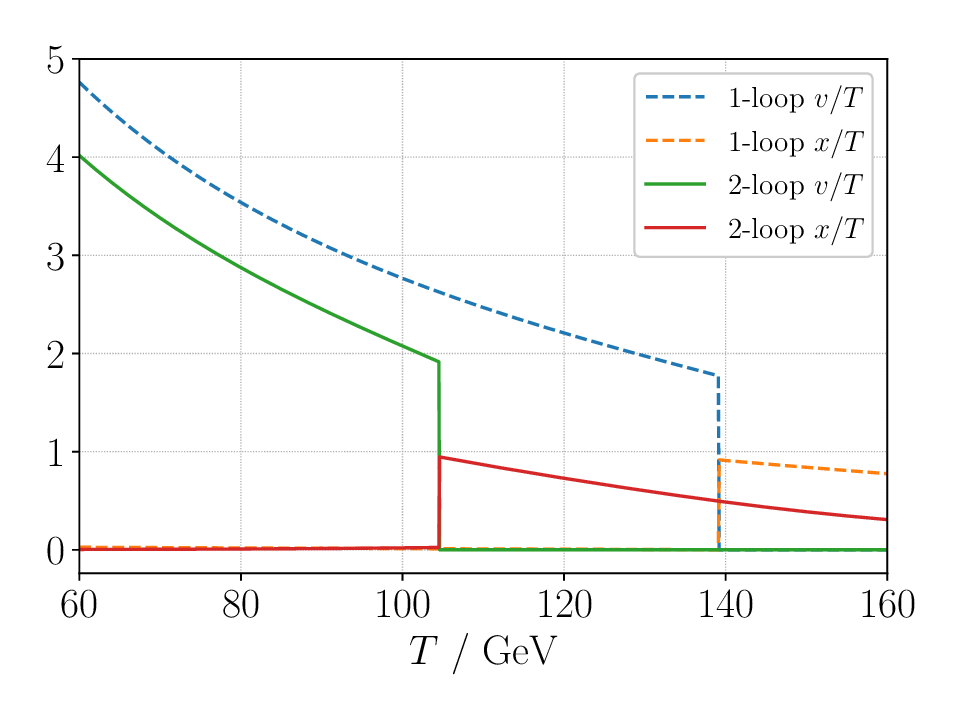} &
    \includegraphics[width=.3333\textwidth]{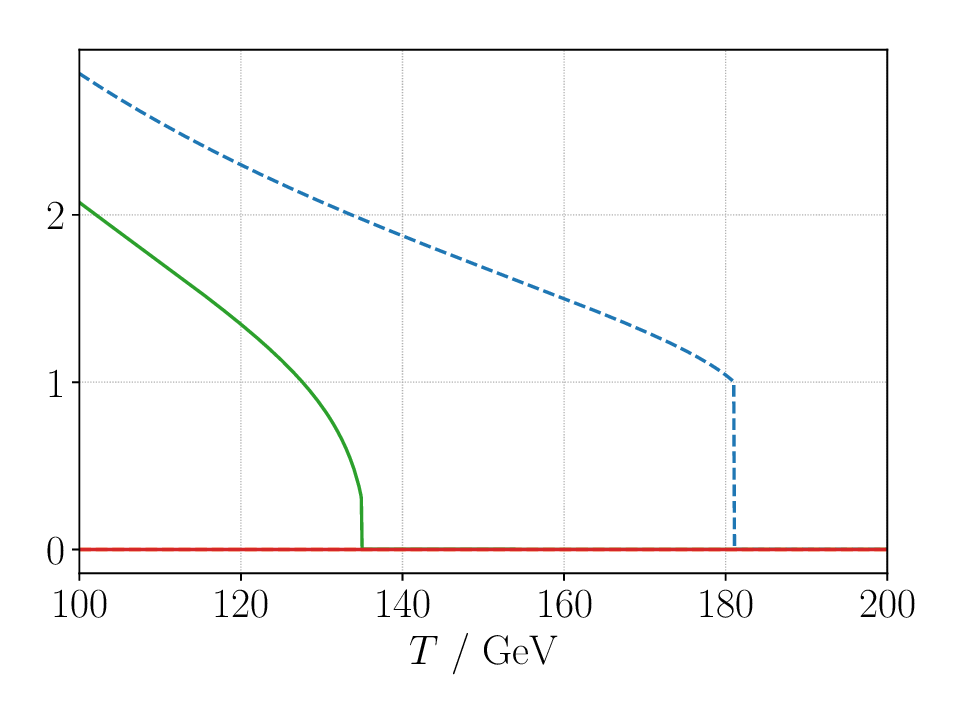} &
    \includegraphics[width=.3333\textwidth]{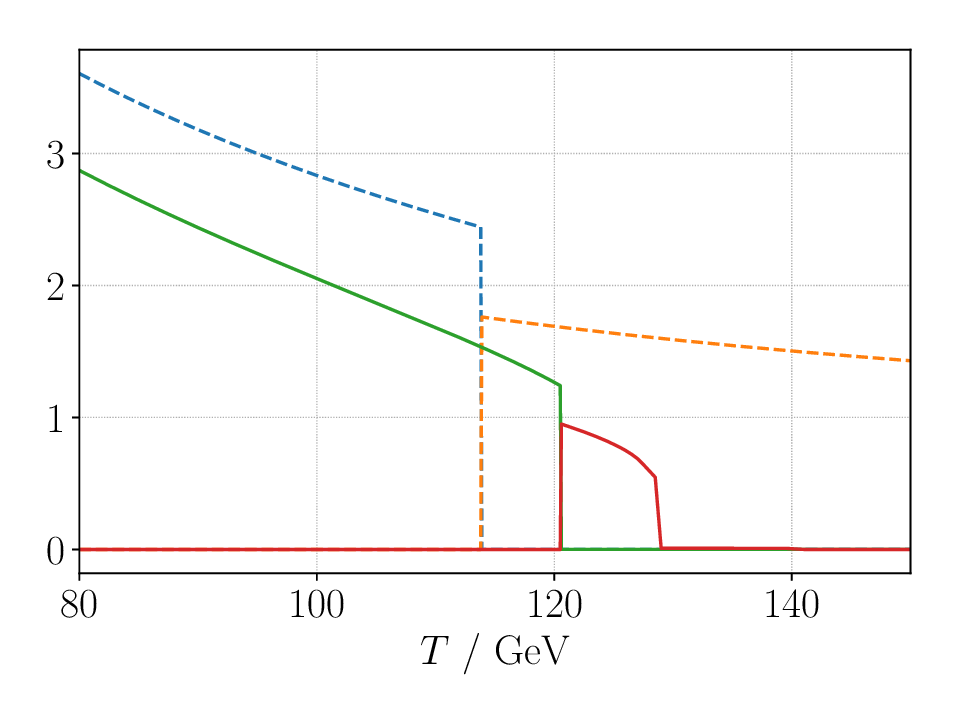} \\
    \small (a) BM2 & \small (b) BM6 & \small (c) BM10
  \end{tabular}
	\caption{Evolution of Higgs and singlet VEVs, $v$ and $x$ respectively, in different types of transitions: (a) symmetric $\rightarrow$ broken transition in the presence of a tree-level barrier, (b) radiatively generated symmetric $\rightarrow$ broken transition in the $Z_2$ symmetric case, (c) two-step phase transition with spontaneously broken $Z_2$ symmetry. The dashed lines are predictions from a one-loop calculation. In BM10 the 2-loop results (solid curves) also show abrupt restoration of the $Z_2$ symmetry associated with the singlet around $T \approx 130$ GeV, while at 1-loop the restoration occurs at much higher $T \approx 300$ GeV, not shown here.
	}
	\label{fig:vevs}
\end{figure*}

To identify phase transitions, we look for temperatures where the effective potential has degenerate local minima. For a set of input parameters {$M_{h_2}, \sin\theta, b_3, b_4, a_2$} and the temperature $T$, the analysis consists of the following steps:
\begin{enumerate}

	\item Solve the renormalized parameters at \MSbar scale $\bar{\mu} = M_Z$ from the one-loop corrected relations (\ref{eq:msbar-1loop-begin})--(\ref{eq:msbar-1loop-end}) in Appendix~\ref{sec:renormalization}.

	\item Run the parameters to the matching scale $\bar{\mu} = 4\pi e^{-\gamma} T$ using one-loop $\beta$-functions (\textit{c.f.}\ Appendix~\ref{sec:renormalization}). This ensures that any leftover $\bar{\mu}$ dependence in the EFT matching is of higher order than the calculation.
	
	 \item Obtain the EFT parameters from eqs.~(\ref{eq:DR-matching-start})--(\ref{eq:DR-matching-end}) and (\ref{eq:A0-matching-start})--(\ref{eq:A0-matching-end}). In the final EFT, we set the renormalization scale to $\bar{\mu}_3 = T$.
	
	\item Obtain the two-loop effective potential from eq.~(\ref{eq:Veff-01}) and Appendix~\ref{sec:Veff-2loop} and minimize $\Veff(\bar{v}, \bar{x})$ with respect to the background fields using a differential evolution algorithm.

\end{enumerate}
Steps 2 through 4 are repeated for different temperatures until a critical temperature $\Tc$ is found (we vary $T$ in steps of $0.1$ GeV). Outside the minimum $\Veff$ can develop an imaginary part, signaling an unstable field configuration \cite{Weinberg:1987vp}. In such cases, we simply discard the imaginary part.

At $\Tc$, we calculate the jumps $\Delta v$ and $\Delta x$ in scalar VEVs across the phase transition, and the latent heat $L = T \Delta (\partial \Veff / \partial T)$. To avoid confusion, we use familiar 4d units when discussing the results: $[v] = [x] = [T]$ and $[\Veff] = [T^4]$, \textit{i.e.}\ $\Veff$ and the fields in 3d get rescaled by approriate powers of $T$. We also compute the discontinuity in the quadratic Higgs condensate, $\Delta \langle \he\phi\phi \rangle = \Delta (\partial \Veff / \partial \bar{m}^2_{\phi,3})$, which provides a gauge-invariant definition of the Higgs VEV through $v_\text{phys} = \sqrt{2 \Delta \langle \he\phi\phi \rangle}$~\cite{Farakos:1994xh} (although our results contain residual gauge dependence, see above). The difference between $\Delta v$ and $v_\text{phys}$ is small in all of our benchmarks. We do not study metastability ranges of the phases nor nucleation processes, so our latent heat will be given at $\Tc$ instead of the nucleation temperature. Therefore our results give a lower bound on the actual transition strength.

We focus mainly on strong transitions with $\Delta v / T_c \gtrsim 1.0$ and examine how the transition changes as the two-loop correction is included. In searching for strong transitions we make use of the one-loop parameter scans appearing in~\cite{Kurup:2017dzf,Gould:2019qek}.
In this paper we concentrate our discussion on selected benchmark (BM) points that represent typical strong-EWPT scenarios, rather than full-fledged scans of the free parameter space. This approach suffices to quantify the effect of two-loop corrections on basic quantities like $\Delta v / T_c$, but it would be interesting to perform systematic two-loop scans over the free parameters in a future study.

Table~\ref{table:1step} collects our results for direct transitions between the EW symmetric and broken minima.
In the non-$Z_2$ symmetric xSM, the minima are separated by a potential barrier already at tree level due to the singlet VEV being non-zero (points BM1-BM5), while in the $Z_2$ symmetric limit the barrier is generated radiatively (BM6 and BM7). In the $Z_2$ symmetric case there are also two-step transitions where the EWPT is preceeded by spontaneous breakdown of the $Z_2$ symmetry, which is restored in the final EW minimum. Here we consider only the second, EW-breaking transition, for which our results in the two-step scenario are in table~\ref{table:2step}. Because the $Z_2$-breaking phase remains metastable even at zero temperature, a realistic two-step scenario necessitates rapid enough nucleation into the EW phase to avoid contradictions with standard cosmology. While this requirement can restrict the parameter space of allowed two-step EWPT considerably~\cite{Kurup:2017dzf}, accounting for it requires calculation of the bubble nucleation rate and will not be considered further in this paper.
Evolution of the background fields in the different types of EWPT are illustrated in fig.~\ref{fig:vevs}.

\begin{figure*}[t]
    \begin{tabular}{c c c}
    \includegraphics[width=.329\textwidth]{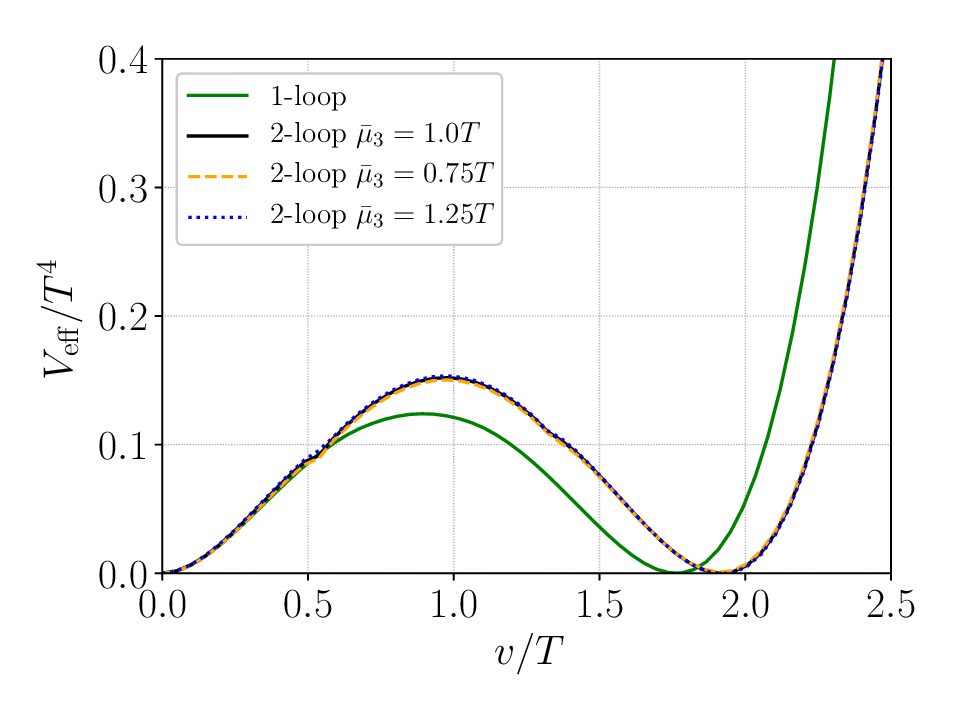} &
    \includegraphics[width=.329\textwidth]{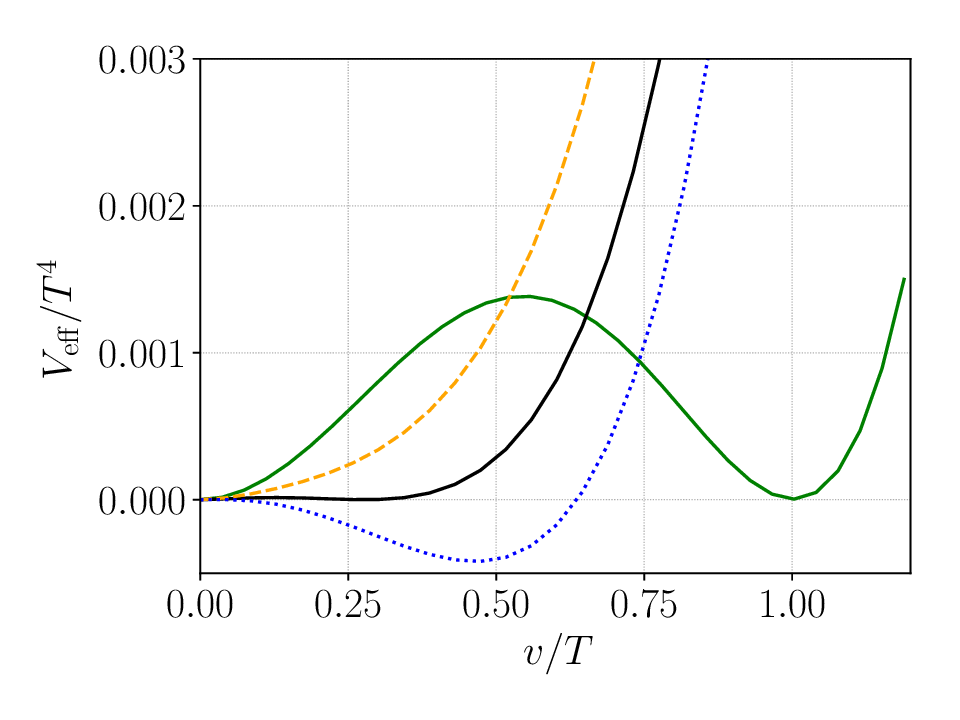} &
    \includegraphics[width=.329\textwidth]{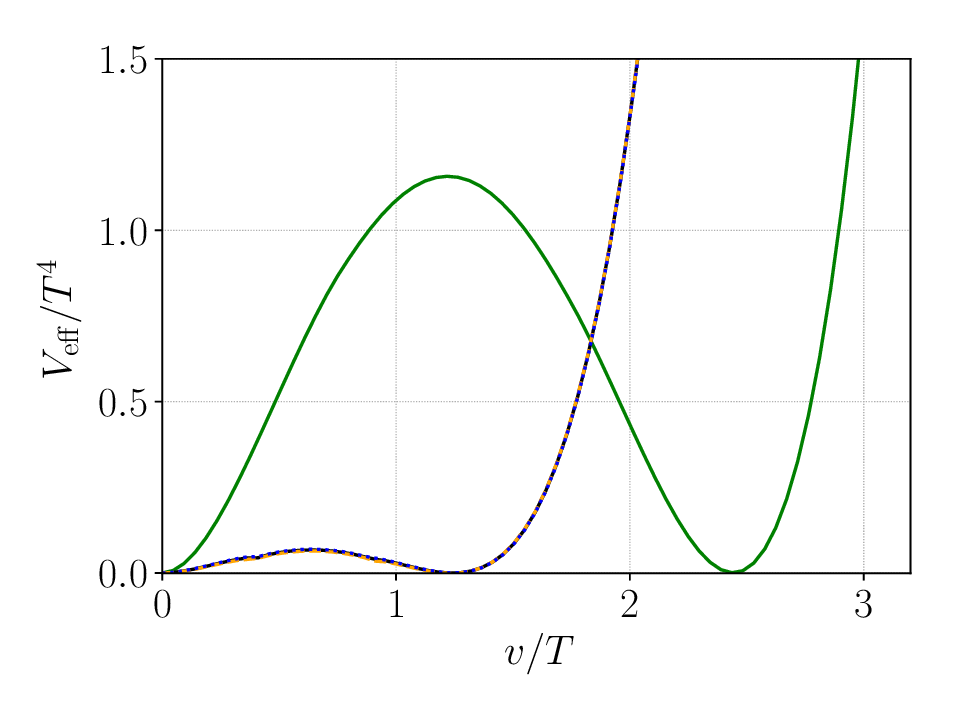} \\
    \small (a) BM2 & \small (b) BM6 & \small (c) BM10
  \end{tabular}
	\caption{Real part of the effective potential, plotted in the Higgs direction at one and two loops at their respective critical temperatures. In (a) and (c) we have accounted for the changing singlet VEV. In (b), the potential barrier vanishes almost completely at two-loop level. The dashed and dotted lines correspond to variations of the RG scale $\bar{\mu}_3$. 
The potential generates an imaginary part near the local maximum as a scalar eigenstate becomes tachyonic, and at specific values of $v/T$ there is a massless mode. In (a) and (c), this results in ``kinks'' in the two-loop potential, since 3d perturbation theory is not well behaved at such points.
This complication does not affect our numerical results, which depend only on properties of the potential in its local minima.
	}
	\label{fig:diffscale}
\end{figure*}

For comparison, we show also the results from a one-loop calculation.  Here "one loop" refers to the one-loop potential in (\ref{eq:Veff-01}) with the EFT parameters matched at $\mathcal{O}(g^2)$, in other words it is the daisy-resummed 4d potential to leading order in high-$T$ expansion (with an additional resummation of $A_0$ loops), correct at $\mathcal{O}(g^3)$. 
``Two loop'' refers to the full $\mathcal{O}(g^4)$ calculation described in detail in the above sections and in appendix~\ref{sec:Veff-2loop}. In both cases we use the same $T=0$ renormalization prescription for input parameters. 

\begin{table}[h]
\begin{center}
	\begin{tabular}{|c |c c c c c c c|}
		\hline
		\textbf{ } & BM1 & BM2 & BM3 & BM4 & BM5 & BM6 & BM7 \\ [0.5ex]
		\hline
		$M_{h_2}$ [GeV] & 300 & 300 & 350 & 370 & 440 & 350 & 450  \\
		\hline
		$\sin\theta$ & 0.05 & 0.05 & -0.08 & 0.07 & 0.07 & 0 & 0  \\
		\hline
		$a_2$ & 2.75 & 3.0 & 3.0 & 3.2 & 3.9 & 3.5 & 4.0  \\
		\hline
		$b_3$ [GeV] & 50 & 50 & 40 & -25 & 70 & 0 & 0  \\
		\hline
		$b_4$ & 0.50 & 0.50 & 0.30 & 0.30 & 0.6 & 0.3 & 0.3  \\
		\hline
		\multicolumn{8}{|l|}{ \hspace{3.75cm} One loop}  \\
		\hline
		$\Tc$ [GeV] & 163.8 & 139.1 & 163.9 & 171.1 & 219.3 & 181.0  & 222.5  \\
		\hline
		$v_\text{phys} / \Tc$ & 1.07 & 1.82 & 1.18  & 1.09 & 1.12 & 0.97 & 0.90   \\
		\hline
		$\Delta v/\Tc$ & 1.08 & 1.77 & 1.19 & 1.11 & 1.14 & 1.00 & 0.93  \\
		\hline
		$\Delta x/\Tc$ & 0.25 & 0.91 & 0.23 & 0.17 & 0.17 & 0 & 0  \\
		\hline
		$L/\Tc^4$ & 0.27 & 0.53 & 0.31 & 0.27 & 0.13 & 0.22 & 0.12  \\
		\hline
		\multicolumn{8}{|l|}{ \hspace{3.75cm} Two loop}  \\
		\hline
		$\Tc$ [GeV] & 128.8 & 104.5 & 123.9 & 127.2 & 119.8 & 134.9 & 138.7  \\
		\hline
		$v_\text{phys} / \Tc$ & 1.02 & 1.97 & 1.20 & 0.99 & 1.09 & 0.29 & 0.28   \\
		\hline
		$\Delta v/\Tc$ & 1.02 & 1.92 & 1.20 & 1.00 & 1.10 & 0.31 & 0.29 \\
		\hline
		$\Delta x/\Tc$ & 0.23 & 0.92 & 0.23 & 0.14 & 0.12 & 0 & 0  \\
		\hline
		$L/\Tc^4$ & 0.34 & 0.83 & 0.49 & 0.35 & 0.48 & 0.042 & 0.037  \\
		\hline
		$\delta_\text{EFT}$ & 0.0031 & 0.0070 & 0.0067 & 0.012 & 0.055 & 0.0016 & 0.0024  \\
		\hline
	\end{tabular}
\end{center}
\caption{\label{tab:1step}
Benchmark scenarios for direct, one-step EWPT. Points BM6 and BM7 are in the $Z_2$-symmetric xSM.
$\delta_\text{EFT}$ estimates the validity of our dimensionally-reduced theory as described around eq.~(\ref{eq:deltaEFT}).
The numbers are obtained using fixed RG scale $\bar{\mu}_3 = T$.
}
\label{table:1step}
\end{table}

With the exception of benchmark points 10 through 12, we find lower critical temperatures at two-loop order and the difference to the one-loop temperature is quite large in almost all points. We expect this shift to originate from the $\mathcal{O}(g^4)$ corrections to thermal masses. For one-step transitions with a tree-level barrier, we observe that transitions tend to be stronger at two loop, and the latent heat is increased by at least $20\%$ relative to the one-loop result. In contrast, two-loop corrections make the transition significantly weaker in the $Z_2$ symmetric xSM. Here, the transition may appear to be strongly first order at one loop, but the radiatively generated potential barrier can be destroyed by higher-order effects if the perturbative expansion convergences slowly. In fact, scanning over $M_{h_2} \in [150, 450]$ GeV and $a_2 \in [1, 5]$ with $b_4 = 0.3$ in the $Z_2$ symmetric case, we found no transitions that would remain strong ($\Delta v / T_c > 1.0$) at the two-loop level. Indeed, one-loop results should be taken with a grain of salt when the coupling $a_2$ is this strong. 
These considerations call for more thorough investigations of higher-order effects in the $Z_2$ symmetric xSM in the future.

Two-step transitions are quite sensitive to the singlet self-interaction and here we concentrate on $b_4 = 0.3$. Contrary to naive expectations, we find the transition strength to be very sensitive to two-loop corrections, despite the presence of a tree-level barrier.
Particularly in BM12 the latent heat is an order of magnitude smaller than the one-loop value, signaling definite breakdown of perturbation theory at least at low orders, probably due to the large value of $a_2$. For points BM8 and BM9 where the coupling is smaller, two-loop effects are moderate but quantitatively important.  We conclude that transitions involving a tree-level barrier in the potential are not protected from large loop corrections. The presence of such a barrier implies that loop effects from the Matsubara zero modes are relatively less important than in radiatively-induced transitions. This argument does not apply to loop corrections involving hard Matsubara loops, whose effect is to modify the long-distance coefficients of the zero-mode EFT. Large loop corrections for two-step transitions were also reported in~\cite{Niemi:2020hto} in a triplet-extended model.

Our analysis was performed at a fixed renormalization scale $\bar{\mu}_3 = T$ and it is important to estimate how sensitive the results are to variations of this scale (as discussed above, the remaining scale dependence is of higher order than the calculation at hand). In our EFT approach the soft IR modes evolve according to the exact RG equations in 3d, so we may expect weaker scale dependence than what is observed in 4d effective potentials. In fig.~\ref{fig:diffscale} we plot the effective potential at different values of $\bar{\mu}_3$ in three benchmark points, together with the one-loop potential for comparison. The scale dependence is seen to be extremely mild when the transition occurs through a tree-level barrier. For the radiatively generated transition in BM6 the scale variations can shift the critical temperature by a few GeV, but the transition itself is so weak that perturbation theory may not be accurate due to non-perturbative gauge loops. 

What is not shown in fig.~\ref{fig:diffscale} is possible sensitivity to the matching scale at which the EFT is constructed, corresponding to residual scale dependence of the hard thermal corrections. By varying the matching scale in the interval $[5T, 9T]$ we have checked that this scale ambiguity affects the two-loop results even less than the residual dependence on the 3d scale $\bar{\mu}_3$. In contrast the one-loop potential is quite sensitive to the matching scale, which is not surprising since we already concluded that two-loop corrections are non-negligible. Dependence on the renormalization scale is further discussed in~\cite{Gould:2021oba}.

\begin{table}[t]
\begin{center}
	\begin{tabular}{|c |c c c c c|}
		\hline
		\textbf{} & BM8 & BM9 & BM10 & BM11 & BM12 \\ [0.5ex]
		\hline
		$M_{h_2}$ [GeV] & 200 & 250 & 325 & 350 & 375 \\
		\hline
		$\sin\theta$ & 0 & 0 & 0 & 0 & 0  \\
		\hline
		$a_2$ & 1.5 & 2.25 & 3.5 & 4.0 & 4.5 \\
		\hline
		$b_3$ [GeV] & 0 & 0 & 0 & 0 & 0 \\
		\hline
		$b_4$ & 0.3 & 0.3 & 0.3 & 0.3 & 0.3 \\
		\hline
		\multicolumn{6}{|l|}{ \hspace{2.8cm} One loop}  \\
		\hline
		$\Tc$ [GeV] & 163.8 & 132.4 & 113.8 & 94.3 & 86.7 \\
		\hline
		$v_\text{phys}/\Tc$ & 0.74 & 1.83 & 2.54 & 3.22 & 3.60 \\
		\hline
		$\Delta v/\Tc$ & 0.68 & 1.75 & 2.44 & 3.11 & 3.49 \\
		\hline
		$\Delta x/\Tc$ & 0.60 & 1.35 & 1.76 & 2.15 & 2.33 \\
		\hline
		$L/\Tc^4$ & 0.068 & 0.47 & 1.12 & 2.16 & 3.24 \\
		\hline
		\multicolumn{6}{|l|}{ \hspace{2.8cm} Two loop}  \\
		\hline
		
		$\Tc$ [GeV] & 135.4 & 108.8 & 120.5 & 97.4 & 96.1  \\
		\hline
		$v_\text{phys} /\Tc$ & 1.07 & 1.91 & 1.37 & 2.23 & 2.23  \\
		\hline
		$\Delta v/\Tc$ & 1.00 & 1.82 & 1.24 & 2.11 & 2.10 \\
		\hline
		$\Delta x/\Tc$ & 0.82 & 1.43 & 0.95 & 1.51 & 1.45  \\
		\hline
		$L/\Tc^4$ & 0.21 & 0.53 & 0.15 & 0.41 & 0.26 \\
		\hline
		$\delta_\text{EFT}$ & 0.0021 & 0.0024 & 0.012 & 0.035 & 0.06 \\
		\hline
	\end{tabular}
\end{center}
\caption{\label{tab:2step}
Collected results for two-step EWPT in chosen benchmark points.
We consider only transitions in the Higgs direction.
}
\label{table:2step}
\end{table}

The validity of our high-$T$ prescription is estimated as follows. After minimizing the two-loop potential at a given $T$, we add to tree-level $\Veff$ the dimension five and six operators from eq.~(\ref{eq:dim6}) that capture the dominant higher-order corrections to our high-$T$ EFT. We then locate the minimum once more using the modified potential and look at the relative change in the Higgs VEV, 
\begin{align}
\label{eq:deltaEFT}
\delta_\text{EFT} = \left| \frac{v - v_\text{mod}}{v} \right| \; .
\end{align}
This quantity estimates the effect of higher-dimensional operators, that were neglected in the main analysis, on properties of the broken phase and particularly on the transition strength. The tables show values of $\delta_\text{EFT}$ at $\Tc$, where $\delta_v$ is evaluated on the $v \neq 0$ side of the transition. We also considered the shift in the singlet VEV, but in all of our points it was smaller than the respective change in $v$. In most of our benchmarks the relative error is small, less than one percent. Hence our 3d EFT should give a very accurate description of at least the broken phase. The excellent performance of high-$T$ approximations has also been demonstrated in \cite{Laine:2017hdk} for a two-Higgs doublet model.

The above accuracy considerations suggest that the uncertainty in our two-loop results should be at most of order $10\%$ and considerably less in most points. One exception is BM12, where the $a_2$ coupling runs rapidly to non-perturbative values. Of course, the simple arguments presented here cannot conclusively rule out the possibility of sizable corrections at higher loop orders or at the non-perturbative level. Considering how much the two-loop terms affect $T_c$ and $L/T_c^4$ in several points, we may even anticipate large higher-order effects in at least some region of the interesting parameter space.

\section{Discussion}
\label{sec:conclusions}

In the paper at hand, we carried out a two-loop 
study of the electroweak phase transition in the singlet-extended Standard model (xSM), quantifying the magnitude of two-loop corrections on equilibrium characteristics of the transition. The study involved radiatively-induced transitions and transitions through a tree-level barrier, as well as two-step phase transitions. In all cases we found considerable deviations from one-loop predictions, in particular the critical temperature and latent heat typically change by at least $20\%$, but particularly for very strong two-step transitions the differences can be $100\%$ or more. 

Although we focused on a limited selection of points in the free parameter space of the xSM, the main conclusion should be clear: one-loop analyses are generally not very accurate in predicting the strength nor the interesting temperature range(s) of the EWPT. In our opinion, this result by itself is not very surprising, because strong transitions in the xSM are typically associated with at least $\mathcal{O}(1)$ coupling constants in the scalar sector, and because the perturbative expansion converges more slowly at high temperature than it does at $T=0$. We consequently anticipate that two-loop corrections 
can significantly impact predictions of, for example, the bubble nucleation rate and other quantities relevant for gravitational waves and baryogenesis. Our results strongly motivate improved studies of the EWPT and its cosmological consequences at the two-loop level, not only in the xSM but also in other scalar extensions.

On the technical side, we wish to emphasize that the dimensionally-reduced approach to the EWPT described in this paper provides a systematic and very intuitive approach to thermal resummations beyond one-loop level. It is also quite straightforward to apply the method to other BSM models and in fact, dimensionally-reduced investigations have already been performed in several such models~\cite{Bodeker:1996pc,Kainulainen:2019kyp,Niemi:2020hto}. The approach is limited to theories in the formal high-$T$ limit, but tests of its validity indicate good accuracy even when there are relatively heavy excitations in the spectrum~\cite{Laine:2017hdk,Kainulainen:2019kyp,Niemi:2020hto}. In the present paper, we found dimensional reduction accurate also in the xSM. This observation motivates one more approach of studying phase transitions in the xSM, namely that of non-perturbative 3d simulations, which allow for a very accurate determination of the relevant thermodynamics at the cost of increased numerical effort.

\textbf{Note added on March 2024.} In previous versions of the paper there was an error in our numerical implementation of eq.~(\ref{eq:deltag}) which affected our benchmark results somewhat. This has been corrected in the present version with figures and tables updated accordingly. In addition, we have corrected misprints in equations (\ref{eq:match_x3}) and (\ref{eq:match_x3p}). See the published Erratum \cite{PhysRevD.109.039902} for further details.

\begin{acknowledgments}
We thank Oliver Gould, Michael J. Ramsey-Musolf, Kari Rummukainen
and Juuso Österman for helpful discussions.
LN is supported
by the Jenny and Antti Wihuri Foundation and
by Academy of Finland grant no.~308791.
PS has been supported
by the European Research Council, grant no.~725369, and
by the Academy of Finland, grant no.~1322507.
\end{acknowledgments}

\onecolumngrid
\newpage
\appendix
\allowdisplaybreaks
\begin{center}{\large \textbf{Appendix}} \end{center}

\section{Renormalization and input parameters}
\label{sec:renormalization}

In this appendix, we discuss one-loop corrections to the parametrization discussed in section~\ref{sec:model}. The goal is to relate the renormalized parameters $g, g', y_t$, and those appearing in the potential (\ref{eq:pot}), to experimental observables such as particle masses. We apply the \MSbar prescription for all counterterms (\textit{i.e.}\ only the $1/\epsilon$ poles are subtracted), but the overall procedure is similar to standard EW precision calculations that often use the on-shell scheme. Our approach is a straightforward generalization of that described in section 5 of~\cite{Kajantie:1995dw}, also in the context of high-$T$ physics. Refs.~\cite{Lopez-Val:2014jva,Bojarski:2015kra,Kanemura:2015fra} discuss renormalization of the xSM in slightly different contexts. In this appendix only, we work in Minkowski space at zero temperature.

\subsection{Counterterms}
\label{sec:counters}

The Lagrangian (\ref{eq:Lag-4d}) is written in terms of bare (non-renormalized) parameters. Working in $D = 4 - 2\epsilon$ dimensions, we renormalize the Higgs doublet as 
\begin{align}
\phi \rightarrow Z^{1/2}_\phi \phi = (1 + \delta Z_\phi)^{1/2} \phi,
\end{align}
and similarly for the other fields ($A_\mu, B_\mu, q_t, t, S$) appearing in the bare Lagrangian.\footnote{Renormalization of the leptonic and QCD sectors can be ignored
  in our discussion.
}
Our renormalized parameters are obtained from the bare quantities by
\begin{align}
  g &\to \mu^\epsilon (g + \delta g) &
  g' &\to \mu^\epsilon (g' + \delta g') &
  \gY &\to Z^{-1/2}_\phi Z^{-1/2}_{q_t} Z^{-1/2}_t  \mu^\epsilon (y_t + \delta y_t)
\nonumber \\
  m^2_\phi &\to m^2_\phi + \delta m^2_\phi &
  m^2_S &\to m^2_S + \delta m^2_S &
  b_1 &\to Z^{-1/2}_S \mu^{-\epsilon} (b_1 + \delta b_1)
\nonumber \\
  \lambda &\to Z^{-2}_\phi\mu^{2\epsilon}(\lambda + \delta \lambda) &
  b_4 &\to Z^{-2}_S \mu^{2\epsilon}(b_4 + \delta b_4) &
  a_2 &\to Z^{-1}_\phi Z^{-1}_S \mu^{2\epsilon}(a_2 + \delta a_2) \nonumber \\
  b_3 &\to Z^{-3/2}_S \mu^\epsilon (b_3 + \delta b_3) &
  a_1 &\to Z^{-1}_\phi Z^{-1/2}_S \mu^\epsilon (a_1 + \delta a_1)
.
\end{align}
Here $\mu$ is the scale associated with dimensional regularization. Running will be given in terms of the \MSbar scale $\bar\mu^2 = 4\pi e^{-\gamma} \mu^2$.

We present the counterterms in a general $R_\xi$ gauge.
At one loop, they read
\begin{align}
\delta Z_\phi & = \frac{1}{(4\pi)^2}\frac{1}{\epsilon} \Big( \frac34 (3-\xi) g^2 + \frac14 (3 - \xi) {g'}^2 - 3 y_t^2 \Big) \\
\delta Z_S & = 0 \\
\delta Z_A & = \frac{g^2}{(4\pi)^2} \frac{1}{\epsilon} \Big( \frac{25}{6} - \frac{4}{3}\nf - \xi \Big) \\
\delta Z_B & = -\frac{{g'}^2}{(4\pi)^2} \frac{1}{\epsilon} \frac{1}{6}\Big(1 + \frac{40}{3} \nf \Big) \\
\delta Z_q & = -\frac{1}{(4\pi)^2} \frac{1}{\epsilon} \Big( \frac12 y_t^2 + \frac34 g^2 \xi + \frac{1}{36} {g'}^2 \xi + \frac43 \gs^2 \xi \Big) \\
\delta Z_t & = -\frac{1}{(4\pi)^2} \frac{1}{\epsilon} \Big( y_t^2 + \frac49 {g'}^2 \xi + \frac43 \gs^2 \xi \Big) \\
\delta g & = -\frac{g^3}{(4\pi)^2} \frac{1}{\epsilon} \Big( \frac{43}{12} - \frac23 \nf \Big) \\
\delta g' & = \frac{{g'}^3}{(4\pi)^2} \frac{1}{\epsilon} \frac{1}{12} \Big( 1 + \frac{40}{3} \nf \Big) \\
\delta y_t & = - \frac{y_t}{(4\pi)^2} \frac{1}{\epsilon} \Big( \frac34 g^2 \xi + \frac13 (1 + \frac{13}{12} \xi) {g'}^2 + 4 (1 + \frac13 \xi) \gs^2  \Big) \\
\delta m^2_\phi & = \frac{1}{(4\pi)^2} \frac{1}{\epsilon} \frac14 \Big( 24\lambda m^2_\phi + 2a_2 m^2_S + a_1^2 - m^2_\phi (3g^2 + {g'}^2)\xi  \Big) \\
\delta m^2_S & = \frac{1}{(4\pi)^2} \frac{1}{\epsilon} \frac12 ( 6 b_4 m^2_S + 4 a_2 m^2_\phi + 4 b_3^2 + a_1^2) \\
\delta b_1 & = \frac{1}{(4\pi)^2} \frac{1}{\epsilon} \Big( b_3 m^2_S + a_1 m^2_\phi \Big) \\
\delta b_3 & = \frac{1}{(4\pi)^2} \frac32 \frac{1}{\epsilon} \Big( 6b_3 b_4 + a_1 a_2 \Big) \\
\delta a_1 & = \frac{1}{(4\pi)^2} \frac{1}{\epsilon} 2 \Big( 3 a_1 \lambda + (b_3 + a_1) a_2 - \frac18 (3g^2 + {g'}^2) \xi a_1 \Big) \\
\delta b_4 & = \frac{1}{(4\pi)^2} \frac{1}{\epsilon} \Big( 9 b_4^2 + a_2^2 \Big) \\
\delta \lambda & = \frac{1}{(4\pi)^2} \frac{1}{\epsilon} \Big( \frac{3}{16} ( 3g^4 + 2 g^2 {g'}^2 + {g'}^4) + 12 \lambda^2 + \frac14 a_2^2 - 3 y_t^4 - \frac12 \lambda (3g^2 + {g'}^2)\xi \Big) \\ 
\delta a_2 & = \frac{1}{(4\pi)^2} \frac{1}{\epsilon} a_2 \Big( 6\lambda + 2 a_2 + 3b_4 - \frac14 (3g^2 + {g'}^2)\xi \Big).
\end{align}
In the above, $\nf = 3$ stands for the number of active fermion generations.
The one-loop counterterms are sufficient to render finite the EFT matching relations at leading order in high-$T$ expansion (section~\ref{sec:DR}) and the zero-temperature self energies in the calculation below. To eliminate dependence on the matching scale in eqs.~(\ref{eq:DR-matching-start}) -- (\ref{eq:DR-matching-end}), $\beta$ functions are needed at one loop. These are obtained in the usual fashion by requiring that the bare parameters are independent of the RG scale. The resulting expressions can be read from section 3.2 of ref.~\cite{Brauner:2016fla} (after converting their scalar potential into our notation).

\subsection{\MSbar parameters in terms of physical observables}

In the EW sector our choice of observables are the Fermi constant $G_\mu$ and pole masses of the top quark, the $W$ and $Z$ bosons, and that of the observed Higgs boson. Numerical values are~\cite{Zyla:2020zbs} $G_\mu = 1.1663787 \times 10^{-5} \text{ GeV}^{-2}$ and 
\begin{align}
( M_t, M_W, M_Z, M_{h_1} ) = (172.76, \; 80.379, \; 91.1876, \; 125.10 ) \; \text{GeV}.
\end{align}
The BSM parameters are fixed by specifying the pole mass of the new scalar, $M_{h_2}$, and by giving $\sin\theta, b_3, b_4, a_2$ directly in the \MSbar scheme. For the QCD coupling, we use the value $\gs^2 = 1.48409$ and neglect its running.

As in section~\ref{sec:model}, we introduce VEVs so that the scalar potential is minimized at tree level, with $\langle S \rangle = 0$. Mixing angle $\theta$ is taken to diagonalize the scalar mass matrix at tree level, and in what follows we identify the lighter eigenstate with the SM Higgs, $M_{h_1} \leq M_{h_2}$. If the lighter $h_1$ is instead taken to be the BSM excitation, the expressions below are valid after interchanging $M_{h_1} \leftrightarrow M_{h_2}$ and $\cos\theta \leftrightarrow \sin\theta$.

The two-point functions are, in terms of the \MSbar renormalized masses $m(\bar\mu)$, of the form
\begin{align}
\langle h_i(-p) h_i(p)\rangle =& i\frac{1}{p^2 - m^2_{h_i}(\bar\mu) + \Pi_{h_i}(p^2)} \nonumber \\
\langle Z_\mu(-p) Z_\nu(p)\rangle =& -i\frac{g_{\mu\nu} - p_\mu p_\nu/p^2}{p^2 - m^2_Z(\bar\mu) + \Pi_Z(p^2) } + \text{longitudinal part} \nonumber \\
\langle \psi_\alpha(p) \bar\psi_\beta(p) \rangle =& i\Big[\frac{1}{p_\mu \gamma^\mu - m_f(\bar\mu) + \Sigma(p)} \Big]_{\alpha\beta}
\end{align}
and the $W$ boson propagator is analogous to that of $Z_\mu$. In addition to the usual 1PI contributions, the self energies contain one-loop diagrams involving 1-particle-reducible $h_1$ or $h_2$ lines (``tadpoles''), because the scalar potential is minimized only at tree level.

The pole-mass conditions for bosons are of the form $m^2(\bar\mu) = M^2 + \RE \Pi(M^2)$, from which the renormalized masses can be solved. We write the fermionic self energies as
\begin{align}
\Sigma(p) = m_f(\bar\mu) \Sigma_s(p^2) + \gamma_\mu p^\mu (p^2) + \gamma_\mu p^\mu \gamma^5 \Sigma_a(p^2),
\end{align}
and treating the top the top quark as an asymptotic state, only the scalar ($\Sigma_s$) and vector ($\Sigma_v$) parts contribute to its pole condition:
\begin{align}
m^2(\bar\mu) = M_t^2 \left( 1 + 2 \RE \left[ \Sigma_s(M_t^2) + \Sigma_v(M_t^2) \right] \right).
\end{align}

The $\gr{SU(2)}_L$ gauge coupling can be fixed by relating it to Fermi constant, which describes muon decay in the low-energy theory obtained by integrating out weak interactions. Conventionally, the matching relation between the full EW theory and the Fermi EFT is written, in the on-shell scheme, as 
\begin{align}
\frac{G_\mu}{\sqrt{2}} = \frac{1}{8M_W^2} \frac{g^2_\text{os}}{1 - \Delta r}.
\end{align}
$\Delta r$ contains contributions from weak gauge bosons and scalars, but not from the QED sector which cancels when performing the matching. The on-shell (OS) coupling is related to our \MSbar gauge coupling by $g^2(\bar\mu) = g^2_\text{os} (1 + \delta g^2_\text{os} / g^2_\text{os})$, where $\delta g^2_\text{os}$ is the regularized OS counterterm (this relation follows because the bare coupling must be equal in both schemes).

In Feynman-t'Hooft gauge and with our sign conventions, $\Delta r$ is given at one-loop by~\cite{Sirlin:1980nh}
\begin{align}
\Delta r =& -\frac{\RE \Pi_W(M_W^2) - \Pi_W(0)}{M_W^2} + \frac{\delta g^2_\text{os}}{g^2_\text{os}} \nonumber \\
& + \frac{g^2_\text{os}}{16\pi^2} \Big[ 4 \ln \frac{\bar\mu^2}{M_W^2} + \left( \frac72 \frac{M_Z^2}{M_Z^2 - M_W^2} - 2 \right) \ln \frac{M_W^2}{M_Z^2} + 6 \Big] + \Delta r_\text{BSM}.
\end{align}
Contributions to muon decay from BSM physics appear both in the self energy $\Pi_W$ and $\Delta r_\text{BSM}$, the latter of which collects diagrams involving interactions of the scalars with light fermions. However, such contributions are suppressed by light fermion masses and are neglected in our analysis; hence we take $\Delta r_\text{BSM} = 0$. We now define 
\begin{align}
g^2_0 \equiv \frac{g^2_\text{os}}{1-\Delta r} = \frac{8 M_W^2 G_\mu}{\sqrt{2}}
\end{align}
and replace $g^2_\text{os} \rightarrow g^2_0$ inside one-loop corrections. The difference is of higher order. Therefore,
\begin{align}
\label{eq:msbar-1loop-begin}
g^2(\bar\mu) =& g^2_\text{0} \left(1 + \frac{\delta g^2}{g^2_0} \right),
\end{align}
where
\begin{align}
\label{eq:deltag}
\frac{\delta g^2}{g^2_0} =& \frac{\RE\Pi_W(M_W^2) - \Pi_W(0)}{M_W^2} - \frac{g^2_0}{16\pi^2} \Big[ 4 \ln \frac{\bar\mu^2}{M_W^2} + \left( \frac72 \frac{M_Z^2}{M_Z^2 - M_W^2} - 2 \right) \ln \frac{M_W^2}{M_Z^2} + 6 \Big].
\end{align}
This fixes the gauge coupling at a given scale; in practice we choose $\bar{\mu} = M_Z$.

Expanding equations (\ref{eq:params-tree1}) -- (\ref{eq:params-tree2}), the defining relations for $g'^2, y_t^2$ and the VEV $v^2 = 4 m_W^2 / g^2$ to linear order in the (regularized) self energies and $\delta g^2 / g_0^2$, the remaining \MSbar parameters read 
\begin{align}
{g'}^2(\bar\mu) =& \frac{g_0^2}{M_W^2} \Big[ \left(M_Z^2 - M_W^2 \right) \Big(1 + \frac{\delta g^2}{g^2_0} \Big) - M_Z^2 \frac{\RE \Pi_W(M_W^2)}{M_W^2} + \RE \Pi_Z(M_Z^2) \Big] \\
y_t^2(\bar\mu) =& \frac{g_0^2}{2}\frac{M_t^2}{M_W^2} \Big[ 1 + \frac{\delta g^2}{g^2_0} - \frac{\RE \Pi_W(M_W^2)}{M_W^2} + 2 \RE \left( \Sigma_s(M_t^2) + \Sigma_v(M_t^2) \right) \Big] \\
m^2_\phi(\bar\mu) =& -\frac14 \Big[ M^2_{h_1} + M^2_{h_2} + \left( M_{h_1}^2 - M_{h_2}^2\right) \cos 2\theta + 2 \RE \Pi_{h_1}(M_{h_1}^2)\cos^2\theta \nonumber \\
&  + 2 \RE \Pi_{h_2}(M_{h_2}^2) \sin^2\theta \Big] \\ 
m^2_S(\bar\mu) =& \frac12 \Big[ M^2_{h_1} + M^2_{h_2} + \Big( M^2_{h_2} - M^2_{h_1}\Big) \cos2\theta + 2 \RE \Pi_{h_1}(M_{h_1}^2)\sin^2\theta  \nonumber \\
& + 2 \RE \Pi_{h_2}(M_{h_2}^2)\cos^2\theta \Big] - a_2\frac{2M_W^2}{g_0^2} \Big[ 1 - \frac{\delta g^2}{g^2_0} + \frac{\RE \Pi_W(M_W^2)}{M_W^2} \Big] \\
b_1(\bar\mu) =& -\frac{M_W}{g_0}\left( M_{h_2}^2 - M_{h_1}^2 \right)\cos\theta \sin\theta \nonumber \\
& \times \Big[ 1 - \frac12 \frac{\delta g^2}{g^2_0} + \frac{\RE \Pi_{h_2}(M_{h_2}^2) - \RE \Pi_{h_1}(M_{h_1}^2)}{M_{h_2}^2 - M_{h_1}^2} + \frac12 \frac{\RE \Pi_W(M_W^2)}{M_W^2} \Big] \\
a_1(\bar\mu) =& \frac{g_0}{M_W} \left( M_{h_2}^2 - M_{h_1}^2 \right)\cos\theta\sin\theta \nonumber \\
& \times \Big[ 1 + \frac12 \frac{\delta g^2}{g^2_0} + \frac{\RE \Pi_{h_2}(M_{h_2}^2) - \RE \Pi_{h_1}(M_{h_1}^2)}{M_{h_2}^2 - M_{h_1}^2} - \frac12 \frac{\RE \Pi_W(M_W^2)}{M_W^2} \Big] \\
\label{eq:msbar-1loop-end}
\lambda(\bar\mu) =& \frac{g_0^2}{16 M_W^2} \Big[ \Big( M_{h_1}^2 + M_{h_2}^2 + \left( M_{h_1}^2 - M_{h_2}^2 \right) \cos2\theta \Big) \Big( 1 + \frac{\delta g^2}{g^2_0} - \frac{\RE \Pi_W(M_W^2)}{M_W^2} \Big)  \nonumber \\
& + 2 \RE \Pi_{h_1}(M_{h_1}^2) \cos^2\theta + 2 \RE \Pi_{h_2}(M_{h_2}^2) \sin^2\theta \Big].
\end{align}
Once the self energies are known, equations (\ref{eq:msbar-1loop-begin}) through (\ref{eq:msbar-1loop-end}) give one-loop corrected relations between the renormalized parameters and physical input.

\subsection{Expressions for the self energies}

The zero-temperature self energies can be evaluated using a standard set of Passarino-Veltman integrals. We denote, in $D=4-2\epsilon$ dimensions,
\begin{align}
A(m) =&  \mu^{2\epsilon} \int \frac{d^D p}{i (2\pi)^D} \frac{1}{p^2 - m^2} = \frac{m^2}{(4\pi)^2} \left( \frac{1}{\epsilon} + 1 + \ln\frac{\bar\mu^2}{m^2} \right)  \\
B_{0; \mu; \mu\nu}(k, m_1, m_2) =& \mu^{2\epsilon} \int \frac{d^D p}{i (2\pi)^D} \frac{1; p_\mu ; p_\mu p_\nu}{(p^2 - m_1^2) \left( (p+k)^2 - m_2^2 \right )}.
\end{align}
Here 
\begin{align}
B_0(k, m_1, m_2) =& \frac{1}{(4\pi)^2} \Big( \frac{1}{\epsilon} + 1 + \frac12 \ln\frac{\bar\mu^2}{m_2^2} + \frac12 \ln\frac{\bar\mu^2}{m_1^2} - \frac12 \frac{m_1^2 + m_2^2}{m_1^2 - m_2^2} \ln \frac{m_1^2}{m_2^2} + F(k, m_1, m_2)  \Big) ,
\end{align}
where the function $F(k, m_1, m_2)$ is defined in Appendix B of ref.~\cite{Bohm:1986rj} and vanishes for $k^2 = 0$.
The vector and tensor integrals have decompositions
\begin{align}
B_\mu(k, m_1, m_2) =& k_\mu B_1(k, m_1, m_2) \\
B_{\mu\nu}(k, m_1, m_2) =& g_{\mu\nu} B_{00}(k, m_1, m_2) + (k_\mu k_\nu \; \text{part}),
\end{align}
with
\begin{align}
B_{00}(k, m_1, m_2) =& \frac{1}{2(D-1)} \Big( 2m_1^2 B_0(k, m_1, m_2) \nonumber \\
& + A(m_2) - (m_2^2 - m_1^2 - k^2) B_1(k, m_1, m_2) \Big) \\ 
B_1(k, m_1, m_2) =& \frac{1}{2k^2} \Big( (m_2^2 - m_1^2 - k^2) B_0(k, m_1, m_2) + A(m_1) - A(m_2) \Big).
\end{align}
We also define
\begin{align}
K(k, m_1, m_2, a, b, c) =& \int_p \frac{ap^2 + b k^\mu p_\mu + c}{(p^2 - m_1^2) ((p+k)^2 - m_2^2)} \nonumber \\
=& a [A(m_2) + m_1^2 B_0(k, m_1, m_2)] + b k^2 B_1(k, m_1, m_2) + c B_0(k, m_1, m_2). 
\end{align}
Several special cases of the $B$ integrals are needed and can be obtained using the expression for $F(k,m_1,m_2)$ in~\cite{Bohm:1986rj} and the definitions of $B_1, B_{00}$, for example 
\begin{align}
B_{00}(0, m_1, m_2) = \frac{1}{D} \left( A(m_2) + m_1^2 B_0(0,m_1,m_2) \right).
\end{align}

We present the self energies in the $\xi = 1$ $R_\xi$ gauge. The resulting expressions for renormalized parameters are gauge invariant because $\Delta r$ and self energies at their respective poles are. Below we denote $\Nc^f = 3$ for quarks and $\Nc^f = 1$ for leptons, and fermions apart from the top quark are taken to be massless in the following expressions. Useful shorthand notation for couplings between $h_1,h_2$ and the would-be Goldstones are
\begin{align}
  c_{111} =& -3 v^{-1} m^2_{h_1} \cos^3\theta - 3v a_2 \cos\theta \sin^2\theta + 2 b_3 \sin^3 \theta \\ 
c_{222} =& -3 v^{-1} m^2_{h_2} \sin^3 \theta - 3v a_2 \sin\theta \cos^2\theta - 2b_3 \cos^3\theta \\
c_{112} =& \frac{\sin\theta}{2v} \Big( v^2 a_2 - 2 m^2_{h_1} - m^2_{h_2} + \left( 3v^2 a_2 - 2 m^2_{h_1} - m^2_{h_2} \right) \cos 2\theta \Big) - 2 b_3 \cos\theta \sin^2\theta  \\
c_{122} =& \frac{\cos\theta}{2v} \Big( v^2 a_2 - m^2_{h_1} - 2m^2_{h_2} - \left( 3v^2 a_2 - m^2_{h_1} - 2 m^2_{h_2} \right) \cos 2\theta \Big) + 2 b_3 \sin\theta \cos^2\theta \\
c_{1111} =& -\frac{3\cos^4\theta}{2v^2}  \Big(m^2_{h_1} + m^2_{h_2} + (m^2_{h_1} - m^2_{h_2}) \cos 2\theta \Big) - 6 \left(b_4 \sin^2\theta  + a_2 \cos^2 \theta \right) \sin^2\theta  \\
c_{2222} =& -\frac{3\sin^4\theta}{2v^2} \Big(m^2_{h_1} + m^2_{h_2} + (m^2_{h_1} - m^2_{h_2}) \cos 2\theta \Big) - 6 \left( b_4 \cos^2\theta + a_2 \sin^2\theta \right) \cos^2\theta \\
c_{1122} =& - \frac{3\sin^2 2\theta}{8v^2} \Big( m^2_{h_1} + m^2_{h_2} + (m^2_{h_1} - m^2_{h_2}) \cos 2\theta \Big) -\frac14 \Big( 3b_4 + a_2 + 3(a_2 - b_4)\cos 4\theta \Big)  \\
c_{h_1 h_1 G G} =& - \frac{\cos^2\theta}{2v^2}  \Big( m^2_{h_1} + m^2_{h_2} + (m^2_{h_1} - m^2_{h_2}) \cos 2\theta \Big) - a_2 \sin^2\theta \\
c_{h_2 h_2 G G} =& - \frac{\sin^2\theta}{2v^2} \Big( m^2_{h_1} + m^2_{h_2} + (m^2_{h_1} - m^2_{h_2}) \cos 2\theta \Big) - a_2 \cos^2\theta .
\end{align}

Tadpole diagrams with an external $h_1$ or $h_2$ leg sum up to
\begin{align}
t_{h_1} =& -4 \cos\theta N^t_c \frac{m_t^2}{v}  A(m_t) + \frac14 (D-1) v \cos\theta \Big( 2g^2 A(m_W) + (g^2 + {g'}^2) A(m_Z) \Big) \nonumber \\
& + \frac{m_{h_1}^2}{2v} \cos\theta \Big( 2 A(m_W) + A(m_Z)\Big) - \frac12 \Big( c_{111} A(m_{h_1}) + c_{122} A(m_{h_2}) \Big)
\end{align}
\begin{align}
t_{h_2} =& -4 \sin\theta \Nc^t \frac{m_t^2}{v} A(m_t) + \frac14 (D-1) v \sin\theta \Big( 2g^2 A(m_W) + (g^2 + {g'}^2) A(m_Z) \Big) \nonumber \\
& + \frac{m_{h_2}^2}{2v} \sin\theta \Big( 2 A(m_W) + A(m_Z)\Big) - \frac12 \Big( c_{112} A(m_{h_1}) + c_{222} A(m_{h_2}) \Big).
\end{align}
The full self energies, including tadpole insertions, are as follows:
\begin{align}
\label{eq:self-h1}
\Pi_{h_1}(k^2) &= \frac{c_{111}}{m_{h_1}^2} t_{h_1} + \frac{c_{112}}{m_{h_2}^2} t_{h_2} -2 \cos^2\theta \Nc^t y_t^2 K(k, m_t, m_t, 1, 1, m_t^2) \nonumber \\ 
& -\frac14 \cos^2\theta \Big( 2g^2 K(k, m_W, m_W, 1, 4, 4k^2) + (g^2 + {g'}^2) K(k, m_Z, m_Z, 1, 4, 4k^2) \Big) \nonumber \\ 
& + \frac{1}{16} \cos^2\theta \Big\{ 8D g^2 A(m_W) + 4D (g^2 + {g'}^2) A(m_Z) \nonumber \\
& + (2D-1)v^2 \Big( 2g^4 B_0(k, m_W, m_W) +(g^2 + {g'}^2)^2 B_0(k, m_Z, m_Z) \Big) \Big\} \nonumber \\
& + \frac{m_{h_1}^4 \cos^2\theta}{v^2} \Big( B_0(k, m_W, m_W) + \frac12 B_0(k, m_Z, m_Z) \Big) \nonumber \\
&+ \frac12 \Big( c_{111}^2 B_0(k, m_{h_1}, m_{h_1}) + 2 c_{112}^2 B_0(k, m_{h_1}, m_{h_2}) + c_{122}^2 B_0(k, m_{h_2}, m_{h_2}) \Big) \nonumber \\ 
& -\frac12 \Big( c_{1111} A(m_{h_1}) + c_{1122} A(m_{h_2}) + 2c_{h_1 h_1 G G} A(m_W) + c_{h_1 h_1 G G} A(m_Z) \Big)
\end{align}
\begin{align}
\Pi_{h_2}(k^2) &= \frac{c_{122}}{m_{h_1}^2} t_{h_1} + \frac{c_{222}}{m_{h_2}^2} t_{h_2} -2 \sin^2\theta \Nc^t y_t^2 K(k, m_t, m_t, 1, 1, m_t^2) \nonumber \\ 
& -\frac14 \sin^2\theta \Big( 2g^2 K(k, m_W, m_W, 1, 4, 4k^2) + (g^2 + {g'}^2) K(k, m_Z, m_Z, 1, 4, 4k^2) \Big) \nonumber \\ 
& + \frac{1}{16} \sin^2\theta \Big\{ 8D g^2 A(m_W) + 4D (g^2 + {g'}^2) A(m_Z) \nonumber \\
& + (2D-1)v^2 \Big( 2g^4 B_0(k, m_W, m_W) +(g^2 + {g'}^2)^2 B_0(k, m_Z, m_Z) \Big) \Big\} \nonumber \\
& + \frac{m_{h_2}^4 \sin^2\theta}{v^2} \Big(B_0(k, m_W, m_W) + \frac12 B_0(k, m_Z, m_Z) \Big) \nonumber \\ 
&+ \frac12 \Big( c_{222}^2 B_0(k, m_{h_2}, m_{h_2}) + 2 c_{122}^2 B_0(k, m_{h_1}, m_{h_2}) + c_{112}^2 B_0(k, m_{h_1}, m_{h_1}) \Big) \nonumber \\ 
& -\frac12 \Big( c_{2222} A(m_{h_2}) + c_{1122} A(m_{h_1}) + 2 c_{h_2 h_2 G G} A(m_W) + c_{h_2 h_2 G G} A(m_Z) \Big).
\end{align}
For gauge bosons,
\begin{align}
\Pi_W(k^2) =& -\frac12 g^2 v  \Big( \frac{\cos\theta}{m_{h_1}^2} t_{h_1} + \frac{ \sin\theta}{m_{h_2}^2} t_{h_2} \Big) + g^2 \sum_\text{families} \Nc^f \left( 2B_{00}(k, 0, m_f) - K(k, 0, m_f, 1, 1, 0)\right) \nonumber \\
& + \frac14 g^2 \Big( \cos^2\theta A(m_{h_1}) + \sin^2\theta A(m_{h_2}) + 2 A(m_W) + A(m_Z) \Big) \nonumber \\
& - g^2 \Big( \cos^2\theta B_{00}(k, m_{h_1}, m_W) + \sin^2\theta B_{00}(k, m_{h_2}, m_W) + B_{00}(k, m_Z, m_W) \Big)  \nonumber \\
& + \frac14 g^4 v^2 \Big( \cos^2\theta B_0(k, m_{h_1}, m_W) + \sin^2\theta B_0(k, m_{h_2}, m_W) \Big) \nonumber \\
& + \frac14 \frac{g^2 {g'}^2}{g^2 + {g'}^2}v^2 \Big( g^2 B_0(k, m_W, 0) + {g'}^2 B_0(k, m_W, m_Z) \Big)  \nonumber \\
& + (D-1) g^2 \Big( A(m_W) + \frac{g^2}{g^2 + {g'}^2} A(m_Z) \Big) \nonumber \\
& - \frac{2g^2}{g^2 + {g'}^2} \Big\{ (2D-3) \Big( g^2  B_{00}(k, m_W, m_Z) + {g'}^2 B_{00}(k, m_W, 0) \Big) - g^2 B_{00}(k, m_W, m_Z) \nonumber \\
& - {g'}^2 B_{00}(k, m_W, 0) + \frac12 g^2 K(k, m_W, m_Z, 2, 2, 5k^2) + \frac12 {g'}^2 K(k, m_W, 0, 2, 2, 5k^2) \Big\} .
\end{align}
Here the sum runs over quark and lepton families (6 total), and $m_f = m_t$ for the top quark and zero otherwise.
\begin{align}
\Pi_Z(k^2) =& -\frac12 v \left(g^2 + {g'}^2\right) \Big( \frac{\cos\theta }{m_{h_1}^2} t_{h_1} + \frac{\sin\theta }{m_{h_2}^2} t_{h_2} \Big) + \frac{2g^4}{g^2 + {g'}^2} B_{00}(k, m_W ,m_W) \nonumber \\ 
& + \frac12 (g^2 + {g'}^2) \sum_{f} \Nc^f (1 + a_f) \Big( 2B_{00}(k, m_f, m_f) - K(k, m_f, m_f, 1,1, -\frac{a_f}{a_f + 1}m_f^2) \Big)  \nonumber \\
& + \frac14 (g^2 + {g'}^2) \Big( \cos^2\theta A(m_{h_1}) + \sin^2\theta A(m_{h_2}) + A(m_Z) \Big) + \frac12 \frac{(g^2 - {g'}^2)^2}{(g^2 + {g'}^2)} A(m_W) \nonumber \\
& - \frac{2g^4}{g^2 + {g'}^2} (1-D) A(m_W) - (g^2 + {g'}^2) \Big( \cos^2\theta B_{00}(k, m_{h_1}, m_Z) + \sin^2\theta B_{00}(k, m_{h_2}, m_Z) \Big) \nonumber \\
& - \frac{(g^2 - {g'}^2)^2}{g^2 + {g'}^2} B_{00}(k, m_W, m_W) + \frac12 \frac{g^2 {g'}^4}{g^2 + {g'}^2} v^2 B_0(k, m_W, m_W) \nonumber \\
& + \frac14 (g^2 + {g'}^2)^2 v^2 \Big( \cos^2\theta B_0(k, m_{h_1}, m_Z) + \sin^2\theta B_0(k, m_{h_2}, m_Z) \Big) \nonumber \\
& - \frac{2g^4}{g^2 + {g'}^2} \Big( (2D-3)B_{00}(k,m_W, m_W) + \frac12 K(k, m_W, m_W, 2, 2, 5k^2) \Big).
\end{align}
The constant $a_f$ appearing in the fermionic part is
\begin{align}
a_f = \frac{4 Q_f {g'}^2}{g^2 + {g'}^2} \left( \frac{2{g'}^2}{g^2 + {g'}^2} Q_f \pm 1 \right)
\end{align}
where $Q_f$ is the electric charge of a fermion in units of proton charge, and the minus sign is chosen for up-type quarks. Sum is over all fermions.

Finally, for the top quark we have
\begin{align}
\label{eq:self-top}
\Sigma_s(k^2) + & \Sigma_v(k^2) = -v^{-1} \Big( \frac{\cos\theta}{m_{h_1}^2} t_{h_1} +  \frac{\sin\theta}{m_{h_2}^2} t_{h_2} \Big) \nonumber \\
& + \frac12 y_t^2 \Big\{ \cos^2\theta \Big( B_0(k, m_t, m_{h_1}) - B_1(k, m_t, m_{h_1}) \Big) \nonumber \\ 
& + \sin^2\theta \Big( B_0(k, m_t, m_{h_2}) - B_1(k, m_t, m_{h_2}) \Big) - \Big( B_0(k, m_t, m_Z) + B_1(k, m_t, m_Z) \Big) \Big\} \nonumber \\
& - \frac12 y_t^2 B_1(k, 0, m_W) - \frac14 (D-2) g^2  B_1(k, 0, m_W)  \nonumber \\ 
&  + \left( \frac{D}{9} \right) \frac{3g^2 - {g'}^2}{g^2 + {g'}^2} {g'}^2 B_0(k, m_t, m_Z) - \left( \frac{D-2}{72} \right) \frac{9g^4 - 6g^2 {g'}^2 + 17 {g'}^4}{g^2 + {g'}^2} B_1(k, m_t, m_Z) \nonumber \\
& - \frac49 \frac{g^2 {g'}^2}{g^2 + {g'}^2} \Big( D B_0(k, m_t, 0) + (D-2) B_1(k, m_t, 0) \Big) \nonumber \\
& - \frac43 \gs^2 \Big( D B_0(k, m_t, 0) + (D-2) B_1(k, m_t, 0) \Big).
\end{align}

The parameters appearing in (\ref{eq:self-h1})--(\ref{eq:self-top}) are the \MSbar renormalized ones. As before, inside loop corrections we replace these with the tree-level values, \textit{i.e.} $g \rightarrow g_0$, $m_i \rightarrow M_i$, $y_t^2 \rightarrow g_0^2 M_t^2 / (2M_W^2), v^2 \rightarrow 4M_W^2 / g_0^2$. In the above, we have not written the counterterm contributions explicitly. It can be checked that all $1/\epsilon$ poles cancel from the self energies once the counterterms from section~\ref{sec:counters} are taken into account. The singlet can be decoupled by taking $\theta, b_3, b_4, a_2 \rightarrow 0$. In this limit, the self energies reduce to their SM values and have been cross-checked against Appendix A.2 of ref.~\cite{Croon:2020cgk} (apart from the $Z$ boson self energy).

\section{Two-loop correction to the 3d effective potential}
\label{sec:Veff-2loop}

\begin{figure}[h]
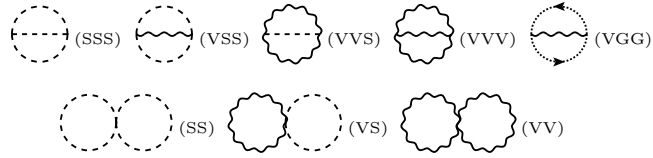

    \centering
    \begin{align*}
        &\ToptVS(\Axx,\Axx,\Lxx)_{\text{(SSS)}}\;
        \ToptVS(\Axx,\Axx,\Lgli)_{\text{(VSS)}}\;
        \ToptVS(\Agli,\Agli,\Lxx)_{\text{(VVS)}}\;
        \ToptVS(\Agli,\Agli,\Lgli)_{\text{(VVV)}}\;
        \ToptVS(\Aghi,\Aghi,\Lgli)_{\text{(VGG)}}
        \\[3mm] & \quad\quad
        \ToptVE(\Axx,\Axx)_{\text{(SS)}}\;
        \ToptVE(\Agli,\Axx)_{\text{(VS)}}\;
        \ToptVE(\Agli,\Agli)_{\text{(VV)}}
    \end{align*}
\caption{
    Diagram topologies contributing to the two-loop effective potential in the 3d EFT.
    Dashed lines denote scalars (S),
    wavy lines denote vector bosons (V) and
    dotted lines refer to ghost fields (G).
}
\label{fig:2loop-veff}
\end{figure}

In this Appendix we compute the required two-loop vacuum diagrams for the 3d effective potential, completing the calculation of Section~\ref{sec:Veff-main}. After introducing the background fields $\bar{v}$ and $\bar{x}$, mass eigenstates of the neutral scalars $h_1, h_2$ are given by 
\begin{align}
\bar{m}^2_{h,1} &= \frac{1}{2} \bigg(
  \Big( 6 \bar{\lambda}_{3} \bar{v}^2 + 2 \bar{m}^2_{\phi,3} + \bar{x}  (\bar{a}_{1,3} + \bar{a}_{2,3} \bar{x}) \Big) \ct^2
+ \Big(  \bar{a}_{2,3}  \bar{v}^2 + 2 \bar{m}^2_{S,3} + 4 \bar{b}_{3,3} \bar{x} + 6 \bar{b}_{4,3}  \bar{x}^2  \Big) \st^2
  \nn &\hphantom{=\frac{1}{2}\bigg(\Big(}
- \bar{v} (\bar{a}_{1,3} + 2 \bar{a}_{2,3} \bar{x}) \sin(2 \bar{\theta})  \bigg)
 \\
\bar{m}^2_{h,2} &= \frac{1}{2} \bigg(
  \Big( 6 \bar{\lambda}_{3} \bar{v}^2 + 2 \bar{m}^2_{\phi,3} + \bar{x}  (\bar{a}_{1,3} + \bar{a}_{2,3} \bar{x}) \Big) \st^2
+ \Big( \bar{a}_{2,3} \bar{v}^2 + 2 \bar{m}^2_{S,3} + 4 \bar{b}_{3,3} \bar{x} + 6 \bar{b}_{4,3} \bar{x}^2  \Big) \ct^2
  \nn &\hphantom{=\frac{1}{2}\bigg(\Big(}
+ \bar{v} (\bar{a}_{1,3} + 2 \bar{a}_{2,3} \bar{x}) \sin(2 \bar{\theta})  \bigg)
,
\end{align}
where
$\ct \equiv \cos(\bar{\theta})$ and
$\st \equiv \sin(\bar{\theta})$.
The field-dependent mixing angle $\bar{\theta}$ is to be solved from
\begin{align}
\bigg(
    \bar{v}^2 (\bar{a}_{2,3}
  - 6 \bar{\lambda}_{3})
  + 2 (\bar{m}^2_{S,3}-\bar{m}^2_{\phi,3})
  &- \bar{x} \Big( \bar{x} ( \bar{a}_{2,3}-6 \bar{b}_{4,3}) - 4 \bar{b}_{3,3} + \bar{a}_{1,3} \Big) \bigg)
  \nn &=
  2 \bar{v} (2 \bar{x} \bar{a}_{2,3} + \bar{a}_{1,3}) \cot(2 \bar{\theta})
  .
\end{align}
The mass eigenstates in the gauge sector are denoted as $W^\pm, Z, A$ with $A$ being massless, and
in addition we need the corresponding ghost fields $c^\pm,c_Z,c_A$. 
We use Landau gauge where the ghosts decouple from the scalars.

The two-loop contribution to effective potential
is obtained from the diagrams in Fig.~\ref{fig:2loop-veff}:
\begin{align}
V_2 &= -\bigg(
    \text{(SSS)}
  + \text{(VSS)}
  + \text{(VVS)}
  + \text{(VVV)}
  + \text{(VGG)}
  + \text{(SS)}
  + \text{(VS)}
  + \text{(VV)} \bigg)
  .
\end{align}
In $d=3-2\epsilon$ dimensions, the diagrams give
\begin{align}
\text{(SSS)} &=
    \frac{1}{12} C^2_{h_1 h_1 h_1}  \; \mathcal{D}_{\text{SSS}}(\bar{m}_{h,1},\bar{m}_{h,1},\bar{m}_{h,1})
  + \frac{1}{12} C^2_{h_2 h_2 h_2} \; \mathcal{D}_{\text{SSS}}(\bar{m}_{h,2},\bar{m}_{h,2},\bar{m}_{h,2})
  \nn &
  + \frac{1}{4} C^2_{h_1 G G} \; \mathcal{D}_{\text{SSS}}(\bar{m}_{h,1},\bar{m}_{G},\bar{m}_{G})
  + \frac{1}{4} C^2_{h_1 h_1 h_2} \; \mathcal{D}_{\text{SSS}}(\bar{m}_{h,1},\bar{m}_{h,1},\bar{m}_{h,2})
  \nn &
  + \frac{1}{4} C^2_{h_2 h_2 h_1} \; \mathcal{D}_{\text{SSS}}(\bar{m}_{h,2},\bar{m}_{h,2},\bar{m}_{h,1})
  + \frac{1}{4} C^2_{G G h_2} \; \mathcal{D}_{\text{SSS}}(\bar{m}_{G},\bar{m}_{G},\bar{m}_{h,2})
  \nn &
  + \frac{1}{2} C^2_{h_1 G^+ G^-} \; \mathcal{D}_{\text{SSS}}(\bar{m}_{h,1},\bar{m}_{G},\bar{m}_{G})
  + \frac{1}{2} C^2_{h_2 G^+ G^-} \; \mathcal{D}_{\text{SSS}}(\bar{m}_{h,2},\bar{m}_{G},\bar{m}_{G})
  \\
\text{(VSS)} &=
  - \frac{1}{2} C^2_{h_1 G Z} \; \mathcal{D}_{\text{VSS}}(\bar{m}_{h,1},\bar{m}_{G},\bar{m}_{Z})
  - \frac{1}{2} C^2_{h_2 G Z} \; \mathcal{D}_{\text{VSS}}(\bar{m}_{h,2},\bar{m}_{G},\bar{m}_{Z})
  \nn &
  + \frac{1}{2} C^2_{G^+ G^- Z} \; \mathcal{D}_{\text{VSS}}(\bar{m}_{G},\bar{m}_{G},\bar{m}_{Z})
  + \frac{1}{2} C^2_{G^+ G^- A} \; \mathcal{D}_{\text{VSS}}(\bar{m}_{G},\bar{m}_{G}, 0)
  \nn &
  - C_{h_1 G^+ W^-} \times C_{h_1 G^- W^+} \;  \mathcal{D}_{\text{VSS}}(\bar{m}_{h,1},\bar{m}_{G},\bar{m}_{W})
  \nn &
  - C_{h_2 G^+ W^-} \times C_{h_2 G^- W^+} \;  \mathcal{D}_{\text{VSS}}(\bar{m}_{h,2},\bar{m}_{G},\bar{m}_{W})
  \nn &
  - C_{G G^+ W^-} \times C_{G G^- W^+} \; \mathcal{D}_{\text{VSS}}(\bar{m}_{G},\bar{m}_{G},\bar{m}_{W})
  \\
\text{(VVS)} &=
    \frac{1}{4} C^2_{Z Z h_1} \; \mathcal{D}_{\text{VVS}}(\bar{m}_{h,1},\bar{m}_{Z},\bar{m}_{Z})
  + \frac{1}{4} C^2_{Z Z h_2} \; \mathcal{D}_{\text{VVS}}(\bar{m}_{h,2},\bar{m}_{Z},\bar{m}_{Z})
  \nn &
  + \frac{1}{2} C^2_{W^+ W^- h_1} \; \mathcal{D}_{\text{VVS}}(\bar{m}_{h,1},\bar{m}_{W},\bar{m}_{W})
  + \frac{1}{2} C^2_{W^+ W^- h_2} \; \mathcal{D}_{\text{VVS}}(\bar{m}_{h,2},\bar{m}_{W},\bar{m}_{W})
  \nn &
  + C_{W^- Z G^+} \times C_{W^+ Z G^-} \; \mathcal{D}_{\text{VVS}}(\bar{m}_{G},\bar{m}_{W},\bar{m}_{Z})
  \nn &
  + C_{W^- A G^+} \times C_{W^+ A G^-} \; \mathcal{D}_{\text{VVS}}(\bar{m}_{G},\bar{m}_{W},0)
  \\
\text{(VVV)} &=
    \frac{1}{2} C^2_{W^+ W^-Z} \; \mathcal{D}_{\text{VVV}}(\bar{m}_{W},\bar{m}_{W},\bar{m}_{Z})
  \nn &
  + \frac{1}{2} C^2_{W^+ W^- A} \; \mathcal{D}_{\text{VVV}}(\bar{m}_{W},\bar{m}_{W},0)
  \;,\\
\text{(VGG)} &=
  - C_{W^+ \bar{c}^- c_Z} \times C_{W^- \bar{c}_Z c^+} \;  \mathcal{D}_{\text{VGG}}(\bar{m}_{W})
  \nn &
  - C_{W^+ \bar{c}_Z c^-} \times C_{W^- \bar{c}^+ c_Z} \; \mathcal{D}_{\text{VGG}}(\bar{m}_{W})
  \nn &
  - C_{W^+ \bar{c}^- c_A} \times C_{W^- \bar{c}_A c^+} \; \mathcal{D}_{\text{VGG}}(\bar{m}_{W})
  \nn &
  - C_{W^+ \bar{c}_A c^-} \times C_{W^- \bar{c}^+ c_A} \; \mathcal{D}_{\text{VGG}}(\bar{m}_{W})
  \nn &
  - \frac{1}{2} C^2_{Z \bar{c}^+ c^-} \;  \mathcal{D}_{\text{VGG}}(\bar{m}_{Z})
  - \frac{1}{2} C^2_{Z \bar{c}^- c^+} \;  \mathcal{D}_{\text{VGG}}(\bar{m}_{Z})
  \\
\text{(SS)} &=
    \frac{1}{8} C_{h_1 h_1 h_1 h_1} \; \Big(I^3_1(\bar{m}_{h,1})\Big)^2
  + \frac{1}{8} C_{h_2 h_2 h_2 h_2} \; \Big(I^3_1(\bar{m}_{h,2})\Big)^2
  \nn &
  + \frac{1}{8} C_{G G G G} \; \Big(I^3_1(\bar{m}_{G})\Big)^2
  + \frac{1}{4} C_{h_1 h_1 h_2 h_2} \; I^3_1(\bar{m}_{h,1}) I^3_1(\bar{m}_{h,2})
  \nn &
  + \frac{1}{4} C_{h_1 h_1 G G}\; I^3_1(\bar{m}_{h,1}) I^3_1(\bar{m}_{G})
  + \frac{1}{4} C_{G G h_2 h_2} \; I^3_1(\bar{m}_{G}) I^3_1(\bar{m}_{h,2})
  \nn &
  + \frac{1}{2} C_{G^+ G^- G^+ G^-} \; \Big(I^3_1(\bar{m}_{G})\Big)^2
  + \frac{1}{2} C_{h_1 h_1 G^+ G^-} \; I^3_1(\bar{m}_{h,1}) I^3_1(\bar{m}_{G})
  \nn &
  + \frac{1}{2} C_{h_2 h_2 G^+ G^-} \; I^3_1(\bar{m}_{h,2}) I^3_1(\bar{m}_{G})
  + \frac{1}{2} C_{G G G^+ G^-} \; \Big(I^3_1(\bar{m}_{G})\Big)^2
  \\
\text{(VS)} &= (d-1) \bigg(
    \frac{1}{4} C_{Z Z h_1 h_1} \; I^3_1(\bar{m}_{h,1}) I^3_1(\bar{m}_{Z})
  + \frac{1}{4} C_{Z Z h_2 h_2} \; I^3_1(\bar{m}_{h,2}) I^3_1(\bar{m}_{Z})
  \nn &
  + \frac{1}{4} C_{Z Z G G} \; I^3_1(\bar{m}_{G}) I^3_1(\bar{m}_{Z})
  + \frac{1}{2} C_{W^+ W^- h_1 h_1} \; I^3_1(\bar{m}_{h,1}) I^3_1(\bar{m}_{W})
  \nn &
  + \frac{1}{2} C_{W^+ W^- h_2 h_2} \; I^3_1(\bar{m}_{h,2}) I^3_1(\bar{m}_{W})
  + \frac{1}{2} C_{W^+ W^- G G} \; I^3_1(\bar{m}_{G}) I^3_1(\bar{m}_{W})
  \nn &
  + \frac{1}{2} C_{Z Z G^+ G^-} \; I^3_1(\bar{m}_{G}) I^3_1(\bar{m}_{Z})
  + C_{W^+ W^- G^+ G^-} \; I^3_1(\bar{m}_{G}) I^3_1(\bar{m}_{W})
  \bigg)
  \\
\text{(VV)} &=
    \frac{1}{2} C_{W^+ W^- W^+ W^-} \; \mathcal{D}_{\text{VV}}(\bar{m}_{W}, \bar{m}_{W})
  - C_{W^+ W^- Z Z} \; \mathcal{D}_{\text{VV}}(\bar{m}_{W}, \bar{m}_{Z})
  .
\end{align}
The integrals $\mathcal{D}$ and $I^3_1$ can be found in the Supplemental Material of ref.~\cite{Niemi:2020hto}.
In the above, $c_{\psi_i \ldots \psi_j}$ denote vertex coefficients for generic fields $\psi_i$ and are defined as minus the respective coefficient in the Lagrangian, including the combinatorial factors arising from contractions. For momentum-dependent vertices the momentum is absorbed inside the integral definition. 
An exhaustive list of required vertex coefficients read
\begin{align}
C_{h_1 h_1 h_1 h_1} &= -6 \Big(
    \bar{\lambda}_{3} \ct^4
  + \bar{a}_{2,3} \ct^2 \st^2
  + \bar{b}_{4,3} \st^4 \Big)
  \\
C_{h_2 h_2 h_2 h_2} &= -6 \Big(
    \bar{b}_{4,3} \ct^4
  + \bar{a}_{2,3} \ct^2 \st^2
  + \bar{\lambda}_{3} \st^4 \Big)
   \\
C_{G G G G} &= -6 \bar{\lambda}_{3}
   \\
C_{h_1 h_1 h_2 h_2} &= \frac{1}{4} \Big(
  - 3 \bar{\lambda}_{3}
  - \bar{a}_{2,3}
  - 3 \bar{b}_{4,3}
  + 3 (\bar{\lambda}_{3} - \bar{a}_{2,3} + \bar{b}_{4,3}) \cos(4 \bar{\theta}) \Big)
  \\
C_{h_1 h_1 G G} &= C_{h_1 h_1 G^+ G^-} = -2 \bar{\lambda}_{3} \ct^2 - \bar{a}_{2,3} \st^2
  \\[2mm]
C_{h_2 h_2 G G} &= C_{h_2 h_2 G^+ G^-} =  - \bar{a}_{2,3} \ct^2 - 2 \bar{\lambda}_{3} \st^2
  \\[2mm]
C_{G^+ G^- G^+ G^-} &= -4 \bar{\lambda}_{3}
  \\[2mm]
C_{G G G^+ G^-} &= -2 \bar{\lambda}_{3}
\end{align}

\begin{align}
C_{Z Z h_1 h_1} &= - \frac{1}{2}(\bar{g}^2_3 + \bar{g}'^2_3) \ct^2 \\
C_{Z Z h_2 h_2} &= - \frac{1}{2}(\bar{g}^2_3 + \bar{g}'^2_3) \st^2 \\
C_{Z Z G G} &= - \frac{1}{2}(\bar{g}^2_3 + \bar{g}'^2_3) \\
C_{W^+ W^- h_1 h_1} &= - \frac{1}{2} \bar{g}^2_3 \ct^2 \\
C_{W^+ W^- h_2 h_2} &= - \frac{1}{2} \bar{g}^2_3 \st^2 \\
C_{W^+ W^- G G} &= c_{W^+ W^- G^+ G^-} = - \frac{1}{2} \bar{g}^2_3 \\
C_{Z Z G^+ G^-} &= - \frac{1}{2} \frac{(\bar{g}^2_3 - \bar{g}'^2_3)^2}{\bar{g}^2_3 + \bar{g}'^2_3} \\
C_{W^+ W^- W^+ W^-} &= - \bar{g}^2_3 \\
C_{W^+ W^- Z Z} &= \frac{\bar{g}^4_3 }{\bar{g}^2_3 + \bar{g}'^2_3}
\end{align}

\begin{align}
C_{h_1 h_1 h_1} &=
  - 6 \bar{v} \bar{\lambda}_{3} \ct^3
  + \frac{3}{2} (\bar{a}_{1,3} + 2 \bar{a}_{2,3} \bar{x}) \ct^2 \st
  - 3 \bar{v} \bar{a}_{2,3} \ct \st^2
  + 2(\bar{b}_{3,3} + 3 \bar{b}_{4,3} \bar{x}) \st^3
  \\
C_{h_2 h_2 h_2} &=
  - 6 \bar{v} \bar{\lambda}_{3} \st^3
  - \frac{3}{2} (\bar{a}_{1,3} + 2 \bar{a}_{2,3} \bar{x}) \st^2 \ct
  - 3 \bar{v} \bar{a}_{2,3} \st \ct^2
  - 2(\bar{b}_{3,3} + 3 \bar{b}_{4,3} \bar{x}) \ct^3
   \\
C_{h_1 G G} &= C_{h_1 G^+ G^-} =
  -2 \bar{v} \bar{\lambda}_{3}  \ct
  + \frac{1}{2}(\bar{a}_{1,3} + 2 \bar{a}_{2,3} \bar{x}) \st
   \\
C_{h_1 h_1 h_2} &=
  - \frac{1}{2} (\bar{a}_{1,3} + 2 \bar{a}_{2,3} \bar{x}) \ct^3
  + 2 \bar{v} (-3\bar{\lambda}_{3} + \bar{a}_{2,3}) \ct^2 \st
  \nn &
  + (-2 \bar{b}_{3,3} + \bar{a}_{1,3} + 2 \bar{a}_{2,3} \bar{x} - 6  \bar{b}_{4,3}  \bar{x}) \ct \st^2
  - \bar{v} \bar{a}_{2,3} \st^3
   \\
C_{h_2 h_2 h_1} &=
    \frac{1}{2} (\bar{a}_{1,3} + 2 \bar{a}_{2,3} \bar{x}) \st^3
  + 2 \bar{v} (-3\bar{\lambda}_{3} + \bar{a}_{2,3}) \st^2 \ct
  \nn &
  - (-2 \bar{b}_{3,3} + \bar{a}_{1,3} + 2 \bar{a}_{2,3} \bar{x} - 6 \bar{b}_{4,3}  \bar{x}) \st \ct^2
  - \bar{v} \bar{a}_{2,3} \ct^3
   \\
C_{G G h_2} &= C_{h_2 G^+ G^-} =
  - 2 \bar{v} \bar{\lambda}_{3} \st
  - \frac{1}{2}(\bar{a}_{1,3} + 2 \bar{a}_{2,3} \bar{x}) \ct
\end{align}

\begin{align}
C_{Z Z h_1} &= - \frac{1}{2}(\bar{g}^2_3 + \bar{g}'^2_3) \bar{v} \ct \\
C_{Z Z h_2} &=  - \frac{1}{2}(\bar{g}^2_3 + \bar{g}'^2_3) \bar{v} \st \\
C_{W^+ W- h_1} &= - \frac{1}{2} \bar{g}^2_3 \bar{v} \ct \\
C_{W^+ W- h_2} &= - \frac{1}{2} \bar{g}^2_3 \bar{v} \st \\
C_{W^- Z G^+} &= C_{W^+ Z G^-} = \frac{\bar{v}}{2} \frac{\bar{g}_3 \bar{g}'^2_3}{\sqrt{\bar{g}^2_3 + \bar{g}'^2_3}}  \\
C_{W^- A G^+} &= C_{W^+ A G^-} = -\frac{\bar{v}}{2} \frac{\bar{g}^2_3 \bar{g}'_3}{\sqrt{\bar{g}^2_3 + \bar{g}'^2_3}}
\end{align}

\begin{align}
C_{h_1 G Z} &= -\frac{i}{2} \sqrt{\bar{g}^2_3 + \bar{g}'^2_3} \ct \\
C_{h_2 G Z} &= -\frac{i}{2} \sqrt{\bar{g}^2_3 + \bar{g}'^2_3} \st \\
C_{G^+ G^- Z} &= \frac{1}{2} \frac{\bar{g}'^2_3-\bar{g}^2_3}{\sqrt{\bar{g}^2_3 + \bar{g}'^2_3}} \\
C_{G^+ G^- A} &= - \frac{\bar{g}_3 \bar{g}'_3}{\sqrt{\bar{g}^2_3 + \bar{g}'^2_3}} \\
C_{h_1 G^+ W^-} &= -C_{h_1 G^- W^+}  = \frac{1}{2} \bar{g}_3 \ct \\
C_{h_2 G^+ W^-} &= -C_{h_2 G^- W^+} = \frac{1}{2} \bar{g}_3 \st \\
C_{G G^+ W^-} &= C_{G G^- W^+} = -\frac{i}{2} \bar{g}_3
\end{align}

\begin{align}
C_{W^+ W^- Z} &= \frac{\bar{g}^2_3}{\sqrt{\bar{g}^2_3 + \bar{g}'^2_3}} \\
C_{W^+ W^- A} &= \frac{\bar{g}_3 \bar{g}'_3}{\sqrt{\bar{g}^2_3 + \bar{g}'^2_3}}
\end{align}

\begin{align}
C_{W^+ \bar{c}^- c_Z} &=
    C_{W^- \bar{c}_Z c^+}
  = -C_{W^+ \bar{c}_Z c^-}
  = -C_{W^- \bar{c}^+ c_Z}
  = - \frac{\bar{g}^2_3}{\sqrt{\bar{g}^2_3 + \bar{g}'^2_3}}
  \\
C_{W^+ \bar{c}^- c_A} &=
    C_{W^- \bar{c}_A c^+}
  = -C_{W^+ \bar{c}_A c^-}
  = -C_{W^- \bar{c}^+ c_A}
  = -\frac{\bar{g}_3 \bar{g}'_3}{\sqrt{\bar{g}^2_3 + \bar{g}'^2_3}}
  \\
C_{Z \bar{c}^+ c^-} &=
  - C_{Z \bar{c}^- c^+} =
  - \frac{\bar{g}^2_3}{\sqrt{\bar{g}^2_3 + \bar{g}'^2_3}}
  .
\end{align}

UV divergences arising from the two-loop diagrams are canceled by introducing counterterms in the tree-level potential (there is also an unphysical vacuum divergence which we ignore here). The required counterterms read
\begin{align}
\delta \bar{m}_{\phi,3}^2 &= -\frac{1}{(4\pi)^2} \frac{1}{4\epsilon} \Big( \frac{51}{16} \bar{g}^4_3 - \frac{9}{8} \bar{g}^2_3 \bar{g}'^2_3 - \frac{5}{16} \bar{g}'^4_3 + 9 \bar{g}_3^2 \bar{\lambda}_3 + 3 \bar{g}'^2_3 \bar{\lambda}_3 - 12 \bar{\lambda}_3^2 - \frac12 \bar{a}_{2,3}^2 \Big) \nn
\delta \bar{m}_{S,3}^2 &= -\frac{1}{(4\pi)^2} \frac{1}{4\epsilon}  \Big( 3\bar{g}^2_3 \bar{a}_{2,3} + \bar{g}'^2_3 \bar{a}_{2,3} - \bar{a}_{2,3}^2 - 6 \bar{b}_{4,3}^2  \Big) \nn
\delta \bar{b}_{1,3} &= -\frac{1}{(4\pi)^2} \frac{1}{4\epsilon}  \Big( -4 \bar{b}_{4,3} \bar{b}_{3,3} - 2 \bar{a}_{2,3}\bar{a}_{1,3} + 3\bar{g}^2_3 \bar{a}_{1,3} + \bar{g}'^2_3 \bar{a}_{1,3}  \Big) .
\end{align}
A simple dimensional analysis in the super-renormalizable 3d theory shows that there are no new divergences at higher loop orders, so these counterterms are exact.
This completes our two-loop computation.

\twocolumngrid
\bibliographystyle{apsrev4-1}
\bibliography{singletrefs}

\begin{thebibliography}{72}%
\makeatletter
\providecommand \@ifxundefined [1]{%
 \@ifx{#1\undefined}
}%
\providecommand \@ifnum [1]{%
 \ifnum #1\expandafter \@firstoftwo
 \else \expandafter \@secondoftwo
 \fi
}%
\providecommand \@ifx [1]{%
 \ifx #1\expandafter \@firstoftwo
 \else \expandafter \@secondoftwo
 \fi
}%
\providecommand \natexlab [1]{#1}%
\providecommand \enquote  [1]{``#1''}%
\providecommand \bibnamefont  [1]{#1}%
\providecommand \bibfnamefont [1]{#1}%
\providecommand \citenamefont [1]{#1}%
\providecommand \href@noop [0]{\@secondoftwo}%
\providecommand \href [0]{\begingroup \@sanitize@url \@href}%
\providecommand \@href[1]{\@@startlink{#1}\@@href}%
\providecommand \@@href[1]{\endgroup#1\@@endlink}%
\providecommand \@sanitize@url [0]{\catcode `\\12\catcode `\$12\catcode
  `\&12\catcode `\#12\catcode `\^12\catcode `\_12\catcode `\%12\relax}%
\providecommand \@@startlink[1]{}%
\providecommand \@@endlink[0]{}%
\providecommand \url  [0]{\begingroup\@sanitize@url \@url }%
\providecommand \@url [1]{\endgroup\@href {#1}{\urlprefix }}%
\providecommand \urlprefix  [0]{URL }%
\providecommand \Eprint [0]{\href }%
\providecommand \doibase [0]{http://dx.doi.org/}%
\providecommand \selectlanguage [0]{\@gobble}%
\providecommand \bibinfo  [0]{\@secondoftwo}%
\providecommand \bibfield  [0]{\@secondoftwo}%
\providecommand \translation [1]{[#1]}%
\providecommand \BibitemOpen [0]{}%
\providecommand \bibitemStop [0]{}%
\providecommand \bibitemNoStop [0]{.\EOS\space}%
\providecommand \EOS [0]{\spacefactor3000\relax}%
\providecommand \BibitemShut  [1]{\csname bibitem#1\endcsname}%
\let\auto@bib@innerbib\@empty
\bibitem [{\citenamefont {Weir}(2018)}]{Weir:2017wfa}%
  \BibitemOpen
  \bibfield  {author} {\bibinfo {author} {\bibfnamefont {D.~J.}\ \bibnamefont
  {Weir}},\ }\href {\doibase 10.1098/rsta.2017.0126} {\bibfield  {journal}
  {\bibinfo  {journal} {Phil. Trans. Roy. Soc. Lond. A}\ }\textbf {\bibinfo
  {volume} {376}},\ \bibinfo {pages} {20170126} (\bibinfo {year} {2018})},\
  \Eprint {http://arxiv.org/abs/1705.01783} {arXiv:1705.01783 [hep-ph]}
  \BibitemShut {NoStop}%
\bibitem [{\citenamefont {Caprini}\ \emph {et~al.}(2020)\citenamefont {Caprini}
  \emph {et~al.}}]{Caprini:2019egz}%
  \BibitemOpen
  \bibfield  {author} {\bibinfo {author} {\bibfnamefont {C.}~\bibnamefont
  {Caprini}} \emph {et~al.},\ }\href {\doibase 10.1088/1475-7516/2020/03/024}
  {\bibfield  {journal} {\bibinfo  {journal} {JCAP}\ }\textbf {\bibinfo
  {volume} {03}},\ \bibinfo {pages} {024} (\bibinfo {year} {2020})},\ \Eprint
  {http://arxiv.org/abs/1910.13125} {arXiv:1910.13125 [astro-ph.CO]}
  \BibitemShut {NoStop}%
\bibitem [{\citenamefont {Kuzmin}\ \emph {et~al.}(1985)\citenamefont {Kuzmin},
  \citenamefont {Rubakov},\ and\ \citenamefont {Shaposhnikov}}]{Kuzmin:1985mm}%
  \BibitemOpen
  \bibfield  {author} {\bibinfo {author} {\bibfnamefont {V.~A.}\ \bibnamefont
  {Kuzmin}}, \bibinfo {author} {\bibfnamefont {V.~A.}\ \bibnamefont {Rubakov}},
  \ and\ \bibinfo {author} {\bibfnamefont {M.~E.}\ \bibnamefont
  {Shaposhnikov}},\ }\href {\doibase 10.1016/0370-2693(85)91028-7} {\bibfield
  {journal} {\bibinfo  {journal} {Phys. Lett.}\ }\textbf {\bibinfo {volume}
  {155B}},\ \bibinfo {pages} {36} (\bibinfo {year} {1985})}\BibitemShut
  {NoStop}%
\bibitem [{\citenamefont {Shaposhnikov}(1987)}]{Shaposhnikov:1987tw}%
  \BibitemOpen
  \bibfield  {author} {\bibinfo {author} {\bibfnamefont {M.~E.}\ \bibnamefont
  {Shaposhnikov}},\ }\href {\doibase 10.1016/0550-3213(87)90127-1} {\bibfield
  {journal} {\bibinfo  {journal} {Nucl. Phys.}\ }\textbf {\bibinfo {volume}
  {B287}},\ \bibinfo {pages} {757} (\bibinfo {year} {1987})}\BibitemShut
  {NoStop}%
\bibitem [{\citenamefont {Morrissey}\ and\ \citenamefont
  {Ramsey-Musolf}(2012)}]{Morrissey:2012db}%
  \BibitemOpen
  \bibfield  {author} {\bibinfo {author} {\bibfnamefont {D.~E.}\ \bibnamefont
  {Morrissey}}\ and\ \bibinfo {author} {\bibfnamefont {M.~J.}\ \bibnamefont
  {Ramsey-Musolf}},\ }\href {\doibase 10.1088/1367-2630/14/12/125003}
  {\bibfield  {journal} {\bibinfo  {journal} {New J. Phys.}\ }\textbf {\bibinfo
  {volume} {14}},\ \bibinfo {pages} {125003} (\bibinfo {year} {2012})},\
  \Eprint {http://arxiv.org/abs/1206.2942} {arXiv:1206.2942 [hep-ph]}
  \BibitemShut {NoStop}%
\bibitem [{\citenamefont {Kajantie}\ \emph
  {et~al.}(1996{\natexlab{a}})\citenamefont {Kajantie}, \citenamefont {Laine},
  \citenamefont {Rummukainen},\ and\ \citenamefont
  {Shaposhnikov}}]{Kajantie:1996mn}%
  \BibitemOpen
  \bibfield  {author} {\bibinfo {author} {\bibfnamefont {K.}~\bibnamefont
  {Kajantie}}, \bibinfo {author} {\bibfnamefont {M.}~\bibnamefont {Laine}},
  \bibinfo {author} {\bibfnamefont {K.}~\bibnamefont {Rummukainen}}, \ and\
  \bibinfo {author} {\bibfnamefont {M.~E.}\ \bibnamefont {Shaposhnikov}},\
  }\href {\doibase 10.1103/PhysRevLett.77.2887} {\bibfield  {journal} {\bibinfo
   {journal} {Phys. Rev. Lett.}\ }\textbf {\bibinfo {volume} {77}},\ \bibinfo
  {pages} {2887} (\bibinfo {year} {1996}{\natexlab{a}})},\ \Eprint
  {http://arxiv.org/abs/hep-ph/9605288} {arXiv:hep-ph/9605288 [hep-ph]}
  \BibitemShut {NoStop}%
\bibitem [{\citenamefont {Csikor}\ \emph {et~al.}(1999)\citenamefont {Csikor},
  \citenamefont {Fodor},\ and\ \citenamefont {Heitger}}]{Csikor:1998eu}%
  \BibitemOpen
  \bibfield  {author} {\bibinfo {author} {\bibfnamefont {F.}~\bibnamefont
  {Csikor}}, \bibinfo {author} {\bibfnamefont {Z.}~\bibnamefont {Fodor}}, \
  and\ \bibinfo {author} {\bibfnamefont {J.}~\bibnamefont {Heitger}},\ }\href
  {\doibase 10.1103/PhysRevLett.82.21} {\bibfield  {journal} {\bibinfo
  {journal} {Phys. Rev. Lett.}\ }\textbf {\bibinfo {volume} {82}},\ \bibinfo
  {pages} {21} (\bibinfo {year} {1999})},\ \Eprint
  {http://arxiv.org/abs/hep-ph/9809291} {arXiv:hep-ph/9809291 [hep-ph]}
  \BibitemShut {NoStop}%
\bibitem [{\citenamefont {Ramsey-Musolf}(2020)}]{Ramsey-Musolf:2019lsf}%
  \BibitemOpen
  \bibfield  {author} {\bibinfo {author} {\bibfnamefont {M.~J.}\ \bibnamefont
  {Ramsey-Musolf}},\ }\href {\doibase 10.1007/JHEP09(2020)179} {\bibfield
  {journal} {\bibinfo  {journal} {JHEP}\ }\textbf {\bibinfo {volume} {09}},\
  \bibinfo {pages} {179} (\bibinfo {year} {2020})},\ \Eprint
  {http://arxiv.org/abs/1912.07189} {arXiv:1912.07189 [hep-ph]} \BibitemShut
  {NoStop}%
\bibitem [{\citenamefont {Arnold}\ and\ \citenamefont
  {Espinosa}(1993)}]{Arnold:1992rz}%
  \BibitemOpen
  \bibfield  {author} {\bibinfo {author} {\bibfnamefont {P.~B.}\ \bibnamefont
  {Arnold}}\ and\ \bibinfo {author} {\bibfnamefont {O.}~\bibnamefont
  {Espinosa}},\ }\href {\doibase 10.1103/physrevd.50.6662.2,
  10.1103/PhysRevD.47.3546} {\bibfield  {journal} {\bibinfo  {journal} {Phys.
  Rev.}\ }\textbf {\bibinfo {volume} {D47}},\ \bibinfo {pages} {3546} (\bibinfo
  {year} {1993})},\ \bibinfo {note} {[Erratum: Phys. Rev.D50,6662(1994)]},\
  \Eprint {http://arxiv.org/abs/hep-ph/9212235} {arXiv:hep-ph/9212235 [hep-ph]}
  \BibitemShut {NoStop}%
\bibitem [{\citenamefont {Profumo}\ \emph {et~al.}(2015)\citenamefont
  {Profumo}, \citenamefont {Ramsey-Musolf}, \citenamefont {Wainwright},\ and\
  \citenamefont {Winslow}}]{Profumo:2014opa}%
  \BibitemOpen
  \bibfield  {author} {\bibinfo {author} {\bibfnamefont {S.}~\bibnamefont
  {Profumo}}, \bibinfo {author} {\bibfnamefont {M.~J.}\ \bibnamefont
  {Ramsey-Musolf}}, \bibinfo {author} {\bibfnamefont {C.~L.}\ \bibnamefont
  {Wainwright}}, \ and\ \bibinfo {author} {\bibfnamefont {P.}~\bibnamefont
  {Winslow}},\ }\href {\doibase 10.1103/PhysRevD.91.035018} {\bibfield
  {journal} {\bibinfo  {journal} {Phys. Rev. D}\ }\textbf {\bibinfo {volume}
  {91}},\ \bibinfo {pages} {035018} (\bibinfo {year} {2015})},\ \Eprint
  {http://arxiv.org/abs/1407.5342} {arXiv:1407.5342 [hep-ph]} \BibitemShut
  {NoStop}%
\bibitem [{\citenamefont {Profumo}\ \emph {et~al.}(2007)\citenamefont
  {Profumo}, \citenamefont {Ramsey-Musolf},\ and\ \citenamefont
  {Shaughnessy}}]{Profumo:2007wc}%
  \BibitemOpen
  \bibfield  {author} {\bibinfo {author} {\bibfnamefont {S.}~\bibnamefont
  {Profumo}}, \bibinfo {author} {\bibfnamefont {M.~J.}\ \bibnamefont
  {Ramsey-Musolf}}, \ and\ \bibinfo {author} {\bibfnamefont {G.}~\bibnamefont
  {Shaughnessy}},\ }\href {\doibase 10.1088/1126-6708/2007/08/010} {\bibfield
  {journal} {\bibinfo  {journal} {JHEP}\ }\textbf {\bibinfo {volume} {08}},\
  \bibinfo {pages} {010} (\bibinfo {year} {2007})},\ \Eprint
  {http://arxiv.org/abs/0705.2425} {arXiv:0705.2425 [hep-ph]} \BibitemShut
  {NoStop}%
\bibitem [{\citenamefont {Ashoorioon}\ and\ \citenamefont
  {Konstandin}(2009)}]{Ashoorioon:2009nf}%
  \BibitemOpen
  \bibfield  {author} {\bibinfo {author} {\bibfnamefont {A.}~\bibnamefont
  {Ashoorioon}}\ and\ \bibinfo {author} {\bibfnamefont {T.}~\bibnamefont
  {Konstandin}},\ }\href {\doibase 10.1088/1126-6708/2009/07/086} {\bibfield
  {journal} {\bibinfo  {journal} {JHEP}\ }\textbf {\bibinfo {volume} {07}},\
  \bibinfo {pages} {086} (\bibinfo {year} {2009})},\ \Eprint
  {http://arxiv.org/abs/0904.0353} {arXiv:0904.0353 [hep-ph]} \BibitemShut
  {NoStop}%
\bibitem [{\citenamefont {Espinosa}\ \emph {et~al.}(2012)\citenamefont
  {Espinosa}, \citenamefont {Konstandin},\ and\ \citenamefont
  {Riva}}]{Espinosa:2011ax}%
  \BibitemOpen
  \bibfield  {author} {\bibinfo {author} {\bibfnamefont {J.~R.}\ \bibnamefont
  {Espinosa}}, \bibinfo {author} {\bibfnamefont {T.}~\bibnamefont
  {Konstandin}}, \ and\ \bibinfo {author} {\bibfnamefont {F.}~\bibnamefont
  {Riva}},\ }\href {\doibase 10.1016/j.nuclphysb.2011.09.010} {\bibfield
  {journal} {\bibinfo  {journal} {Nucl. Phys. B}\ }\textbf {\bibinfo {volume}
  {854}},\ \bibinfo {pages} {592} (\bibinfo {year} {2012})},\ \Eprint
  {http://arxiv.org/abs/1107.5441} {arXiv:1107.5441 [hep-ph]} \BibitemShut
  {NoStop}%
\bibitem [{\citenamefont {Curtin}\ \emph {et~al.}(2014)\citenamefont {Curtin},
  \citenamefont {Meade},\ and\ \citenamefont {Yu}}]{Curtin:2014jma}%
  \BibitemOpen
  \bibfield  {author} {\bibinfo {author} {\bibfnamefont {D.}~\bibnamefont
  {Curtin}}, \bibinfo {author} {\bibfnamefont {P.}~\bibnamefont {Meade}}, \
  and\ \bibinfo {author} {\bibfnamefont {C.-T.}\ \bibnamefont {Yu}},\ }\href
  {\doibase 10.1007/JHEP11(2014)127} {\bibfield  {journal} {\bibinfo  {journal}
  {JHEP}\ }\textbf {\bibinfo {volume} {11}},\ \bibinfo {pages} {127} (\bibinfo
  {year} {2014})},\ \Eprint {http://arxiv.org/abs/1409.0005} {arXiv:1409.0005
  [hep-ph]} \BibitemShut {NoStop}%
\bibitem [{\citenamefont {Fuyuto}\ and\ \citenamefont
  {Senaha}(2014)}]{Fuyuto:2014yia}%
  \BibitemOpen
  \bibfield  {author} {\bibinfo {author} {\bibfnamefont {K.}~\bibnamefont
  {Fuyuto}}\ and\ \bibinfo {author} {\bibfnamefont {E.}~\bibnamefont
  {Senaha}},\ }\href {\doibase 10.1103/PhysRevD.90.015015} {\bibfield
  {journal} {\bibinfo  {journal} {Phys. Rev. D}\ }\textbf {\bibinfo {volume}
  {90}},\ \bibinfo {pages} {015015} (\bibinfo {year} {2014})},\ \Eprint
  {http://arxiv.org/abs/1406.0433} {arXiv:1406.0433 [hep-ph]} \BibitemShut
  {NoStop}%
\bibitem [{\citenamefont {Kozaczuk}(2015)}]{Kozaczuk:2015owa}%
  \BibitemOpen
  \bibfield  {author} {\bibinfo {author} {\bibfnamefont {J.}~\bibnamefont
  {Kozaczuk}},\ }\href {\doibase 10.1007/JHEP10(2015)135} {\bibfield  {journal}
  {\bibinfo  {journal} {JHEP}\ }\textbf {\bibinfo {volume} {10}},\ \bibinfo
  {pages} {135} (\bibinfo {year} {2015})},\ \Eprint
  {http://arxiv.org/abs/1506.04741} {arXiv:1506.04741 [hep-ph]} \BibitemShut
  {NoStop}%
\bibitem [{\citenamefont {Chala}\ \emph {et~al.}(2016)\citenamefont {Chala},
  \citenamefont {Nardini},\ and\ \citenamefont {Sobolev}}]{Chala:2016ykx}%
  \BibitemOpen
  \bibfield  {author} {\bibinfo {author} {\bibfnamefont {M.}~\bibnamefont
  {Chala}}, \bibinfo {author} {\bibfnamefont {G.}~\bibnamefont {Nardini}}, \
  and\ \bibinfo {author} {\bibfnamefont {I.}~\bibnamefont {Sobolev}},\ }\href
  {\doibase 10.1103/PhysRevD.94.055006} {\bibfield  {journal} {\bibinfo
  {journal} {Phys. Rev. D}\ }\textbf {\bibinfo {volume} {94}},\ \bibinfo
  {pages} {055006} (\bibinfo {year} {2016})},\ \Eprint
  {http://arxiv.org/abs/1605.08663} {arXiv:1605.08663 [hep-ph]} \BibitemShut
  {NoStop}%
\bibitem [{\citenamefont {Beniwal}\ \emph {et~al.}(2017)\citenamefont
  {Beniwal}, \citenamefont {Lewicki}, \citenamefont {Wells}, \citenamefont
  {White},\ and\ \citenamefont {Williams}}]{Beniwal:2017eik}%
  \BibitemOpen
  \bibfield  {author} {\bibinfo {author} {\bibfnamefont {A.}~\bibnamefont
  {Beniwal}}, \bibinfo {author} {\bibfnamefont {M.}~\bibnamefont {Lewicki}},
  \bibinfo {author} {\bibfnamefont {J.~D.}\ \bibnamefont {Wells}}, \bibinfo
  {author} {\bibfnamefont {M.}~\bibnamefont {White}}, \ and\ \bibinfo {author}
  {\bibfnamefont {A.~G.}\ \bibnamefont {Williams}},\ }\href {\doibase
  10.1007/JHEP08(2017)108} {\bibfield  {journal} {\bibinfo  {journal} {JHEP}\
  }\textbf {\bibinfo {volume} {08}},\ \bibinfo {pages} {108} (\bibinfo {year}
  {2017})},\ \Eprint {http://arxiv.org/abs/1702.06124} {arXiv:1702.06124
  [hep-ph]} \BibitemShut {NoStop}%
\bibitem [{\citenamefont {Kurup}\ and\ \citenamefont
  {Perelstein}(2017)}]{Kurup:2017dzf}%
  \BibitemOpen
  \bibfield  {author} {\bibinfo {author} {\bibfnamefont {G.}~\bibnamefont
  {Kurup}}\ and\ \bibinfo {author} {\bibfnamefont {M.}~\bibnamefont
  {Perelstein}},\ }\href {\doibase 10.1103/PhysRevD.96.015036} {\bibfield
  {journal} {\bibinfo  {journal} {Phys. Rev.}\ }\textbf {\bibinfo {volume}
  {D96}},\ \bibinfo {pages} {015036} (\bibinfo {year} {2017})},\ \Eprint
  {http://arxiv.org/abs/1704.03381} {arXiv:1704.03381 [hep-ph]} \BibitemShut
  {NoStop}%
\bibitem [{\citenamefont {Chen}\ \emph {et~al.}(2017)\citenamefont {Chen},
  \citenamefont {Kozaczuk},\ and\ \citenamefont {Lewis}}]{Chen:2017qcz}%
  \BibitemOpen
  \bibfield  {author} {\bibinfo {author} {\bibfnamefont {C.-Y.}\ \bibnamefont
  {Chen}}, \bibinfo {author} {\bibfnamefont {J.}~\bibnamefont {Kozaczuk}}, \
  and\ \bibinfo {author} {\bibfnamefont {I.~M.}\ \bibnamefont {Lewis}},\ }\href
  {\doibase 10.1007/JHEP08(2017)096} {\bibfield  {journal} {\bibinfo  {journal}
  {JHEP}\ }\textbf {\bibinfo {volume} {08}},\ \bibinfo {pages} {096} (\bibinfo
  {year} {2017})},\ \Eprint {http://arxiv.org/abs/1704.05844} {arXiv:1704.05844
  [hep-ph]} \BibitemShut {NoStop}%
\bibitem [{\citenamefont {Chiang}\ \emph {et~al.}(2019)\citenamefont {Chiang},
  \citenamefont {Li},\ and\ \citenamefont {Senaha}}]{Chiang:2018gsn}%
  \BibitemOpen
  \bibfield  {author} {\bibinfo {author} {\bibfnamefont {C.-W.}\ \bibnamefont
  {Chiang}}, \bibinfo {author} {\bibfnamefont {Y.-T.}\ \bibnamefont {Li}}, \
  and\ \bibinfo {author} {\bibfnamefont {E.}~\bibnamefont {Senaha}},\ }\href
  {\doibase 10.1016/j.physletb.2018.12.017} {\bibfield  {journal} {\bibinfo
  {journal} {Phys. Lett. B}\ }\textbf {\bibinfo {volume} {789}},\ \bibinfo
  {pages} {154} (\bibinfo {year} {2019})},\ \Eprint
  {http://arxiv.org/abs/1808.01098} {arXiv:1808.01098 [hep-ph]} \BibitemShut
  {NoStop}%
\bibitem [{\citenamefont {Carena}\ \emph {et~al.}(2020)\citenamefont {Carena},
  \citenamefont {Liu},\ and\ \citenamefont {Wang}}]{Carena:2019une}%
  \BibitemOpen
  \bibfield  {author} {\bibinfo {author} {\bibfnamefont {M.}~\bibnamefont
  {Carena}}, \bibinfo {author} {\bibfnamefont {Z.}~\bibnamefont {Liu}}, \ and\
  \bibinfo {author} {\bibfnamefont {Y.}~\bibnamefont {Wang}},\ }\href {\doibase
  10.1007/JHEP08(2020)107} {\bibfield  {journal} {\bibinfo  {journal} {JHEP}\
  }\textbf {\bibinfo {volume} {08}},\ \bibinfo {pages} {107} (\bibinfo {year}
  {2020})},\ \Eprint {http://arxiv.org/abs/1911.10206} {arXiv:1911.10206
  [hep-ph]} \BibitemShut {NoStop}%
\bibitem [{\citenamefont {Papaefstathiou}\ and\ \citenamefont
  {White}(2021)}]{Papaefstathiou:2020iag}%
  \BibitemOpen
  \bibfield  {author} {\bibinfo {author} {\bibfnamefont {A.}~\bibnamefont
  {Papaefstathiou}}\ and\ \bibinfo {author} {\bibfnamefont {G.}~\bibnamefont
  {White}},\ }\href {\doibase 10.1007/JHEP05(2021)099} {\bibfield  {journal}
  {\bibinfo  {journal} {JHEP}\ }\textbf {\bibinfo {volume} {05}},\ \bibinfo
  {pages} {099} (\bibinfo {year} {2021})},\ \Eprint
  {http://arxiv.org/abs/2010.00597} {arXiv:2010.00597 [hep-ph]} \BibitemShut
  {NoStop}%
\bibitem [{\citenamefont {Espinosa}(1996)}]{Espinosa:1996qw}%
  \BibitemOpen
  \bibfield  {author} {\bibinfo {author} {\bibfnamefont {J.~R.}\ \bibnamefont
  {Espinosa}},\ }\href {\doibase 10.1016/0550-3213(96)00297-0} {\bibfield
  {journal} {\bibinfo  {journal} {Nucl. Phys. B}\ }\textbf {\bibinfo {volume}
  {475}},\ \bibinfo {pages} {273} (\bibinfo {year} {1996})},\ \Eprint
  {http://arxiv.org/abs/hep-ph/9604320} {arXiv:hep-ph/9604320} \BibitemShut
  {NoStop}%
\bibitem [{\citenamefont {B{\"o}deker}\ \emph {et~al.}(1997)\citenamefont
  {B{\"o}deker}, \citenamefont {John}, \citenamefont {Laine},\ and\
  \citenamefont {Schmidt}}]{Bodeker:1996pc}%
  \BibitemOpen
  \bibfield  {author} {\bibinfo {author} {\bibfnamefont {D.}~\bibnamefont
  {B{\"o}deker}}, \bibinfo {author} {\bibfnamefont {P.}~\bibnamefont {John}},
  \bibinfo {author} {\bibfnamefont {M.}~\bibnamefont {Laine}}, \ and\ \bibinfo
  {author} {\bibfnamefont {M.~G.}\ \bibnamefont {Schmidt}},\ }\href {\doibase
  10.1016/S0550-3213(97)00252-6} {\bibfield  {journal} {\bibinfo  {journal}
  {Nucl. Phys. B}\ }\textbf {\bibinfo {volume} {497}},\ \bibinfo {pages} {387}
  (\bibinfo {year} {1997})},\ \Eprint {http://arxiv.org/abs/hep-ph/9612364}
  {arXiv:hep-ph/9612364} \BibitemShut {NoStop}%
\bibitem [{\citenamefont {Laine}\ and\ \citenamefont
  {Rummukainen}(2001)}]{Laine:2000rm}%
  \BibitemOpen
  \bibfield  {author} {\bibinfo {author} {\bibfnamefont {M.}~\bibnamefont
  {Laine}}\ and\ \bibinfo {author} {\bibfnamefont {K.}~\bibnamefont
  {Rummukainen}},\ }\href {\doibase 10.1016/S0550-3213(00)00736-7} {\bibfield
  {journal} {\bibinfo  {journal} {Nucl. Phys. B}\ }\textbf {\bibinfo {volume}
  {597}},\ \bibinfo {pages} {23} (\bibinfo {year} {2001})},\ \Eprint
  {http://arxiv.org/abs/hep-lat/0009025} {arXiv:hep-lat/0009025} \BibitemShut
  {NoStop}%
\bibitem [{\citenamefont {Laine}\ \emph {et~al.}(2017)\citenamefont {Laine},
  \citenamefont {Meyer},\ and\ \citenamefont {Nardini}}]{Laine:2017hdk}%
  \BibitemOpen
  \bibfield  {author} {\bibinfo {author} {\bibfnamefont {M.}~\bibnamefont
  {Laine}}, \bibinfo {author} {\bibfnamefont {M.}~\bibnamefont {Meyer}}, \ and\
  \bibinfo {author} {\bibfnamefont {G.}~\bibnamefont {Nardini}},\ }\href
  {\doibase 10.1016/j.nuclphysb.2017.04.023} {\bibfield  {journal} {\bibinfo
  {journal} {Nucl. Phys.}\ }\textbf {\bibinfo {volume} {B920}},\ \bibinfo
  {pages} {565} (\bibinfo {year} {2017})},\ \Eprint
  {http://arxiv.org/abs/1702.07479} {arXiv:1702.07479 [hep-ph]} \BibitemShut
  {NoStop}%
\bibitem [{\citenamefont {Kainulainen}\ \emph {et~al.}(2019)\citenamefont
  {Kainulainen}, \citenamefont {Keus}, \citenamefont {Niemi}, \citenamefont
  {Rummukainen}, \citenamefont {Tenkanen},\ and\ \citenamefont
  {Vaskonen}}]{Kainulainen:2019kyp}%
  \BibitemOpen
  \bibfield  {author} {\bibinfo {author} {\bibfnamefont {K.}~\bibnamefont
  {Kainulainen}}, \bibinfo {author} {\bibfnamefont {V.}~\bibnamefont {Keus}},
  \bibinfo {author} {\bibfnamefont {L.}~\bibnamefont {Niemi}}, \bibinfo
  {author} {\bibfnamefont {K.}~\bibnamefont {Rummukainen}}, \bibinfo {author}
  {\bibfnamefont {T.~V.~I.}\ \bibnamefont {Tenkanen}}, \ and\ \bibinfo {author}
  {\bibfnamefont {V.}~\bibnamefont {Vaskonen}},\ }\href {\doibase
  10.1007/JHEP06(2019)075} {\bibfield  {journal} {\bibinfo  {journal} {JHEP}\
  }\textbf {\bibinfo {volume} {06}},\ \bibinfo {pages} {075} (\bibinfo {year}
  {2019})},\ \Eprint {http://arxiv.org/abs/1904.01329} {arXiv:1904.01329
  [hep-ph]} \BibitemShut {NoStop}%
\bibitem [{\citenamefont {Niemi}\ \emph {et~al.}(2021)\citenamefont {Niemi},
  \citenamefont {Ramsey-Musolf}, \citenamefont {Tenkanen},\ and\ \citenamefont
  {Weir}}]{Niemi:2020hto}%
  \BibitemOpen
  \bibfield  {author} {\bibinfo {author} {\bibfnamefont {L.}~\bibnamefont
  {Niemi}}, \bibinfo {author} {\bibfnamefont {M.~J.}\ \bibnamefont
  {Ramsey-Musolf}}, \bibinfo {author} {\bibfnamefont {T.~V.~I.}\ \bibnamefont
  {Tenkanen}}, \ and\ \bibinfo {author} {\bibfnamefont {D.~J.}\ \bibnamefont
  {Weir}},\ }\href {\doibase 10.1103/PhysRevLett.126.171802} {\bibfield
  {journal} {\bibinfo  {journal} {Phys. Rev. Lett.}\ }\textbf {\bibinfo
  {volume} {126}},\ \bibinfo {pages} {171802} (\bibinfo {year} {2021})},\
  \Eprint {http://arxiv.org/abs/2005.11332} {arXiv:2005.11332 [hep-ph]}
  \BibitemShut {NoStop}%
\bibitem [{\citenamefont {Appelquist}\ and\ \citenamefont
  {Pisarski}(1981)}]{Appelquist:1981vg}%
  \BibitemOpen
  \bibfield  {author} {\bibinfo {author} {\bibfnamefont {T.}~\bibnamefont
  {Appelquist}}\ and\ \bibinfo {author} {\bibfnamefont {R.~D.}\ \bibnamefont
  {Pisarski}},\ }\href {\doibase 10.1103/PhysRevD.23.2305} {\bibfield
  {journal} {\bibinfo  {journal} {Phys. Rev.}\ }\textbf {\bibinfo {volume}
  {D23}},\ \bibinfo {pages} {2305} (\bibinfo {year} {1981})}\BibitemShut
  {NoStop}%
\bibitem [{\citenamefont {Ginsparg}(1980)}]{Ginsparg:1980ef}%
  \BibitemOpen
  \bibfield  {author} {\bibinfo {author} {\bibfnamefont {P.~H.}\ \bibnamefont
  {Ginsparg}},\ }\href {\doibase 10.1016/0550-3213(80)90418-6} {\bibfield
  {journal} {\bibinfo  {journal} {Nucl. Phys.}\ }\textbf {\bibinfo {volume}
  {B170}},\ \bibinfo {pages} {388} (\bibinfo {year} {1980})}\BibitemShut
  {NoStop}%
\bibitem [{\citenamefont {Farakos}\ \emph {et~al.}(1995)\citenamefont
  {Farakos}, \citenamefont {Kajantie}, \citenamefont {Rummukainen},\ and\
  \citenamefont {Shaposhnikov}}]{Farakos:1994xh}%
  \BibitemOpen
  \bibfield  {author} {\bibinfo {author} {\bibfnamefont {K.}~\bibnamefont
  {Farakos}}, \bibinfo {author} {\bibfnamefont {K.}~\bibnamefont {Kajantie}},
  \bibinfo {author} {\bibfnamefont {K.}~\bibnamefont {Rummukainen}}, \ and\
  \bibinfo {author} {\bibfnamefont {M.~E.}\ \bibnamefont {Shaposhnikov}},\
  }\href {\doibase 10.1016/0550-3213(95)80129-4} {\bibfield  {journal}
  {\bibinfo  {journal} {Nucl. Phys.}\ }\textbf {\bibinfo {volume} {B442}},\
  \bibinfo {pages} {317} (\bibinfo {year} {1995})},\ \Eprint
  {http://arxiv.org/abs/hep-lat/9412091} {arXiv:hep-lat/9412091 [hep-lat]}
  \BibitemShut {NoStop}%
\bibitem [{\citenamefont {Robens}\ and\ \citenamefont
  {Stefaniak}(2015)}]{Robens:2015gla}%
  \BibitemOpen
  \bibfield  {author} {\bibinfo {author} {\bibfnamefont {T.}~\bibnamefont
  {Robens}}\ and\ \bibinfo {author} {\bibfnamefont {T.}~\bibnamefont
  {Stefaniak}},\ }\href {\doibase 10.1140/epjc/s10052-015-3323-y} {\bibfield
  {journal} {\bibinfo  {journal} {Eur. Phys. J. C}\ }\textbf {\bibinfo {volume}
  {75}},\ \bibinfo {pages} {104} (\bibinfo {year} {2015})},\ \Eprint
  {http://arxiv.org/abs/1501.02234} {arXiv:1501.02234 [hep-ph]} \BibitemShut
  {NoStop}%
\bibitem [{\citenamefont {Adhikari}\ \emph {et~al.}(2021)\citenamefont
  {Adhikari}, \citenamefont {Lewis},\ and\ \citenamefont
  {Sullivan}}]{Adhikari:2020vqo}%
  \BibitemOpen
  \bibfield  {author} {\bibinfo {author} {\bibfnamefont {S.}~\bibnamefont
  {Adhikari}}, \bibinfo {author} {\bibfnamefont {I.~M.}\ \bibnamefont {Lewis}},
  \ and\ \bibinfo {author} {\bibfnamefont {M.}~\bibnamefont {Sullivan}},\
  }\href {\doibase 10.1103/PhysRevD.103.075027} {\bibfield  {journal} {\bibinfo
   {journal} {Phys. Rev. D}\ }\textbf {\bibinfo {volume} {103}},\ \bibinfo
  {pages} {075027} (\bibinfo {year} {2021})},\ \Eprint
  {http://arxiv.org/abs/2003.10449} {arXiv:2003.10449 [hep-ph]} \BibitemShut
  {NoStop}%
\bibitem [{\citenamefont {Gould}\ \emph {et~al.}(2019)\citenamefont {Gould},
  \citenamefont {Kozaczuk}, \citenamefont {Niemi}, \citenamefont
  {Ramsey-Musolf}, \citenamefont {Tenkanen},\ and\ \citenamefont
  {Weir}}]{Gould:2019qek}%
  \BibitemOpen
  \bibfield  {author} {\bibinfo {author} {\bibfnamefont {O.}~\bibnamefont
  {Gould}}, \bibinfo {author} {\bibfnamefont {J.}~\bibnamefont {Kozaczuk}},
  \bibinfo {author} {\bibfnamefont {L.}~\bibnamefont {Niemi}}, \bibinfo
  {author} {\bibfnamefont {M.~J.}\ \bibnamefont {Ramsey-Musolf}}, \bibinfo
  {author} {\bibfnamefont {T.~V.~I.}\ \bibnamefont {Tenkanen}}, \ and\ \bibinfo
  {author} {\bibfnamefont {D.~J.}\ \bibnamefont {Weir}},\ }\href {\doibase
  10.1103/PhysRevD.100.115024} {\bibfield  {journal} {\bibinfo  {journal}
  {Phys. Rev. D}\ }\textbf {\bibinfo {volume} {100}},\ \bibinfo {pages}
  {115024} (\bibinfo {year} {2019})},\ \Eprint
  {http://arxiv.org/abs/1903.11604} {arXiv:1903.11604 [hep-ph]} \BibitemShut
  {NoStop}%
\bibitem [{\citenamefont {Farakos}\ \emph {et~al.}(1994)\citenamefont
  {Farakos}, \citenamefont {Kajantie}, \citenamefont {Rummukainen},\ and\
  \citenamefont {Shaposhnikov}}]{Farakos:1994kx}%
  \BibitemOpen
  \bibfield  {author} {\bibinfo {author} {\bibfnamefont {K.}~\bibnamefont
  {Farakos}}, \bibinfo {author} {\bibfnamefont {K.}~\bibnamefont {Kajantie}},
  \bibinfo {author} {\bibfnamefont {K.}~\bibnamefont {Rummukainen}}, \ and\
  \bibinfo {author} {\bibfnamefont {M.~E.}\ \bibnamefont {Shaposhnikov}},\
  }\href {\doibase 10.1016/0550-3213(94)90173-2} {\bibfield  {journal}
  {\bibinfo  {journal} {Nucl. Phys.}\ }\textbf {\bibinfo {volume} {B425}},\
  \bibinfo {pages} {67} (\bibinfo {year} {1994})},\ \Eprint
  {http://arxiv.org/abs/hep-ph/9404201} {arXiv:hep-ph/9404201 [hep-ph]}
  \BibitemShut {NoStop}%
\bibitem [{\citenamefont {Kajantie}\ \emph
  {et~al.}(1996{\natexlab{b}})\citenamefont {Kajantie}, \citenamefont {Laine},
  \citenamefont {Rummukainen},\ and\ \citenamefont
  {Shaposhnikov}}]{Kajantie:1995dw}%
  \BibitemOpen
  \bibfield  {author} {\bibinfo {author} {\bibfnamefont {K.}~\bibnamefont
  {Kajantie}}, \bibinfo {author} {\bibfnamefont {M.}~\bibnamefont {Laine}},
  \bibinfo {author} {\bibfnamefont {K.}~\bibnamefont {Rummukainen}}, \ and\
  \bibinfo {author} {\bibfnamefont {M.~E.}\ \bibnamefont {Shaposhnikov}},\
  }\href {\doibase 10.1016/0550-3213(95)00549-8} {\bibfield  {journal}
  {\bibinfo  {journal} {Nucl. Phys.}\ }\textbf {\bibinfo {volume} {B458}},\
  \bibinfo {pages} {90} (\bibinfo {year} {1996}{\natexlab{b}})},\ \Eprint
  {http://arxiv.org/abs/hep-ph/9508379} {arXiv:hep-ph/9508379 [hep-ph]}
  \BibitemShut {NoStop}%
\bibitem [{\citenamefont {Jackiw}\ and\ \citenamefont
  {Templeton}(1981)}]{Jackiw:1980kv}%
  \BibitemOpen
  \bibfield  {author} {\bibinfo {author} {\bibfnamefont {R.}~\bibnamefont
  {Jackiw}}\ and\ \bibinfo {author} {\bibfnamefont {S.}~\bibnamefont
  {Templeton}},\ }\href {\doibase 10.1103/PhysRevD.23.2291} {\bibfield
  {journal} {\bibinfo  {journal} {Phys. Rev. D}\ }\textbf {\bibinfo {volume}
  {23}},\ \bibinfo {pages} {2291} (\bibinfo {year} {1981})}\BibitemShut
  {NoStop}%
\bibitem [{\citenamefont {Carrington}(1992)}]{Carrington:1991hz}%
  \BibitemOpen
  \bibfield  {author} {\bibinfo {author} {\bibfnamefont {M.}~\bibnamefont
  {Carrington}},\ }\href {\doibase 10.1103/PhysRevD.45.2933} {\bibfield
  {journal} {\bibinfo  {journal} {Phys. Rev. D}\ }\textbf {\bibinfo {volume}
  {45}},\ \bibinfo {pages} {2933} (\bibinfo {year} {1992})}\BibitemShut
  {NoStop}%
\bibitem [{\citenamefont {Linde}(1980)}]{Linde:1980ts}%
  \BibitemOpen
  \bibfield  {author} {\bibinfo {author} {\bibfnamefont {A.~D.}\ \bibnamefont
  {Linde}},\ }\href {\doibase 10.1016/0370-2693(80)90769-8} {\bibfield
  {journal} {\bibinfo  {journal} {Phys. Lett.}\ }\textbf {\bibinfo {volume}
  {96B}},\ \bibinfo {pages} {289} (\bibinfo {year} {1980})}\BibitemShut
  {NoStop}%
\bibitem [{\citenamefont {Braaten}\ and\ \citenamefont
  {Nieto}(1995)}]{Braaten:1995cm}%
  \BibitemOpen
  \bibfield  {author} {\bibinfo {author} {\bibfnamefont {E.}~\bibnamefont
  {Braaten}}\ and\ \bibinfo {author} {\bibfnamefont {A.}~\bibnamefont
  {Nieto}},\ }\href {\doibase 10.1103/PhysRevD.51.6990} {\bibfield  {journal}
  {\bibinfo  {journal} {Phys. Rev.}\ }\textbf {\bibinfo {volume} {D51}},\
  \bibinfo {pages} {6990} (\bibinfo {year} {1995})},\ \Eprint
  {http://arxiv.org/abs/hep-ph/9501375} {arXiv:hep-ph/9501375 [hep-ph]}
  \BibitemShut {NoStop}%
\bibitem [{\citenamefont {Landsman}(1989)}]{Landsman:1989be}%
  \BibitemOpen
  \bibfield  {author} {\bibinfo {author} {\bibfnamefont {N.~P.}\ \bibnamefont
  {Landsman}},\ }\href {\doibase 10.1016/0550-3213(89)90424-0} {\bibfield
  {journal} {\bibinfo  {journal} {Nucl. Phys. B}\ }\textbf {\bibinfo {volume}
  {322}},\ \bibinfo {pages} {498} (\bibinfo {year} {1989})}\BibitemShut
  {NoStop}%
\bibitem [{\citenamefont {de~Vries}\ \emph {et~al.}(2018)\citenamefont
  {de~Vries}, \citenamefont {Postma}, \citenamefont {van~de Vis},\ and\
  \citenamefont {White}}]{deVries:2017ncy}%
  \BibitemOpen
  \bibfield  {author} {\bibinfo {author} {\bibfnamefont {J.}~\bibnamefont
  {de~Vries}}, \bibinfo {author} {\bibfnamefont {M.}~\bibnamefont {Postma}},
  \bibinfo {author} {\bibfnamefont {J.}~\bibnamefont {van~de Vis}}, \ and\
  \bibinfo {author} {\bibfnamefont {G.}~\bibnamefont {White}},\ }\href
  {\doibase 10.1007/JHEP01(2018)089} {\bibfield  {journal} {\bibinfo  {journal}
  {JHEP}\ }\textbf {\bibinfo {volume} {01}},\ \bibinfo {pages} {089} (\bibinfo
  {year} {2018})},\ \Eprint {http://arxiv.org/abs/1710.04061} {arXiv:1710.04061
  [hep-ph]} \BibitemShut {NoStop}%
\bibitem [{\citenamefont {Kajantie}\ \emph
  {et~al.}(1996{\natexlab{c}})\citenamefont {Kajantie}, \citenamefont {Laine},
  \citenamefont {Rummukainen},\ and\ \citenamefont
  {Shaposhnikov}}]{Kajantie:1995kf}%
  \BibitemOpen
  \bibfield  {author} {\bibinfo {author} {\bibfnamefont {K.}~\bibnamefont
  {Kajantie}}, \bibinfo {author} {\bibfnamefont {M.}~\bibnamefont {Laine}},
  \bibinfo {author} {\bibfnamefont {K.}~\bibnamefont {Rummukainen}}, \ and\
  \bibinfo {author} {\bibfnamefont {M.~E.}\ \bibnamefont {Shaposhnikov}},\
  }\href {\doibase 10.1016/0550-3213(96)00052-1} {\bibfield  {journal}
  {\bibinfo  {journal} {Nucl. Phys.}\ }\textbf {\bibinfo {volume} {B466}},\
  \bibinfo {pages} {189} (\bibinfo {year} {1996}{\natexlab{c}})},\ \Eprint
  {http://arxiv.org/abs/hep-lat/9510020} {arXiv:hep-lat/9510020 [hep-lat]}
  \BibitemShut {NoStop}%
\bibitem [{\citenamefont {Blaizot}\ \emph {et~al.}(2003)\citenamefont
  {Blaizot}, \citenamefont {Iancu},\ and\ \citenamefont
  {Rebhan}}]{Blaizot:2003iq}%
  \BibitemOpen
  \bibfield  {author} {\bibinfo {author} {\bibfnamefont {J.~P.}\ \bibnamefont
  {Blaizot}}, \bibinfo {author} {\bibfnamefont {E.}~\bibnamefont {Iancu}}, \
  and\ \bibinfo {author} {\bibfnamefont {A.}~\bibnamefont {Rebhan}},\ }\href
  {\doibase 10.1103/PhysRevD.68.025011} {\bibfield  {journal} {\bibinfo
  {journal} {Phys. Rev. D}\ }\textbf {\bibinfo {volume} {68}},\ \bibinfo
  {pages} {025011} (\bibinfo {year} {2003})},\ \Eprint
  {http://arxiv.org/abs/hep-ph/0303045} {arXiv:hep-ph/0303045} \BibitemShut
  {NoStop}%
\bibitem [{\citenamefont {Laine}\ and\ \citenamefont
  {Schr{\"o}der}(2006)}]{Laine:2006cp}%
  \BibitemOpen
  \bibfield  {author} {\bibinfo {author} {\bibfnamefont {M.}~\bibnamefont
  {Laine}}\ and\ \bibinfo {author} {\bibfnamefont {Y.}~\bibnamefont
  {Schr{\"o}der}},\ }\href {\doibase 10.1103/PhysRevD.73.085009} {\bibfield
  {journal} {\bibinfo  {journal} {Phys. Rev. D}\ }\textbf {\bibinfo {volume}
  {73}},\ \bibinfo {pages} {085009} (\bibinfo {year} {2006})},\ \Eprint
  {http://arxiv.org/abs/hep-ph/0603048} {arXiv:hep-ph/0603048} \BibitemShut
  {NoStop}%
\bibitem [{\citenamefont {Kajantie}\ \emph {et~al.}(1998)\citenamefont
  {Kajantie}, \citenamefont {Laine}, \citenamefont {Rummukainen},\ and\
  \citenamefont {Shaposhnikov}}]{Kajantie:1997ky}%
  \BibitemOpen
  \bibfield  {author} {\bibinfo {author} {\bibfnamefont {K.}~\bibnamefont
  {Kajantie}}, \bibinfo {author} {\bibfnamefont {M.}~\bibnamefont {Laine}},
  \bibinfo {author} {\bibfnamefont {K.}~\bibnamefont {Rummukainen}}, \ and\
  \bibinfo {author} {\bibfnamefont {M.~E.}\ \bibnamefont {Shaposhnikov}},\
  }\href {\doibase 10.1016/S0370-2693(97)01584-0} {\bibfield  {journal}
  {\bibinfo  {journal} {Phys. Lett. B}\ }\textbf {\bibinfo {volume} {423}},\
  \bibinfo {pages} {137} (\bibinfo {year} {1998})},\ \Eprint
  {http://arxiv.org/abs/hep-ph/9710538} {arXiv:hep-ph/9710538} \BibitemShut
  {NoStop}%
\bibitem [{\citenamefont {Collins}\ and\ \citenamefont
  {Vermaseren}(2016)}]{Collins:2016aya}%
  \BibitemOpen
  \bibfield  {author} {\bibinfo {author} {\bibfnamefont {J.~C.}\ \bibnamefont
  {Collins}}\ and\ \bibinfo {author} {\bibfnamefont {J.~A.~M.}\ \bibnamefont
  {Vermaseren}},\ }\href@noop {} {\  (\bibinfo {year} {2016})},\ \Eprint
  {http://arxiv.org/abs/1606.01177} {arXiv:1606.01177 [cs.OH]} \BibitemShut
  {NoStop}%
\bibitem [{\citenamefont {Gorda}\ \emph {et~al.}(2019)\citenamefont {Gorda},
  \citenamefont {Helset}, \citenamefont {Niemi}, \citenamefont {Tenkanen},\
  and\ \citenamefont {Weir}}]{Gorda:2018hvi}%
  \BibitemOpen
  \bibfield  {author} {\bibinfo {author} {\bibfnamefont {T.}~\bibnamefont
  {Gorda}}, \bibinfo {author} {\bibfnamefont {A.}~\bibnamefont {Helset}},
  \bibinfo {author} {\bibfnamefont {L.}~\bibnamefont {Niemi}}, \bibinfo
  {author} {\bibfnamefont {T.~V.~I.}\ \bibnamefont {Tenkanen}}, \ and\ \bibinfo
  {author} {\bibfnamefont {D.~J.}\ \bibnamefont {Weir}},\ }\href {\doibase
  10.1007/JHEP02(2019)081} {\bibfield  {journal} {\bibinfo  {journal} {JHEP}\
  }\textbf {\bibinfo {volume} {02}},\ \bibinfo {pages} {081} (\bibinfo {year}
  {2019})},\ \Eprint {http://arxiv.org/abs/1802.05056} {arXiv:1802.05056
  [hep-ph]} \BibitemShut {NoStop}%
\bibitem [{\citenamefont {Schicho}\ \emph {et~al.}(2021)\citenamefont
  {Schicho}, \citenamefont {Tenkanen},\ and\ \citenamefont
  {\"Osterman}}]{Schicho:2021gca}%
  \BibitemOpen
  \bibfield  {author} {\bibinfo {author} {\bibfnamefont {P.~M.}\ \bibnamefont
  {Schicho}}, \bibinfo {author} {\bibfnamefont {T.~V.~I.}\ \bibnamefont
  {Tenkanen}}, \ and\ \bibinfo {author} {\bibfnamefont {J.}~\bibnamefont
  {\"Osterman}},\ }\href {\doibase 10.1007/JHEP06(2021)130} {\bibfield
  {journal} {\bibinfo  {journal} {JHEP}\ }\textbf {\bibinfo {volume} {06}},\
  \bibinfo {pages} {130} (\bibinfo {year} {2021})},\ \Eprint
  {http://arxiv.org/abs/2102.11145} {arXiv:2102.11145 [hep-ph]} \BibitemShut
  {NoStop}%
\bibitem [{\citenamefont {Gould}(2021)}]{Gould:2021dzl}%
  \BibitemOpen
  \bibfield  {author} {\bibinfo {author} {\bibfnamefont {O.}~\bibnamefont
  {Gould}},\ }\href {\doibase 10.1007/JHEP04(2021)057} {\bibfield  {journal}
  {\bibinfo  {journal} {JHEP}\ }\textbf {\bibinfo {volume} {04}},\ \bibinfo
  {pages} {057} (\bibinfo {year} {2021})},\ \Eprint
  {http://arxiv.org/abs/2101.05528} {arXiv:2101.05528 [hep-ph]} \BibitemShut
  {NoStop}%
\bibitem [{\citenamefont {Arnold}\ and\ \citenamefont
  {Zhai}(1994)}]{Arnold:1994ps}%
  \BibitemOpen
  \bibfield  {author} {\bibinfo {author} {\bibfnamefont {P.~B.}\ \bibnamefont
  {Arnold}}\ and\ \bibinfo {author} {\bibfnamefont {C.-X.}\ \bibnamefont
  {Zhai}},\ }\href {\doibase 10.1103/PhysRevD.50.7603} {\bibfield  {journal}
  {\bibinfo  {journal} {Phys. Rev. D}\ }\textbf {\bibinfo {volume} {50}},\
  \bibinfo {pages} {7603} (\bibinfo {year} {1994})},\ \Eprint
  {http://arxiv.org/abs/hep-ph/9408276} {arXiv:hep-ph/9408276} \BibitemShut
  {NoStop}%
\bibitem [{\citenamefont {Croon}\ \emph {et~al.}(2021)\citenamefont {Croon},
  \citenamefont {Gould}, \citenamefont {Schicho}, \citenamefont {Tenkanen},\
  and\ \citenamefont {White}}]{Croon:2020cgk}%
  \BibitemOpen
  \bibfield  {author} {\bibinfo {author} {\bibfnamefont {D.}~\bibnamefont
  {Croon}}, \bibinfo {author} {\bibfnamefont {O.}~\bibnamefont {Gould}},
  \bibinfo {author} {\bibfnamefont {P.}~\bibnamefont {Schicho}}, \bibinfo
  {author} {\bibfnamefont {T.~V.~I.}\ \bibnamefont {Tenkanen}}, \ and\ \bibinfo
  {author} {\bibfnamefont {G.}~\bibnamefont {White}},\ }\href {\doibase
  10.1007/JHEP04(2021)055} {\bibfield  {journal} {\bibinfo  {journal} {JHEP}\
  }\textbf {\bibinfo {volume} {04}},\ \bibinfo {pages} {055} (\bibinfo {year}
  {2021})},\ \Eprint {http://arxiv.org/abs/2009.10080} {arXiv:2009.10080
  [hep-ph]} \BibitemShut {NoStop}%
\bibitem [{\citenamefont {Nishimura}\ and\ \citenamefont
  {Schroder}(2012)}]{Nishimura:2012ee}%
  \BibitemOpen
  \bibfield  {author} {\bibinfo {author} {\bibfnamefont {M.}~\bibnamefont
  {Nishimura}}\ and\ \bibinfo {author} {\bibfnamefont {Y.}~\bibnamefont
  {Schroder}},\ }\href {\doibase 10.1007/JHEP09(2012)051} {\bibfield  {journal}
  {\bibinfo  {journal} {JHEP}\ }\textbf {\bibinfo {volume} {09}},\ \bibinfo
  {pages} {051} (\bibinfo {year} {2012})},\ \Eprint
  {http://arxiv.org/abs/1207.4042} {arXiv:1207.4042 [hep-ph]} \BibitemShut
  {NoStop}%
\bibitem [{\citenamefont {Niemi}\ \emph {et~al.}(2019)\citenamefont {Niemi},
  \citenamefont {Patel}, \citenamefont {Ramsey-Musolf}, \citenamefont
  {Tenkanen},\ and\ \citenamefont {Weir}}]{Niemi:2018asa}%
  \BibitemOpen
  \bibfield  {author} {\bibinfo {author} {\bibfnamefont {L.}~\bibnamefont
  {Niemi}}, \bibinfo {author} {\bibfnamefont {H.~H.}\ \bibnamefont {Patel}},
  \bibinfo {author} {\bibfnamefont {M.~J.}\ \bibnamefont {Ramsey-Musolf}},
  \bibinfo {author} {\bibfnamefont {T.~V.~I.}\ \bibnamefont {Tenkanen}}, \ and\
  \bibinfo {author} {\bibfnamefont {D.~J.}\ \bibnamefont {Weir}},\ }\href
  {\doibase 10.1103/PhysRevD.100.035002} {\bibfield  {journal} {\bibinfo
  {journal} {Phys. Rev. D}\ }\textbf {\bibinfo {volume} {100}},\ \bibinfo
  {pages} {035002} (\bibinfo {year} {2019})},\ \Eprint
  {http://arxiv.org/abs/1802.10500} {arXiv:1802.10500 [hep-ph]} \BibitemShut
  {NoStop}%
\bibitem [{\citenamefont {Ruijl}\ \emph {et~al.}(2017)\citenamefont {Ruijl},
  \citenamefont {Ueda},\ and\ \citenamefont {Vermaseren}}]{Ruijl:2017dtg}%
  \BibitemOpen
  \bibfield  {author} {\bibinfo {author} {\bibfnamefont {B.}~\bibnamefont
  {Ruijl}}, \bibinfo {author} {\bibfnamefont {T.}~\bibnamefont {Ueda}}, \ and\
  \bibinfo {author} {\bibfnamefont {J.}~\bibnamefont {Vermaseren}},\
  }\href@noop {} {\bibfield  {journal} {\bibinfo  {journal} {{}}\ } (\bibinfo
  {year} {2017})},\ \Eprint {http://arxiv.org/abs/1707.06453} {arXiv:1707.06453
  [hep-ph]} \BibitemShut {NoStop}%
\bibitem [{\citenamefont {Laporta}(2000)}]{Laporta:2001dd}%
  \BibitemOpen
  \bibfield  {author} {\bibinfo {author} {\bibfnamefont {S.}~\bibnamefont
  {Laporta}},\ }\href {\doibase 10.1016/S0217-751X(00)00215-7} {\bibfield
  {journal} {\bibinfo  {journal} {Int. J. Mod. Phys.\ A}\ }\textbf {\bibinfo
  {volume} {15}},\ \bibinfo {pages} {5087} (\bibinfo {year} {2000})},\ \Eprint
  {http://arxiv.org/abs/hep-ph/0102033} {arXiv:hep-ph/0102033} \BibitemShut
  {NoStop}%
\bibitem [{\citenamefont {Fukuda}\ and\ \citenamefont
  {Kugo}(1976)}]{Fukuda:1975di}%
  \BibitemOpen
  \bibfield  {author} {\bibinfo {author} {\bibfnamefont {R.}~\bibnamefont
  {Fukuda}}\ and\ \bibinfo {author} {\bibfnamefont {T.}~\bibnamefont {Kugo}},\
  }\href {\doibase 10.1103/PhysRevD.13.3469} {\bibfield  {journal} {\bibinfo
  {journal} {Phys. Rev. D}\ }\textbf {\bibinfo {volume} {13}},\ \bibinfo
  {pages} {3469} (\bibinfo {year} {1976})}\BibitemShut {NoStop}%
\bibitem [{\citenamefont {Patel}\ and\ \citenamefont
  {Ramsey-Musolf}(2011)}]{Patel:2011th}%
  \BibitemOpen
  \bibfield  {author} {\bibinfo {author} {\bibfnamefont {H.~H.}\ \bibnamefont
  {Patel}}\ and\ \bibinfo {author} {\bibfnamefont {M.~J.}\ \bibnamefont
  {Ramsey-Musolf}},\ }\href {\doibase 10.1007/JHEP07(2011)029} {\bibfield
  {journal} {\bibinfo  {journal} {JHEP}\ }\textbf {\bibinfo {volume} {07}},\
  \bibinfo {pages} {029} (\bibinfo {year} {2011})},\ \Eprint
  {http://arxiv.org/abs/1101.4665} {arXiv:1101.4665 [hep-ph]} \BibitemShut
  {NoStop}%
\bibitem [{\citenamefont {Laine}(1995)}]{Laine:1994zq}%
  \BibitemOpen
  \bibfield  {author} {\bibinfo {author} {\bibfnamefont {M.}~\bibnamefont
  {Laine}},\ }\href {\doibase 10.1103/PhysRevD.51.4525} {\bibfield  {journal}
  {\bibinfo  {journal} {Phys. Rev.}\ }\textbf {\bibinfo {volume} {D51}},\
  \bibinfo {pages} {4525} (\bibinfo {year} {1995})},\ \Eprint
  {http://arxiv.org/abs/hep-ph/9411252} {arXiv:hep-ph/9411252 [hep-ph]}
  \BibitemShut {NoStop}%
\bibitem [{\citenamefont {Laine}(1994)}]{Laine:1994bf}%
  \BibitemOpen
  \bibfield  {author} {\bibinfo {author} {\bibfnamefont {M.}~\bibnamefont
  {Laine}},\ }\href {\doibase 10.1016/0370-2693(94)91409-5} {\bibfield
  {journal} {\bibinfo  {journal} {Phys. Lett. B}\ }\textbf {\bibinfo {volume}
  {335}},\ \bibinfo {pages} {173} (\bibinfo {year} {1994})},\ \Eprint
  {http://arxiv.org/abs/hep-ph/9406268} {arXiv:hep-ph/9406268} \BibitemShut
  {NoStop}%
\bibitem [{\citenamefont {Martin}\ and\ \citenamefont
  {Patel}(2018)}]{Martin:2018emo}%
  \BibitemOpen
  \bibfield  {author} {\bibinfo {author} {\bibfnamefont {S.~P.}\ \bibnamefont
  {Martin}}\ and\ \bibinfo {author} {\bibfnamefont {H.~H.}\ \bibnamefont
  {Patel}},\ }\href {\doibase 10.1103/PhysRevD.98.076008} {\bibfield  {journal}
  {\bibinfo  {journal} {Phys. Rev. D}\ }\textbf {\bibinfo {volume} {98}},\
  \bibinfo {pages} {076008} (\bibinfo {year} {2018})},\ \Eprint
  {http://arxiv.org/abs/1808.07615} {arXiv:1808.07615 [hep-ph]} \BibitemShut
  {NoStop}%
\bibitem [{\citenamefont {Weinberg}\ and\ \citenamefont
  {Wu}(1987)}]{Weinberg:1987vp}%
  \BibitemOpen
  \bibfield  {author} {\bibinfo {author} {\bibfnamefont {E.~J.}\ \bibnamefont
  {Weinberg}}\ and\ \bibinfo {author} {\bibfnamefont {A.-q.}\ \bibnamefont
  {Wu}},\ }\href {\doibase 10.1103/PhysRevD.36.2474} {\bibfield  {journal}
  {\bibinfo  {journal} {Phys. Rev. D}\ }\textbf {\bibinfo {volume} {36}},\
  \bibinfo {pages} {2474} (\bibinfo {year} {1987})}\BibitemShut {NoStop}%
\bibitem [{\citenamefont {Gould}\ and\ \citenamefont
  {Tenkanen}(2021)}]{Gould:2021oba}%
  \BibitemOpen
  \bibfield  {author} {\bibinfo {author} {\bibfnamefont {O.}~\bibnamefont
  {Gould}}\ and\ \bibinfo {author} {\bibfnamefont {T.~V.~I.}\ \bibnamefont
  {Tenkanen}},\ }\href {\doibase 10.1007/JHEP06(2021)069} {\bibfield  {journal}
  {\bibinfo  {journal} {JHEP}\ }\textbf {\bibinfo {volume} {06}},\ \bibinfo
  {pages} {069} (\bibinfo {year} {2021})},\ \Eprint
  {http://arxiv.org/abs/2104.04399} {arXiv:2104.04399 [hep-ph]} \BibitemShut
  {NoStop}%
\bibitem [{\citenamefont {Niemi}\ \emph {et~al.}(2024)\citenamefont {Niemi},
  \citenamefont {Schicho},\ and\ \citenamefont
  {Tenkanen}}]{PhysRevD.109.039902}%
  \BibitemOpen
  \bibfield  {author} {\bibinfo {author} {\bibfnamefont {L.}~\bibnamefont
  {Niemi}}, \bibinfo {author} {\bibfnamefont {P.}~\bibnamefont {Schicho}}, \
  and\ \bibinfo {author} {\bibfnamefont {T.~V.~I.}\ \bibnamefont {Tenkanen}},\
  }\href {\doibase 10.1103/PhysRevD.109.039902} {\bibfield  {journal} {\bibinfo
   {journal} {Phys. Rev. D}\ }\textbf {\bibinfo {volume} {109}},\ \bibinfo
  {pages} {039902} (\bibinfo {year} {2024})}\BibitemShut {NoStop}%
\bibitem [{\citenamefont {L\'opez-Val}\ and\ \citenamefont
  {Robens}(2014)}]{Lopez-Val:2014jva}%
  \BibitemOpen
  \bibfield  {author} {\bibinfo {author} {\bibfnamefont {D.}~\bibnamefont
  {L\'opez-Val}}\ and\ \bibinfo {author} {\bibfnamefont {T.}~\bibnamefont
  {Robens}},\ }\href {\doibase 10.1103/PhysRevD.90.114018} {\bibfield
  {journal} {\bibinfo  {journal} {Phys. Rev. D}\ }\textbf {\bibinfo {volume}
  {90}},\ \bibinfo {pages} {114018} (\bibinfo {year} {2014})},\ \Eprint
  {http://arxiv.org/abs/1406.1043} {arXiv:1406.1043 [hep-ph]} \BibitemShut
  {NoStop}%
\bibitem [{\citenamefont {Bojarski}\ \emph {et~al.}(2016)\citenamefont
  {Bojarski}, \citenamefont {Chalons}, \citenamefont {Lopez-Val},\ and\
  \citenamefont {Robens}}]{Bojarski:2015kra}%
  \BibitemOpen
  \bibfield  {author} {\bibinfo {author} {\bibfnamefont {F.}~\bibnamefont
  {Bojarski}}, \bibinfo {author} {\bibfnamefont {G.}~\bibnamefont {Chalons}},
  \bibinfo {author} {\bibfnamefont {D.}~\bibnamefont {Lopez-Val}}, \ and\
  \bibinfo {author} {\bibfnamefont {T.}~\bibnamefont {Robens}},\ }\href
  {\doibase 10.1007/JHEP02(2016)147} {\bibfield  {journal} {\bibinfo  {journal}
  {JHEP}\ }\textbf {\bibinfo {volume} {02}},\ \bibinfo {pages} {147} (\bibinfo
  {year} {2016})},\ \Eprint {http://arxiv.org/abs/1511.08120} {arXiv:1511.08120
  [hep-ph]} \BibitemShut {NoStop}%
\bibitem [{\citenamefont {Kanemura}\ \emph {et~al.}(2016)\citenamefont
  {Kanemura}, \citenamefont {Kikuchi},\ and\ \citenamefont
  {Yagyu}}]{Kanemura:2015fra}%
  \BibitemOpen
  \bibfield  {author} {\bibinfo {author} {\bibfnamefont {S.}~\bibnamefont
  {Kanemura}}, \bibinfo {author} {\bibfnamefont {M.}~\bibnamefont {Kikuchi}}, \
  and\ \bibinfo {author} {\bibfnamefont {K.}~\bibnamefont {Yagyu}},\ }\href
  {\doibase 10.1016/j.nuclphysb.2016.04.005} {\bibfield  {journal} {\bibinfo
  {journal} {Nucl. Phys. B}\ }\textbf {\bibinfo {volume} {907}},\ \bibinfo
  {pages} {286} (\bibinfo {year} {2016})},\ \Eprint
  {http://arxiv.org/abs/1511.06211} {arXiv:1511.06211 [hep-ph]} \BibitemShut
  {NoStop}%
\bibitem [{\citenamefont {Brauner}\ \emph {et~al.}(2017)\citenamefont
  {Brauner}, \citenamefont {Tenkanen}, \citenamefont {Tranberg}, \citenamefont
  {Vuorinen},\ and\ \citenamefont {Weir}}]{Brauner:2016fla}%
  \BibitemOpen
  \bibfield  {author} {\bibinfo {author} {\bibfnamefont {T.}~\bibnamefont
  {Brauner}}, \bibinfo {author} {\bibfnamefont {T.~V.~I.}\ \bibnamefont
  {Tenkanen}}, \bibinfo {author} {\bibfnamefont {A.}~\bibnamefont {Tranberg}},
  \bibinfo {author} {\bibfnamefont {A.}~\bibnamefont {Vuorinen}}, \ and\
  \bibinfo {author} {\bibfnamefont {D.~J.}\ \bibnamefont {Weir}},\ }\href
  {\doibase 10.1007/JHEP03(2017)007} {\bibfield  {journal} {\bibinfo  {journal}
  {JHEP}\ }\textbf {\bibinfo {volume} {03}},\ \bibinfo {pages} {007} (\bibinfo
  {year} {2017})},\ \Eprint {http://arxiv.org/abs/1609.06230} {arXiv:1609.06230
  [hep-ph]} \BibitemShut {NoStop}%
\bibitem [{\citenamefont {Zyla}\ \emph {et~al.}(2020)\citenamefont {Zyla} \emph
  {et~al.}}]{Zyla:2020zbs}%
  \BibitemOpen
  \bibfield  {author} {\bibinfo {author} {\bibfnamefont {P.}~\bibnamefont
  {Zyla}} \emph {et~al.} (\bibinfo {collaboration} {Particle Data Group}),\
  }\href {\doibase 10.1093/ptep/ptaa104} {\bibfield  {journal} {\bibinfo
  {journal} {PTEP}\ }\textbf {\bibinfo {volume} {2020}},\ \bibinfo {pages}
  {083C01} (\bibinfo {year} {2020})}\BibitemShut {NoStop}%
\bibitem [{\citenamefont {Sirlin}(1980)}]{Sirlin:1980nh}%
  \BibitemOpen
  \bibfield  {author} {\bibinfo {author} {\bibfnamefont {A.}~\bibnamefont
  {Sirlin}},\ }\href {\doibase 10.1103/PhysRevD.22.971} {\bibfield  {journal}
  {\bibinfo  {journal} {Phys. Rev. D}\ }\textbf {\bibinfo {volume} {22}},\
  \bibinfo {pages} {971} (\bibinfo {year} {1980})}\BibitemShut {NoStop}%
\bibitem [{\citenamefont {Bohm}\ \emph {et~al.}(1986)\citenamefont {Bohm},
  \citenamefont {Spiesberger},\ and\ \citenamefont {Hollik}}]{Bohm:1986rj}%
  \BibitemOpen
  \bibfield  {author} {\bibinfo {author} {\bibfnamefont {M.}~\bibnamefont
  {Bohm}}, \bibinfo {author} {\bibfnamefont {H.}~\bibnamefont {Spiesberger}}, \
  and\ \bibinfo {author} {\bibfnamefont {W.}~\bibnamefont {Hollik}},\ }\href
  {\doibase 10.1002/prop.19860341102} {\bibfield  {journal} {\bibinfo
  {journal} {Fortsch. Phys.}\ }\textbf {\bibinfo {volume} {34}},\ \bibinfo
  {pages} {687} (\bibinfo {year} {1986})}\BibitemShut {NoStop}%
\end{thebibliography}%

\end{document}